\documentclass[12pt]{elsart}
\usepackage{graphicx}
\usepackage{amssymb}
\usepackage{amsmath}
\usepackage{amsfonts}
\usepackage{graphics}
\usepackage[breaklinks,bookmarks,colorlinks,linkcolor=darkblue,citecolor=darkblue]{hyperref}
\usepackage{color}
\usepackage{xcolor}
\usepackage{ragged2e}
\usepackage{epstopdf}
\usepackage{natbib}
\usepackage{bm}
\usepackage{txfonts} 

\definecolor{darkblue}{rgb}{0,0,.8}

\journal{Advances in Atomic, Molecular and Optical Physics}

\begin{document}

\begin{frontmatter}

\title{Excitons and Cavity Polaritons for Optical Lattice Ultracold Atoms}

\author{Hashem Zoubi and Helmut Ritsch}

\address{Institut f\"{u}r Theoretische Physik, Universit\"{a}t Innsbruck, Technikerstrasse 25, A-6020 Innsbruck, Austria}

\tableofcontents

\begin{abstract}
Ultracold atoms uniformly filling an optical lattice can be treated like an artificial crystal. An implementation including the atomic occupation of a single excited atomic state can be represented by a two-component Bose-Hubbard model. Its phase diagram at zero temperature exhibits a quantum phase transition from a superfluid to a Mott insulator phase. The dynamics of electronic excitations governed by electrostatic dipole-dipole interactions in the ordered regime can be well described by wave-like collective excitations called excitons. Here we present an extensive study of such excitons for a wide range of optical lattice geometries and of different dimensionality including boundary effects in finite lattices. As they are coupled to the free space radiation field, their decay depends on the lattice geometry, polarization and lattice constant. Their lifetimes can vary over many orders of magnitude from metastable propagation to superradiant decay. Particularly strong effects occur in one dimensional atomic chains coupled to tapered optical fibers where free space emission can be completely suppressed and only a superradiant interaction with the fiber modes takes place.  We show that coherent transfer of excitons among spatially separated optical lattices can be controlled and represents a promising candidate for quantum information transfer. For an optical lattice within a cavity the excitons are coupled to cavity photons and the resulting collective cavity QED model can be efficiently formulated in terms of polaritons as elementary excitations.  Their properties are explicitly calculated for different lattices and they constitute a non-destructive monitoring tool for important system properties as e.g. the atomic quantum statistics. Even the formation of bound states and molecules in optical lattices manifests itself in modified polariton properties as e.g. an anisotropic optical spectrum. Partial dissipation of the exciton energy in the lattice leads to heating, which can be microscopically understood through a mechanism transferring atoms into higher Bloch bands via a resonant excitation transfer among neighboring lattice sites. The presence of lattice defects like vacancies in the Mott insulator induces a characteristic scattering of polaritons, which can be optically observed to monitor the lattice integrity. Our models can be applied to simulate and understand corresponding collective phenomena in solid crystals, where many effects are often masked by noise and disorder.
\end{abstract}

\begin{keyword}
Ultracold Atoms \sep Optical Lattices \sep Quantum Phase Transitions \sep
Bose-Hubbard Model \sep Mott Insulator Phase \sep Excitons \sep Dark States
\sep Superradiance \sep Cavity QED \sep Cavity Polaritons \sep Tapered
Nanofibers

\PACS 37.10.JK \sep 42.50.-p \sep 71.35.-y \sep 71.36.+c

H Zoubi moved recently to: {\it Max-Planck Institute for the Physics of Complex
Systems, Noethnitzer Str. 38, 01187 Dresden, Germany.}

\end{keyword}

\end{frontmatter}

\section{Introduction}

Trapping and manipulating ultracold neutral atoms in an array of optical potentials opened a vast new research area in the field of quantum fluids and solid state modeling \citep{Bloch2008}. Demonstrating the superfluid to Mott insulator quantum phase transition of interacting bosons in an optical lattice laid the foundation to study and simulate strongly correlated quantum many-body systems with a big impact on condensed matter physics \citep{Lewenstein2007}. Ultracold atoms in optical lattices are now one of the most flourishing directions of experimental and theoretical quantum physics \citep{Jaksch1998,Greiner2002,Morsch2006}. Besides fundamental physics issues on entanglement, measurement and decoherence \citep{Mekhov2012}, more and more applications appear for the quantum simulations for a wide range of puzzling effects in solid state physics.

A BEC of a dilute bosons, in which a very large number of atoms occupy the same quantum state of motion, was experimentally realized by several groups \citep{Dalfovo1999}. Optical control of the quantum state of such BEC opened the door to a new generation of experiments \citep{Metcalf1999}. In particular the loading of the BEC into a standing wave formed by pairs of counter propagating far detuned laser beams creates a synthetic controllable periodic lattice \citep{Greiner2002,Spielman2007}. Trapped atoms still can tunnel from site to site, while interacting and repelling each other when occupying the same site. The stronger the laser field, the deeper the lattice and the slower the hopping rate of the atoms, while at the same time the on-site interaction between the atoms becomes stronger. Thus the laser light allows to switch between a weakly interacting gas, where atoms form a superfluid, to a strongly correlated regime, where atoms avoid each other and localize into individual sites forming a perfectly regular array called Mott insulator phase.  As this crossover happens at $T=0$ as a function of an external parameter, it is called a quantum phase transition \citep{Fisher1989,Sheshadri1993,Sachdev1999}.  It is theoretically well described by a Bose-Hubbard model \citep{Jaksch1998} and soon was experimentally demonstrated \citep{Greiner2002}. An increasing degree of control of experimental parameters in optical lattices has been demonstrated up to the level of addressing and controlling single atoms at a specific lattice site \citep{Sherson2010,Bakr2010,Weitenberg2011}.

In a parallel development the dynamics of the electromagnetic field modes in a high-Q optical cavity resonantly coupled to quantized excitations of atoms has been extensively  studied. Many phenomena discussed  as Gedanken-experiments to exhibit the quantum physics of coupled systems and measurements were actually performed experimentally and lead to the recent 2012 physics Nobel prize \citep{Haroche2006,Walther2006}. The first experimental realizations of cavity QED used single Rydberg atoms in superconducting microwave resonators. Later ground state alkali atoms in high-Q optical resonators allowed to reach the strong coupling limit as well, where light matter coupling dominates decoherence \citep{Ye1999,Pinkse2000}. Other implementations using solid state technology followed more recently \citep{Kavokin2003}.

It is of course a natural development to combine these two threads and investigate ultracold atomic lattices within optical resonators \citep{Maschler2005,Zoubi2007,Mekhov2007,Larson2008,Ritter2009}. As a major experimental step in this direction, several groups \citep{Brennecke2007,Colombe2007,Slama2007} achieved operations of a BEC of trapped ultracold atoms within an optical high-Q cavity deeply in the strong resonant coupling regime. It seems thus that no major technical obstacles prevent the implementation of an optical lattice within an optical cavity. An alternative route for realizing an optical lattice within a cavity is based on a co-planar waveguide microwave resonator resonantly coupled to the collective hyperfine state of a BEC of ultracold atoms in a wire surface trap fabricated on a superconducting atom chip \citep{Folman2002}. This system is predicted to allow the experimental study of magnetic dipole interactions in the strong coupling regime \citep{Verdu2009,Imamoglu2009}.

Placing ultracold atoms within optical cavities leads to a wealth of interesting dynamical phenomena already in the dispersive regime, where the light frequency is far detuned from the atomic resonance. This involves a dynamic interplay of light forces, non-local couplings and field dynamics \citep{Ritsch2012}. In this review instead we concentrate on resonant interaction of quantized light modes and trapped atoms in the Mott phase, which we treat as a new type of matter closely related to an atomic crystal with widely controllable lattice geometry \citep{Zoubi2007}.  While in some general cases we consider coupling of the light to motional degrees of freedom as well, our main emphasis is on collective internal excitations dynamics. This reduced description already allows to study many solid state effects in a very generic form. The wide and precise controllability of different lattices and coupling parameters allows to change the number of atoms per site, the geometry, and the symmetry of the system. These facts give the present system a decisive advantage over conventional solid state crystals where parameters are hardly tunable and particular effects cannot be easily singled out.  Such toy implementations  provide to us a more detailed and deeper understanding of solid state effects.  Mathematically one extends the common Bose-Hubbard atom model to include atoms in excited states formally described as a second kind of bosons within the optical lattice. The system is then represented by a two-component Bose-Hubbard model exhibiting a much richer phase diagram \citep{Zoubi2009d}.

Optical lattices allow us to go beyond the weakly interacting regime of Bose gas into strongly correlated systems of solid state physics \citep{Bloch2008,Lewenstein2007}.  Including the internal atomic level structure bears a strong analogy to the excitonic dynamics of molecular crystals, described by ``Frenkel excitons'' \citep{Agranovich2009,Davydov1971}. The electronic excitations delocalize in the lattice due to electrostatic interactions, e.g. resonant dipole-dipole interactions.  By exploiting the lattice symmetry one is lead to introduce wave-like collective electronic excitations called ``excitons'' as fundamental system excitations \citep{Zoubi2005a}. For this expansion to make sense, the transfer time has to be faster than the exciton lifetime, so that collective states of electronic excitations dominate the electrical and optical system properties. Excitons thus provide an ideal basis to describe resonant effects in optical lattices. In fact the atomic recoil heating due to emission and absorption of a photon is significantly suppressed as the momentum is shared among all atoms in the lattice and strongly suppress diffusion and heating in the Mott insulator. We present studies of the properties and dynamics of excitons for several generic lattice types of different geometries and dimensionality, starting from finite chains, infinite chains to stacks of two dimensional planar lattices \citep{Zoubi2007,Zoubi2009c,Zoubi2010b,Zoubi2011a,Zoubi2011b}.  Boundary conditions of finite systems give rise to standing wave excitons, while extended systems produce propagating wave solutions. As a special case a Mott insulator with two atoms per site can also possess dark on-site localized excitons decoupled from the remaining lattice sites \citep{Zoubi2008b}.

The energy delocalization of excitons among many particles amounts to the occurrence of intrinsic entanglement. To generate and preserve significant entanglement requires sufficient isolation from the environment in order to bring decoherence and dissipation to a minimum, which is a big challenge for optically excited systems.  Here spontaneous emission due to the coupling to free space radiation is a major source of dissipation and decoherence \citep{Loudon2000}. Excitons in multi-atom systems exhibit significantly modified spontaneous emission compared to a single atom \citep{Ficek2002}, which we have extensively studied for different geometries and dimensions \citep{Zoubi2010b,Zoubi2011a,Zoubi2012a,Ostermann2012}. While often excitons exhibit enhanced coupling to excitation lasers and vacuum modes and fast or superradiant decay, in many cases one can get long lived or even  dark and metastable excitations, which can only decay optically into a narrow set of wave-vectors analogous to Bragg angles. If the lattice spacing is smaller than the transition wavelength, the exciton wave vector can be larger than the free space photon wave-vector, so that no optical decay is possible. Nevertheless one can still engineer coherent transfer of such excitons among different lattice regions or different optical lattice planes \citep{Zoubi2010c,Zoubi2011b,Zoubi2012b}. This effect should have important applications in optical information storage and transfer based on large ensembles.  Interestingly, while the individual dipole-dipole energy transfer has an inverse cubic dependence on the distance, collective states within one lattice plane only couple with an exponentially decreasing strength to neighboring planes, which generally allows simplification to a nearest neighbor coupling model.

Tailored light-matter interaction is of great importance to fundamental physics tests as well as high end technical applications \citep{Cohen1992,Metcalf1999,Meystre2001} and in particular for the development of quantum information technologies \citep{Pinkse2000,Ye1999}. New frontiers of cavity QED involving ultracold atoms trapped in intra-cavity optical lattices form a natural, but only very recently opened, research direction \citep{Zoubi2007,Zoubi2008b,Zoubi2009c}. As above, we concentrate here on atoms with an internal electronic transition close to resonance with a cavity mode, while the optical lattice is generated by external classical fields which are off-resonance to any of the internal atomic transitions forming the dynamical part of the system. The case of cavity generated optical potentials also exhibits a wealth of interesting phenomena \citep{Ritsch2012}. Here we generalize the two-component Bose-Hubbard model to include cavity photons and their coupling to electronic excitations of the cold atoms \citep{Zoubi2009d}. The phase diagram including superfluid and Mott regions is affected by the coupling to cavity photons which mediate long range interactions and enhance tunneling via field fluctuations. Even though using a rather simple model we see that the quantum phase transition to a perfectly ordered Mott state occurs at much deeper optical lattices than in free space.

In cavities, excitons and photons are coherently coupled to form new quasi-particles called polaritons \citep{Agranovich2009,Kavokin2003,Zoubi2005a}. As the cavity mirrors exhibit a finite transition probability, the detection of the emitted light provides us with detailed information about the system properties \citep{Zoubi2007}. The corresponding optical spectra give a largely nondestructive observation tool for different quantum phases with minimal back-action \citep{Mekhov2012}. As an additional example the formation of molecules in the optical lattice introduces optical anisotropy in our artificial crystal. This manifests itself in the spectra through the cavity photon polarization mixing \citep{Zoubi2009b}. Again our studies range from finite lattices inside optical cavities of spherical mirrors up to two dimensional lattices between planar cavity mirrors. As an interesting property for the search of lower temperatures, the elastic scattering of excitons and polaritons by the defects or vacancies in the Mott insulator phase can be observed via cavity transmission spectra without destroying the lattice state as it occurs in high aperture microscope detection via fluorescence \citep{Zoubi2008a}. Similarly atoms excited to higher Bloch bands change the dynamics of excitons and polaritons \citep{Zoubi2009a,Zoubi2010a}.

Optical properties of a gas of ultracold atoms with coherently dressed two-level and three-level systems in a Mott insulator phase of an optical lattice have been studied by \citet{Bariani2008,Carusotto2008}. Cooperative effects of ultracold molecules in optical lattices are also treated \citep{Kuznetsova2012}. Recently Rydberg atoms become of large interest for optical lattices. Rydberg excitations in Bose-Einstein condensates of Rubidium atoms loaded into quasi-one-dimensional traps and in optical lattices experimentally have been realized \citep{Viteau2011,Anderson2011}. Moreover, the formation of excitons in a regular and flexible chain of Rydberg atoms as introduced through the interplay between excitonic and atomic motion was proposed theoretically by \citet{Wuster2010}.

Free space optical lattices have been formed in many geometries from one dimensional to three dimensional. In a parallel development also evanescent fields have been used to trap atoms and create new lattice geometries. In a particularly interesting and potentially important example,  tapered optical fibers were used to trap neutral atoms \citep{Bhagwat2008,Christensen2008,Bajcsy2009}. These are coupled to the propagating fiber modes via evanescent fields surrounding the fiber \citep{Nayak2007}. Such a hybrid combination of atoms and a solid state devices allows trapping and, simultaneously,  optically interfacing \citep{Hammerer2010}. Fibers with diameters smaller than the wavelength exhibit strong transverse confinement and a pronounced evanescent field, which by using two colors can be designed to create an array of deep optical micro-traps. It was recently realized with Cesium atoms, where it was shown that the atoms can be efficiently trapped and interrogated with a resonant light field sent through the nanofiber \citep{Vetsch2010,Goban2012}. This setup provides an ideal test system to study one dimensional excitons, for which we developed a microscopic theory. The strong collective coupling between atoms and field modes naturally leads to the formations polaritons in this setup \citep{Zoubi2010c}. As one of the most interesting consequences we could show that one can create polaritons with a strong fiber mode contributions but with virtually no coupling to the free space radiation modes. this should constitute an ideal photon-atom ensemble interface for coherent quantum information storage and processing.

The review is organized as follows. In section 2 we extend the Bose-Hubbard model to exploit electronically excited atoms. In section 3 we introduce excitons into a system of ultracold atoms in optical lattices. We discuss their formation, dispersion and life times for different geometries and dimensionality. Section 4 is opened by discussing the quantum phase transition within a cavity followed by introducing cavity polaritons in the strong coupling regime, where various possible physical realizations are studied. Defects of atoms excited into higher Bloch bands or vacancies in the Mott insulator phase are treated in section 5. The conclusions are contained in section 6.

\section{Ultracold Atoms in an Optical Lattice as Artificial Crystals}

\subsection{Superfluid to Mott-Insulator Transitions}

Optical lattices created by off resonance laser fields constitute a regular array of micro-traps into which a dilute gas of ultracold atoms can be loaded \citep{Greiner2002,Bloch2008}. The atoms experience an optical lattice potential with lattice constant of half the wave length of the lattice laser with different depth for each internal atomic state.  At ultra low temperatures, $T\approx 0$, atoms are localized at the lowest Bloch band of the lattice. In such structures, atoms can hop among different sites through quantum tunneling and interact locally via scattering. The tunneling process is mainly among nearest neighbor sites with the hopping parameter $J$, and the atom-atom interactions are efficient only for atoms at the same lattice site with an effective strength $U$,  which is taken to be a repulsive determined by the $s$-wave scattering length. This system is well described by a Bose-Hubbard model \citep{Jaksch1998}  represented by the Hamiltonian
\begin{equation}
H=-J\sum_{\langle i,j\rangle}b_i^{\dagger}b_j+\sum_i(\varepsilon_i-\mu)\ b_i^{\dagger}b_i+\frac{U}{2}\sum_ib_i^{\dagger}b_i^{\dagger}b_ib_i,
\end{equation}
where $b_i^{\dagger}$ and $b_i$ are the creation and annihilation operators of an atom at site $i$, and $\varepsilon_i$ is the atom on-site energy, which includes the external trap potential, $\mu$ is the chemical potential accounting for a possible exchange of atoms with an external atomic reservoir. The $i$ summation is over the lattice sites, and the bracket indicates a possible hopping only between nearest neighbor sites.

\begin{figure}
\begin{center}
\leavevmode
\includegraphics[width=85mm]{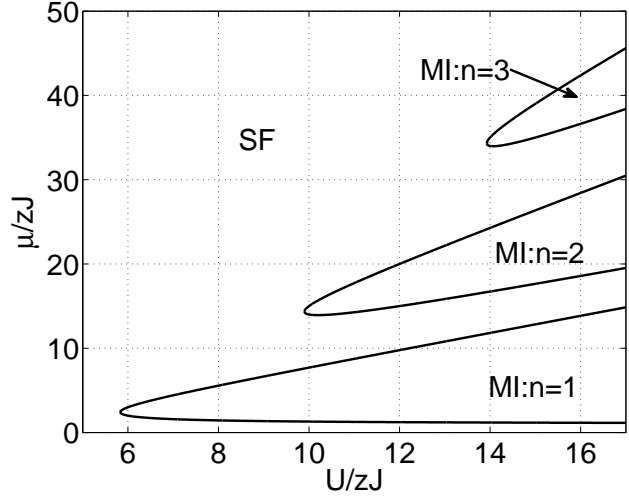}
\caption{Phase diagram: $(\mu/zJ)$ vs. $(U/zJ)$. The Superfluid SF and the three Mott insulator MI regions, for $n=1,2,3$, are shown. Figure reprinted with permission from~\citet{Zoubi2009d} Copyright 2009 by American Physical Society.}
\label{1}
\end{center}
\end{figure}

The Bose-Hubbard model predicts a quantum phase transition between the superfluid and the Mott insulator phase by changing the external laser field intensity, that is changing the kinetic versus the repulsion energy of the quantum gas. In the limit of $J\gg U$ the atom hopping among the lattice sites dominates the dynamics and the atomic states are spread out over the whole lattice and one gets a superfluid. In the opposite limit $J\ll U$, the on-site atom-atom interaction dominates and the atoms localize deterministically at individual sites, so that we have a Mott insulator.  The corresponding phase diagram can be calculated in the mean-field approach \citep{Sheshadri1993,Sachdev1999,vanOosten2001}. In Fig.~\ref{1} the phase diagram is plotted in the plane $(\mu/zJ)-(U/zJ)$ which is scaled by $zJ$, where $z$ is the number of nearest neighbors. The Superfluid (SF) phase appears outside the shaded three Mott Insulator (MI) phase regions, which are for one, two, and three atoms per site, that is $n=1,2,3$, where we assumed $\varepsilon_i=0$. In the mean field approximation the systems dimensionality enters only through $z$, where in a hyper-cubic lattice for the one dimension $d=1$ we have $z=2$, for $d=2$ we have $z=4$, and for $d=3$ we have $z=6$.

\subsection{Mott Insulator for a Two-Component Bose-Hubbard Model}

\begin{figure}
\begin{center}
\leavevmode
\includegraphics[width=85mm]{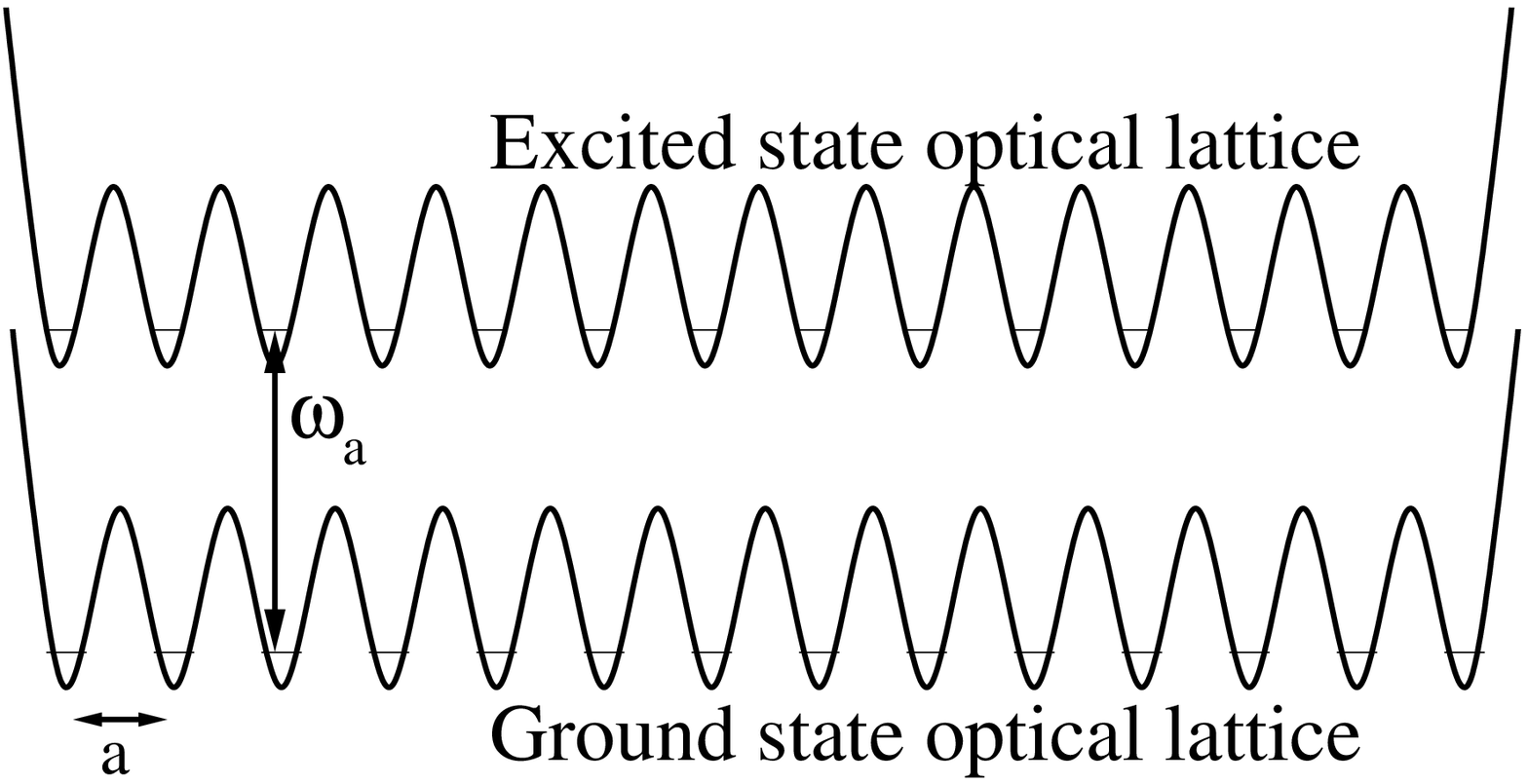}
\caption{Schematic plot of the ground and excited state optical lattice potentials.}
\label{2}
\end{center}
\end{figure}

In the present review we study collective electronic excitations of an atomic Mott insulator. Hence we have to introduce an excited atom state, which modifies the above phase diagram as we now have two-level atoms treated here as two kinds of atoms, namely ground state atoms and atoms excited to a fixed internal state. When loaded on an optical lattice due to different polarizability and Stark shift of the ground and excited states each group of atoms experiences a different optical lattice potential, but with ground and excited potentials showing extrema at the same positions, as schematized in Fig.~\ref{2}. For certain, magic, laser wave lengths, light shift  yield nevertheless identical potentials. This is important e.g. to create an optical lattice clock \citep{Takamoto2005}. Close to $T=0$ the ground and the excited atoms will again occupy only their lowest Bloch band.

We consider the two groups of atoms with different internal states, hyperfine or electronic, as two kinds of bosons. Therefore, we generalize the single-component Bose-Hubbard model into two-component Bose Hubbard model \citep{Jaksch1998,Chen2003,Zoubi2009d}, with the Hamiltonian
\begin{eqnarray}
H&=&-J_g\sum_{\langle i,j\rangle}b_i^{\dagger}b_j-J_e\sum_{\langle i,j\rangle}c_i^{\dagger}c_j+\sum_i(\varepsilon_i^g-\mu_g)\ b_i^{\dagger}b_i+\sum_i(\varepsilon_i^e-\mu_e)\ c_i^{\dagger}c_i \nonumber \\
&+&\frac{U_g}{2}\sum_ib_i^{\dagger}b_i^{\dagger}b_ib_i+\frac{U_e}{2}\sum_ic_i^{\dagger}c_i^{\dagger}c_ic_i+U_{eg}\sum_ib_i^{\dagger}c_i^{\dagger}b_ic_i,
\end{eqnarray}
where $b_i^{\dagger}$ and $b_i$ are the creation and annihilation operators of a ground state atom at site $i$, respectively, with the on-site energy $\varepsilon_i^g$, and chemical potential $\mu_g$; $c_i^{\dagger}$ and $c_i$ are the creation and annihilation operators of an excited state atom at site $i$, respectively, with on-site energy $\varepsilon_i^e$, and chemical potential $\mu_e$; where $\varepsilon_i^e=\varepsilon_i^g+\hbar\omega_a$, with $\omega_a$ being the effective atomic transition frequency, which can include AC Stark-shifts relative to the free atomic transition frequency. Here $J_g$ and $J_e$ are the hopping parameters of ground and excited state atoms, respectively. $U_g$ and $U_e$ are the on-site ground and excited state atom-atom interactions, respectively, and $U_{eg}$ is the on-site ground-excited atom interaction related to scattering between ground and excited state atoms. This extended model leads to a much richer phase diagram.

\begin{figure}
\begin{center}
\leavevmode
\includegraphics[width=85mm]{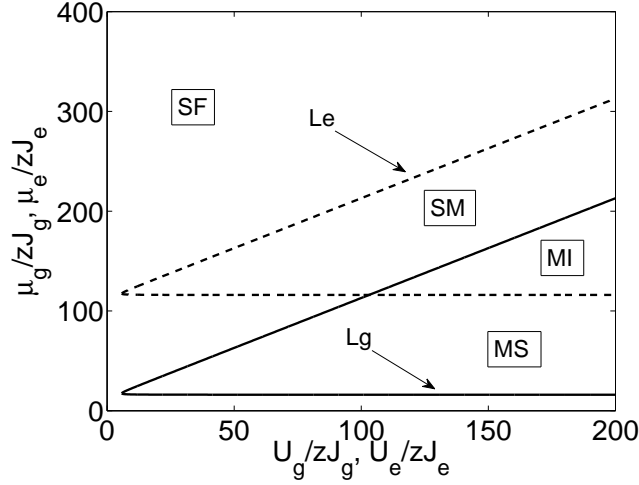}
\caption{Phase diagram: $(\mu_g/zJ_g)$ vs. $(U_g/zJ_g)$, and $(\mu_e/zJ_e)$ vs. $(U_e/zJ_e)$. We used $U_{eg}/zJ_g=U_{eg}/zJ_e=15$, with $\varepsilon_i^g/zJ_g=0$ and $\varepsilon_i^e/zJ_e=100$, for $n_g=n_e=1$. Beside the SF and the MI regions, the SM region is associated to the Mott insulator of excited atoms and superfluid phase of ground atoms, and the MS region to the Mott insulator of ground atoms and superfluid phase of excited atoms. The full-line Lg determines the SF-MI transition of the ground atoms, and the dashed-line Le  that transition for the excited atoms. Figure reprinted with permission from~\citet{Zoubi2009d} Copyright 2009 by American Physical Society.}
\label{3}
\end{center}
\end{figure}

Now we present one typical example of the phase diagram, which is calculated in using the mean-field approach. In figures \ref{3} we plot the phase diagram for the plane $(\mu_g/zJ_g)$ vs. $(U_g/zJ_g)$, and $(\mu_e/zJ_e)$ vs. $(U_e/zJ_e)$, which are scaled by $zJ_g$ and $zJ_e$. We used for the energies $\varepsilon_i^g/zJ_g=0$ and $\varepsilon_i^e/zJ_e=100$, and for the scaled atom-atom coupling the number $U_{eg}/zJ_g=U_{eg}/zJ_e=15$. Here the excited and ground phase transition lines split, the transition line (Le) is for excited state atoms, and (Lg) for ground state atoms. Beside the superfluid phase, we get three other regions. The Mott insulator region with one ground state and one excited state atom per site, that is $n_g=n_e=1$. The SM region, where the excited state atoms are in the Mott insulator phase with one atom per site, and the ground state atoms are in the superfluid phase. The MS region, where the ground state atoms are in the Mott insulator phase with one atom per site, and the excited state atoms are in the superfluid phase. We like to mention here that more complex quantum phases are suppressed under the present mean field theory, e.g., that of the anti-ferromagnetic type in which the Mott insular includes an excited atom at one site and a ground one in the nearest neighbor sites \citep{Altman2003}.

\section{Excitons in Optical Lattices}

As shown above even for two state atoms one can prepare a perfectly filled Mott phase for the atoms, which  we will use as the basis of our further considerations involving resonant excitations.  We focus on a single atom per site, but eventually allow for two atoms per site as in Subsection 3.5. The Mott insulator constitutes an artificial crystal with a lattice constant much larger than the average dimension of the atom. Therefore, each atom retains its identity except from the internal shifts and there are no overlaps among electronic wave functions of atoms setting at different sites as in conventional solids. We conclude that our artificial lattice is similar to molecular or noble atom crystals but allow for the wide controlablity of the system parameters, as the artificial lattice is formed by variable external laser field. Solid molecular crystals, however,  are fixed by the chemical bonds, e.g. van der Waals forces, and can be only slightly controlled in applying external fields. Frenkel excitons play a central role in optical and electrical properties of molecular crystals \citep{Davydov1971,Agranovich2009,Zoubi2005a}. Here we ask the question related to the possibility of the formation of Frenkel like-excitons for optical lattices. First we discuss the possibility of energy transfer among atoms at different sites induced by electrostatic interactions or resonant dipole-dipole interactions, which is the main mechanism behind the formation of excitons.

\subsection{Resonance Dipole-Dipole Interactions}

 Lets take the simplest case of two atoms setting on a lattice at nearest neighbor sites such that they are separated by the lattice constant $a$. The atoms are considered to be two-level systems with transition energy $E_A=\hbar\omega_A$. If one atom is excited and the second is in the ground state, can this excitation be transferred among the two atoms, in which the excited atom decays to the ground state and the ground state one becomes excited. Here we consider only energy transfer without charge transfer? This process is possible through the resonance dipole-dipole interaction \citep{Craig1984} with the energy transfer parameter
\begin{eqnarray}\label{Exact}
J\left(q_Aa\right)&=&\frac{3}{4}\Gamma_A\left\{\left[\frac{\sin\left(q_Aa\right)}{\left(q_Aa\right)^2}+\frac{\cos\left(q_Aa\right)}{\left(q_Aa\right)^3}\right]\left(1-3\cos^2\theta\right)\right. \nonumber \\
&-&\left.\frac{\cos\left(q_Aa\right)}{q_Aa}\left(1-\cos^2\theta\right)\right\},
\end{eqnarray}
where $q_A$ is the atomic transition wave number, as $E_A=\hbar q_A$, and $\Gamma_A$ is the single excited atom damping rate \citep{Loudon2000}, which is given by $\Gamma_A=\omega^3_A\mu^2(3\pi\epsilon_0\hbar c^3)^{-1}$, with $\mu$ the transition dipole. Here $\theta$ is the angle between the transition dipole and vector distance between the two atoms.

\begin{figure}
\begin{center}
\leavevmode
\includegraphics[width=65mm]{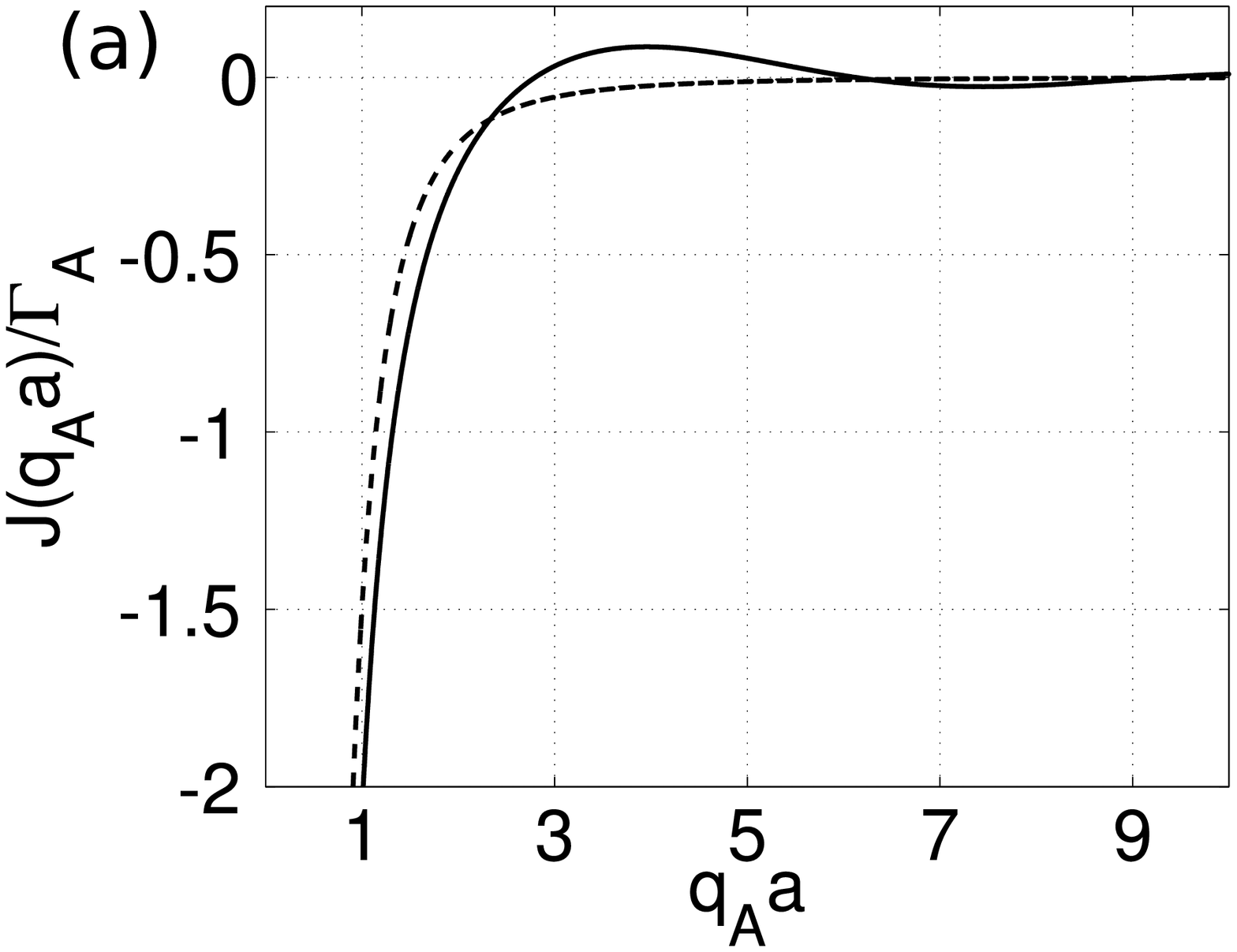}\ \ \ \includegraphics[width=65mm]{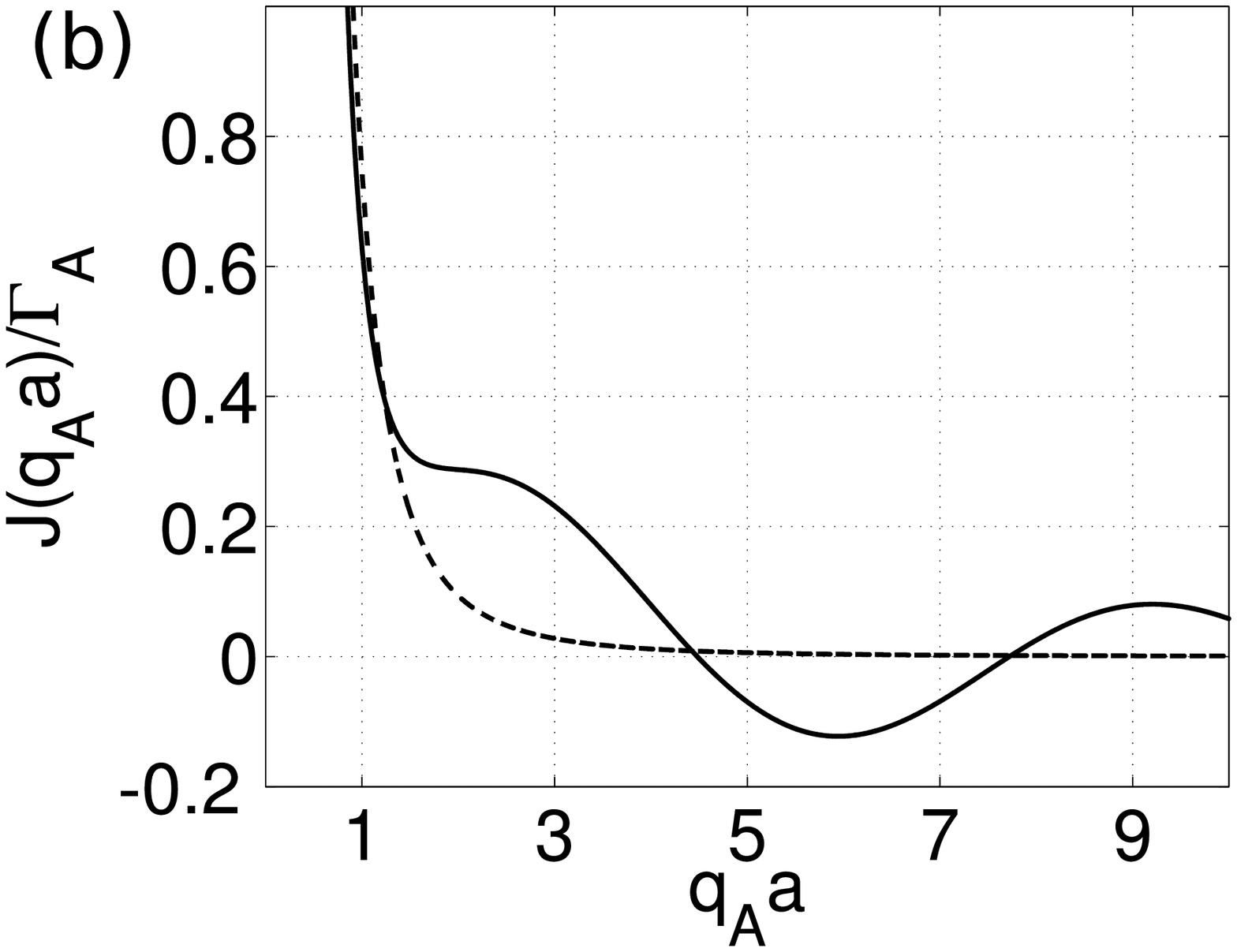}
\caption{The scaled interaction $J\left(q_Aa\right)/\Gamma_A$ vs. $q_Aa$, for (a) $\theta=0^{\circ}$ and (b) $\theta=90^{\circ}$. The full line is for equation (\ref{Exact}), and the dashed line for equation (\ref{Appr}). Figure reprinted with permission from~\citet{Zoubi2012a} Copyright 2012 by European Physical Society.}
\label{4}
\end{center}
\end{figure}

The energy transfer is efficient in the limit of $|J\left(q_Aa\right)|/\Gamma_A>1$, where the electronic excitation can transfer among the atoms before it spontaneously decays, and this limit is achievable up to $q_Aa\approx 1$. For a typical optical lattice we have, e.g., $E_A=1\ eV$, with $q_A\approx 4\times10^{-4}\ \AA^{-1}$. For $a=1000\ \AA$ we get $q_Aa\approx 0.5$. For $\theta=0^{\circ}$ we obtain $J(0.5)/\Gamma_A\approx -13.4$, and for $\theta=90^{\circ}$ we obtain $J(0.5)/\Gamma_A\approx 5.4$, in which the energy transfer is possible. Note that in the limit $q_Aa<1$, with efficient energy transfer, the electrostatic interaction limit is applicable \citep{Zoubi2012a}. In this limit and in neglecting the radiative corrections, we get
\begin{equation}\label{Appr}
J\approx\frac{3}{4}\frac{\Gamma_A}{(q_Aa)^3}\left(1-3\cos^2\theta\right).
\end{equation}
Using the previous numbers, $\theta=0^{\circ}$ yields $J(0.5)/\Gamma_A\approx -12$, and $\theta=90^{\circ}$ yields $J(0.5)/\Gamma_A\approx 6$, which are slightly different from the above exact results. For smaller $q_Aa$ we get much better agreement. Figure \ref{4}(a) reports  $J\left(q_Aa\right)/\Gamma_A$ as a function of $q_Aa$ for Eqs. (\ref{Exact}) and (\ref{Appr}) at the polarization direction $\theta=0^{\circ}$, and in (b) for $\theta=90^{\circ}$. The results justify the use of electrostatic dipole-dipole interactions for optical lattices when $q_Aa<1$.

Energy transfer of electronic excitations among close lattice sites induced by dipole-dipole interaction tends to delocalize the excitation coherently in the lattice. Exploiting the lattice symmetry we can construct collective excitations in such systems, which are called excitons. In momentum space such de-localized electronic excitations can be represented by propagating excitation waves. In the following we explicitly calculate these excitons for various optical lattices with different geometries and dimensionality.

We concentrate now in the Mott insulator phase with one atom per site \citep{Zoubi2007}. The atoms as before are assumed to be two-level systems, and the electronic excitation can transfer among atoms at different sites, e.g. among sites $n$ and $m$, with the transfer parameter $J_{nm}$. As discussed above, such a process can be efficient in the limit of $|J_{nm}|>\Gamma_A$, where an electronic excitation could transfer before the decay of the excited atom. The excitation Hamiltonian in general can be written in the second quantized form as
\begin{equation} \label{excitons}
H=\sum_nE_A\ B_n^{\dagger}B_n+\sum_{nm}\hbar J_{nm}\ B_n^{\dagger}B_m,
\end{equation}
where $B_n^{\dagger}$ and $B_n$ are the creation and annihilation operators of an electronic excitation at site $n$. The first term is for the on-site excitation and the second for the energy transfer. We neglect terms that do not conserve the number of excitations. The operators are of the spin-half type, and they behave on-site as fermions, and as bosons among different sites. But for a single excitation, or for very small number of excitations relative to the number of lattice sites, saturation effects can be neglected, and then the possibility of having two or more excitations to appear at the same lattice atom is negligible \citep{Agranovich2009,Zoubi2005b}. Therefore, from this point on we assume the excitations to behave as bosons with the commutation relation $[B_n,B_m^{\dagger}]=\delta_{nm}$.

\subsection{One-Dimensional Atomic Chains}

\begin{figure}
\begin{center}
\leavevmode
\includegraphics[width=85mm]{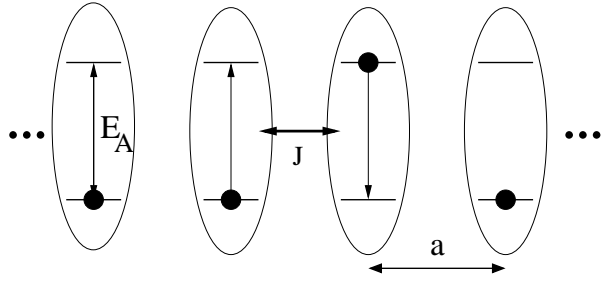}
\caption{Schematic plot of a one dimensional atomic chain of two-level atoms of transition energy $E_A$, lattice constant $a$, and energy transfer parameter $J$.}
\label{5}
\end{center}
\end{figure}

We treat first a long one dimensional optical lattice \citep{Zoubi2010b}, as in Fig. \ref{5}. The lattice length is enough long to neglect edge effects, which allows the use of the periodic boundary condition. The previous Hamiltonian can be easily diagonalized in applying the Fourier transform
\begin{equation}
B_k=\frac{1}{\sqrt{N}}\sum_ne^{-ikz_n}B_n,
\end{equation}
where $N$ is the number of lattice sites, and $z_n=an$ is the position of an atom at site $n$. Here $k$ is a wave number that in using periodic boundary conditions takes the values $k=\frac{2\pi}{L}p$, with $(p=0,\pm1,\cdots,\pm N/2)$, and $L=Na$ is the lattice length. This solution holds also for a ring of an atomic chain. In applying the transformation the diagonal Hamiltonian reads
\begin{equation}
H=\sum_kE_k\ B_k^{\dagger}B_k,
\end{equation}
where the exciton dispersion is given by $E_k=E_A+J_k$. As the energy transfer is a function of the interatomic distance, $R=a(n-m)$, the exciton dynamical matrix is given by $E_k=\sum_RJ_k(R)e^{ikR}$. Note that in place of a discrete energy level of the excited atom we get an energy band for the excitons.

The single exciton state of wave number $k$ is obtained from the vacuum state, $|0\rangle=|\{g_n\}\rangle$ in which all atoms are in the ground state, in applying the exciton creation operator, namely $|1_k\rangle=B_k^{\dagger}|\{g_n\}\rangle$, where $|1_k\rangle=\frac{1}{\sqrt{N}}\sum_n|e_n\rangle e^{-ikan}$. Here $|e_n\rangle$ stands for a state where only atom $n$ is excited and all the others are in the ground state. The single exciton is a coherent state built of a superposition of electronic excitations at all sites, where the probabilities to find an excitation at each site are equal. This state is a collective electronic excitation and is considered as a wave that propagates to the left or the right in the lattice with a fixed wave number $k$. A superposition of such states can be used to build a more physical state of a wave packet of excitons.

We assume now energy transfer only among nearest neighbor sites. In the $q_Aa<1$ limit, the transfer parameter $J$ is given by  Eq.~(\ref{Appr}) and  shows a strong dependence on the $\theta$ angle  between the directions of the transition dipole and of the lattice. The transfer parameter changes from being negative (attractive) at $\theta=0$ to positive (repulsive) at $\theta=\pi/2$. The dispersion relation now reads
\begin{equation}\label{OneDispersion}
E_k=E_A+2J\cos(ka),
\end{equation}
with a bandwidth of $4J$. For small wave numbers, that is in the limit of $ka\ll 1$, or for long wavelength excitons, we get the dispersion $E_k\approx E_A+2J+\frac{\hbar^2 k^2}{2m_{ex}}$, where we defined the exciton effective mass $m_{ex}=\frac{-\hbar^2}{2Ja^2}$. The exciton in this limit is a quasi-particle that propagates in the lattice with a given wave number and an effective mass. At $k\approx 0$ we get $E_k\approx E_A+2J$, with a shift produced by  the dipole-dipole interaction. In this limit of long wavelength excitons, beyond the nearest neighbor interactions the long range dipole-dipole interaction yields the correction $E_k\approx E_A+2\zeta(3)J$, where $\zeta(3)\approx 1.2$ \citep{Zoubi2010c}. Such a shift has a large influence on optical lattice clock states \citep{Takamoto2005}, but can be controlled through the direction of the transition dipole \citep{Chang2004}, and even can be brought to zero at the angle of $\theta\approx 54.7$.

\begin{figure}
\begin{center}
\leavevmode
\includegraphics[width=85mm]{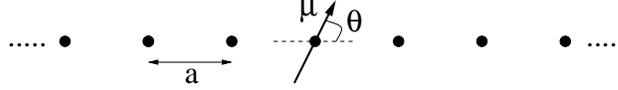}
\caption{A one-dimensional infinite lattice of lattice constant $a$, and with one atom per site. The transition dipole $\mu$ is plotted which makes an angle $\theta$ with the lattice axis. For a finite optical lattice of $N$ sites with one atom per site, the lattice length is $L=a(N+1)$,}
\label{7}
\end{center}
\end{figure}

Next we treat a finite optical lattice chain of $N$ two-level atoms \citep{Zoubi2009c,Zoubi2012b}, as schematized  in Fig. \ref{7}. The lattice length is taken to be smaller than the atomic transition wavelength. In considering energy transfer only between nearest neighbor atoms, the Hamiltonian can be diagonalized by introducing standing wave excitons
\begin{equation}
B_n=\sqrt{\frac{2}{N+1}}\sum_k\sin\left(\frac{\pi nk}{N+1}\right)B_k,
\end{equation}
with $(k=1,\cdots,N)$.  As a mathematical trick to find appropriate solutions, we add two additional empty sites at the two lattice edges, at $n=0$ and $n=N+1$, where the collective electronic excitation wave function vanishes. The exciton dispersion then reads
\begin{equation}\label{FinDis}
E_k=E_A+2J\cos\left(\frac{\pi k}{N+1}\right).
\end{equation}
The single electronic transition energy splits into $N$ discrete collective electronic excitation energies. The modes are divided into two groups, symmetric and antisymmetric modes. In the next section we will show that the antisymmetric modes are dark, and the symmetric ones are bright. Furthermore the first bright mode is node less and found to be super radiant and dominates the optical properties.

\begin{figure}
\begin{center}
\leavevmode
\includegraphics[width=70mm]{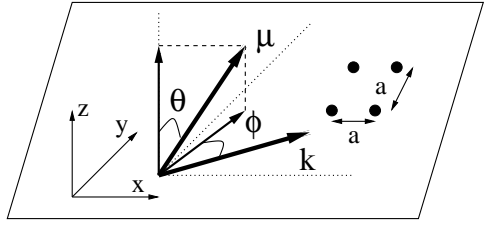}
\caption{A planar square optical lattice with lattice constant $a$. The directions of the in-plane wave vector ${\bf k}$ and the transition dipole \boldmath$\mu$ are indicated.}
\label{10}
\end{center}
\end{figure}

\subsection{Two-Dimensional Planar Optical Lattices}

Let us consider the case of a two dimensional planar optical lattice \citep{Zoubi2007}. For simplicity we treat the case of a square symmetry with lattice constant $a$, and one two-level atom per site, as in Fig. \eqref{10}. The position of a lattice site is defined by the in-plane vector ${\bf n}=(n_x,n_y)$, where $n_i=a(0,\pm 1,\cdots,\pm N_i/2)$, with $N=N_x\times N_y$ the number of lattice sites. The electronic excitation Hamiltonian is given by
\begin{equation}\label{Planar}
H_{ex}=\sum_{\bf n}E_A\ B_{\bf n}^{\dagger}B_{\bf n}+\sum_{\bf nm}J_{\bf nm}\ B_{\bf n}^{\dagger}B_{\bf m},
\end{equation}
which can be diagonalized by applying the transformation
\begin{equation}
B_{\bf n}=\frac{1}{\sqrt{N}}\sum_{\bf k}e^{i{\bf k}\cdot{\bf n}}B_{\bf k},
\end{equation}
to get the diagonal exciton Hamiltonian
\begin{equation}
H_{ex}=\sum_kE_{ex}({\bf k})\ B_{\bf k}^{\dagger}B_{\bf k},
\end{equation}
with the energy dispersion
\begin{equation}\label{TwoDis}
E_{ex}({\bf k})=E_A+\sum_{\bf R}J({\bf R})e^{i{\bf k}\cdot{\bf R}},
\end{equation}
where we defined the distance between two sites by ${\bf R}={\bf n-m}$ and introduced as previously the $J$ transfer parameter.

The exciton is now represented by a wave that propagates in the lattice with in-plane wave vector ${\bf k}$, where we have $k_i=\frac{2\pi}{N_ia}\left(0,\pm 1,\cdots,\pm N_i/2\right)$. The nearest neighbor interaction yields the dispersion  with  $E_{ex}({\bf k})=E_A+2J[\cos(k_xa)+\cos(k_ya)]$, which for small wave numbers  that is $ka\ll 1$ with $k=|{\bf k}|$, reads $E_k\approx E_A+4J+\frac{\hbar^2 k^2}{2m_{ex}}$, with the exciton effective mass $m_{ex}=\frac{-\hbar^2}{2Ja^2}$.  At $k\approx 0$ we get $E_k\approx E_A+4J$ with the dipole-dipole interaction shift. In the limit of long wavelength excitons, beyond the nearest neighbor interactions the long range dipole-dipole interaction yields the correction $E_k\approx E_A+2FJ$, where in using the Ewald's summation method \citep{Zoubi2011b} we get $F\approx 9/2$.

\subsection{Radiative Decay of Excitons}

The exciton damping rate through the spontaneous emission into free space can be calculated in using Fermi's golden rule \citep{Zoubi2010b,Zoubi2011a,Zoubi2012b,Zoubi2012a,Ostermann2012} and shows significant deviation from a single excited atom \citep{Loudon2000}. For a finite atomic chain the standing wave excitons are divided into two groups of dark and bright modes, which exhibit metastable or superradiant propagating excitons. Let us mention here an important point concerning the atomic recoil effect via the emission of a photon. As the exciton is a collective electronic excitation delocalized in the lattice, where the exciton momentum is shared among all lattice atoms, the emitted photon inherits the momentum and energy of the exciton, which requires the photon wave vector component parallel to the lattice to be equal to the exciton wave vector. In this process the whole lattice reacts and absorbs the momentum collectively and not only a single atom. Hence, the recoil heating is significantly suppressed, and we can conclude that excitonic interactions will not heat the lattice significantly and destroy the Mott insulator.

We start with a long optical lattice with one atom per site \citep{Zoubi2010b} as  in Fig \ref{7}. We assume that a single exciton is excited in the lattice with a fixed wave number $k$, and a given polarization $\theta$ for the transition dipole direction relative to the lattice axis. The damping rate of the exciton into the radiation field of the free space is given by
\begin{equation} \label{Damping}
\Gamma_k=\frac{\mu^2E_{k}^2}{4\epsilon_0a\hbar^3c^2}\left\{1+\cos^2\theta-\frac{(\hbar ck)^2}{E_{k}^2}\left(3\cos^2\theta-1\right)\right\},
\end{equation}
where $E_k$ is the previous one dimensional exciton dispersion, e.g. as for the nearest neighbor interactions in Eq.~(\ref{OneDispersion}). For some wave numbers $k$ and certain angles $\theta$ the damping rate can be much larger than the single excited atom one, as we get a superradiant enhancement analogous to the Dicke model. However, for other $k$-s and $\theta$-s the damping rate can be much smaller than the single excited atom one, or even completely vanish. In general the damping expression shows the existence of a critical wave vector $k_c$ that is given by $E_{ex}(k_c)=\hbar ck_c$, beyond which the damping rate is zero and the excitations can no longer decay radiatively. For $\theta=0^o$ the damping rate at the critical wave vector $k_c$ is zero, and for $\theta=90^o$ the damping rate at $k_c$ is finite and equal to $\Gamma(k_c)=\mu^2E_{ex}^2(k_c)(2\epsilon_0a\hbar^3c^2)^{-1}$. We conclude that Eq.~(\ref{Damping}) holds up to the critical wave number, $k_c$, where beyond $k_c$ no free space photon exists which conserves both energy and momentum component parallel to the lattice.

\begin{figure}
\begin{center}
\leavevmode
\includegraphics[width=65mm]{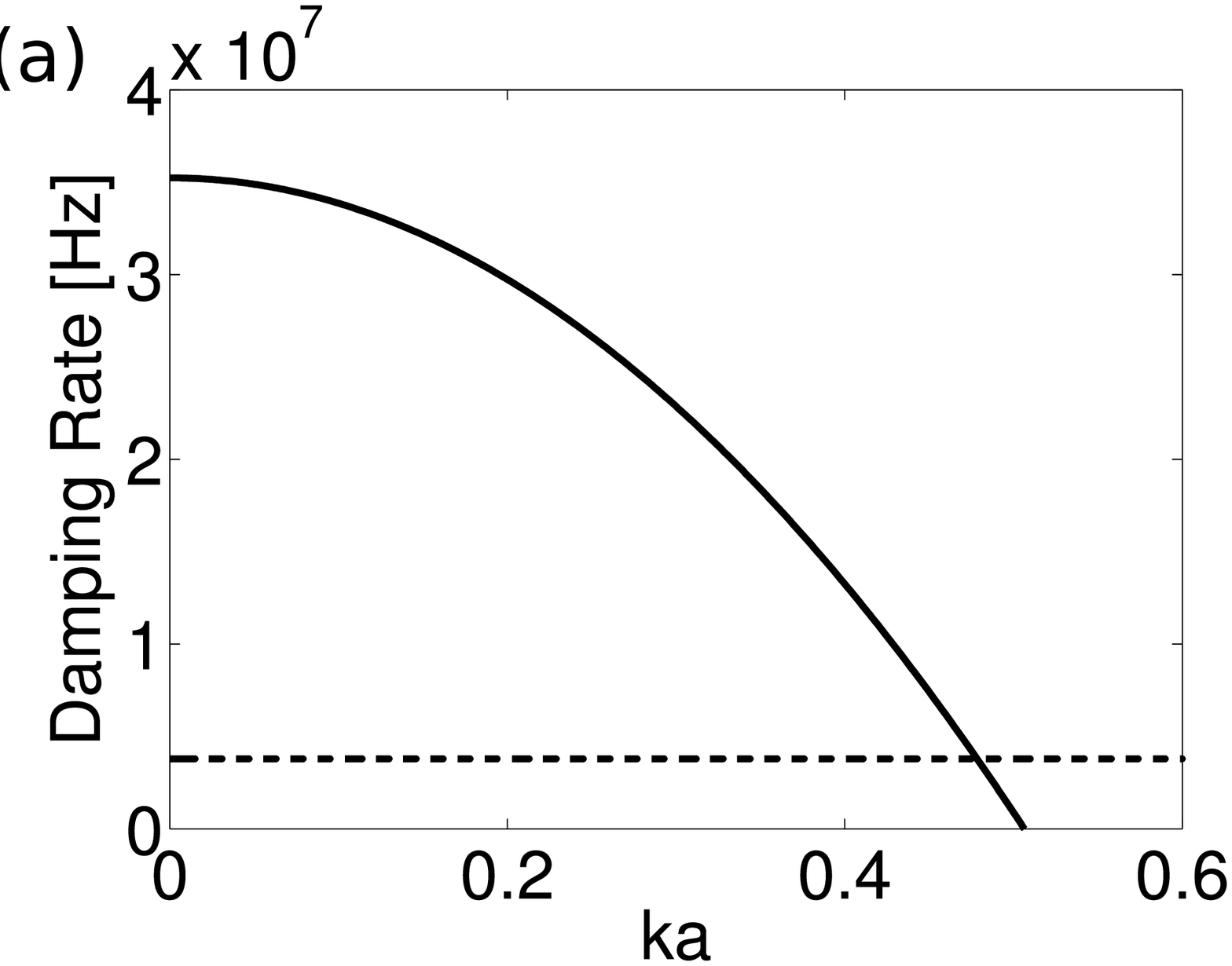}\ \ \ \includegraphics[width=65mm]{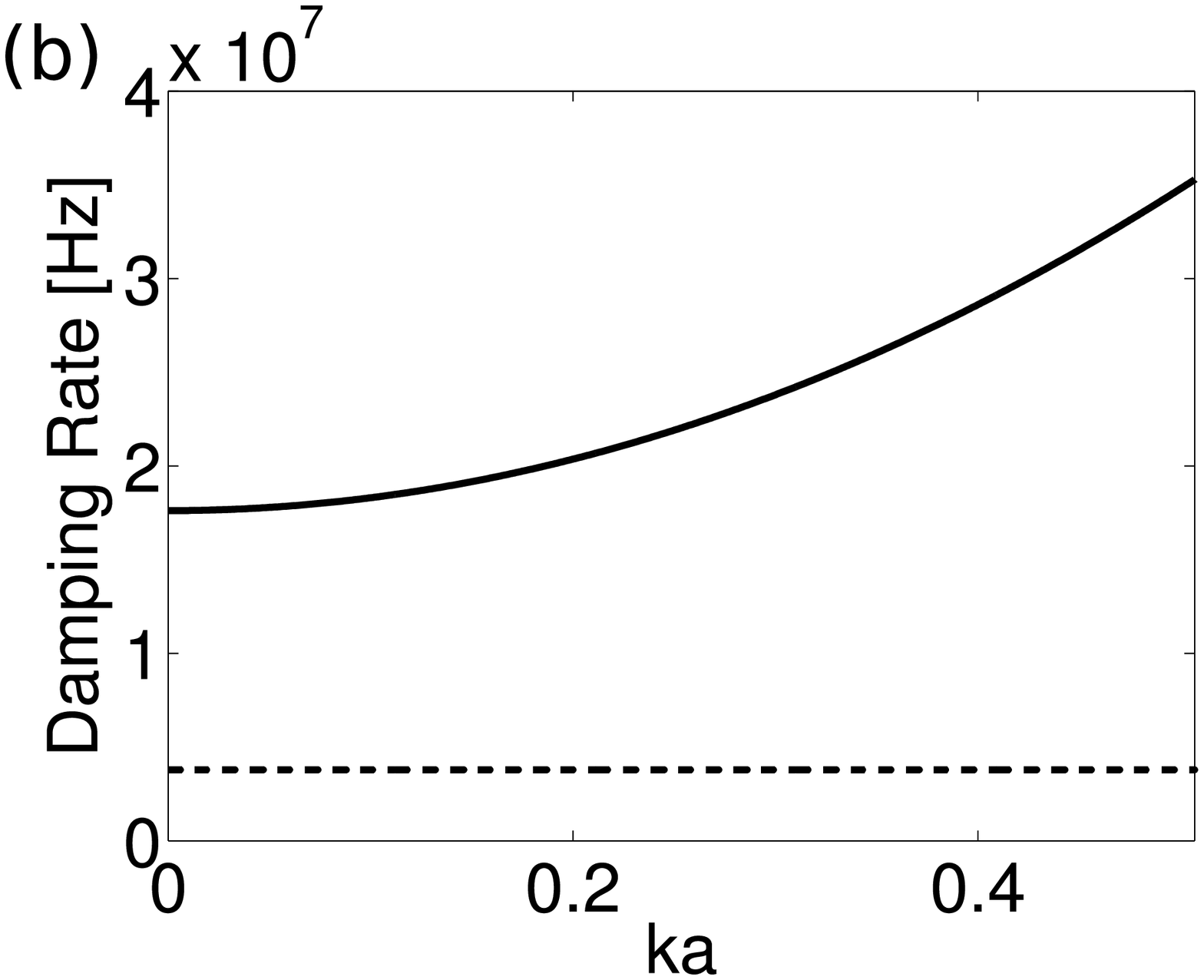}
\caption{The damping rate vs. $ka$, for (a) $\theta=0$ and (b) $\theta=90$. The damping rate is zero beyond the critical wave number $k_ca\approx 0.507$.  The dashed-line is for a single excited atom damping rate. Figure reprinted with permission from~\citet{Zoubi2010b} Copyright 2010 by European Physical Society.}
\label{8}
\end{center}
\end{figure}

We exhibit this behavior in two plots of the damping rate using the following parameters: transition energy $E_A=1\ eV$, lattice constant $a=1000\ A$, and transition dipole $\mu=1\ eA$. In Fig. \ref{8}(a) we plot  as a function of $ka$ for $\theta=0^o$ the damping rate compared to the damping rate of a single atom. Note that for small wave numbers, $ka\sim 0$, the exciton damping rate is much larger than the free atom damping rate, and these states are superradiant states. With increasing wave numbers the damping rate drops and reaches the single atom rate, and beyond a critical wave number $k_c$ the damping rate becomes zero and no damping is obtained beyond $k_c$. The critical wave number is obtained here at $k_ca\approx 0.507$. In Fig. \ref{8}(b) the same plot is for $\theta=90^o$. Here the damping rates are larger than the single atom one for small wave numbers, $ka\sim 0$, but now the damping rate increases with increasing the wave number. Namely, the large wave number states became more superradiant states. The maximum damping rate is obtained at the critical wave number $k_c$, and the damping rate is plotted up to that value, as beyond this point the excitons are metastable. Note that excitons can appear up to the boundary of the Brillouin zone at $k_Ba=\pi$, and hence excitons in the region $k_c<k<k_B$ are metastable. These excitons are not dark, only they are decoupled from the free space photons, but they can be coupled to confined photons, as presented in Subsection 4.5 through the coupling to tapered nanofiber photons \citep{Zoubi2010c}.

\begin{figure}
\begin{center}
\leavevmode
\includegraphics[width=85mm]{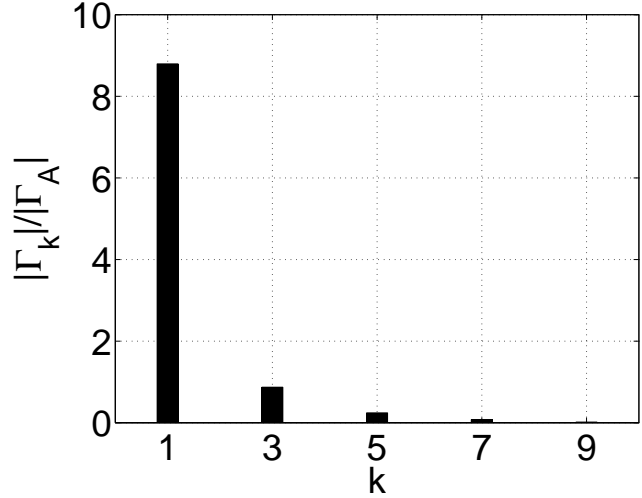}
\caption{The $\Gamma_k/\Gamma_A$ collective electronic excitation damping rate rapported to that of a single excited atom  $k$, for $N=10$. Figure reprinted with permission from~\citet{Zoubi2012b} Copyright 2012 by European Physical Society.}
\label{9}
\end{center}
\end{figure}

The results are drastically changed for a finite one dimensional atomic chain of only a few atoms \citep{Zoubi2009c,Zoubi2012b,Zoubi2012a,Ostermann2012} which is short relative to the atomic transition wave length. In this case we obtained that the exciton states are divided into two groups of symmetric and antisymmetric standing waves. For each collective mode we can define a collective transition dipole, which is assumed to be localized at the center of the lattice, for the limit of short chains. The antisymmetric excitons with even wave numbers $(k=2,4,\cdots)$ are dark with zero damping rates, that is $\Gamma_k=0$. The symmetric excitons with odd wave numbers $(k=1,3,\cdots)$ are bright with the damping rates
\begin{equation}\label{DampingFinite}
\Gamma_k=\frac{\mu^2E_k^3}{3\pi\epsilon_0\hbar^4c^3}\left(\frac{2a}{L}\right)\cot^2\left(\frac{\pi ka}{2L}\right),
\end{equation}
where the lattice length is $L=a(N+1)$, and the exciton dispersion is given in Eq.~(\ref{FinDis}). The first mode $(k=1)$, the one without nodes, is superradiant, with a damping rate nine times larger than the second bright mode $(k=3)$. In Fig. \ref{9} the collective electronic excitation damping rate rapported to that of a single excited atom, $\Gamma_k/\Gamma_A$ is plotted versus $k$, using the previous parameters, $\theta=0$ and $N=10$.

Interesting results can be also obtained for planar optical lattices \citep{Zoubi2011a}. We deal here, as before, with two dimensional large optical lattice with one atom per site and of square symmetry, as plotted schematically in Fig. \ref{10}. We calculate the damping rate into free space of an exciton with in-plane wave vector ${\bf k}$. Now the dipole makes an angle $\theta$ with the normal to the lattice plane, and an additional angle $\phi$ is needed that is between the in-plane component of the dipole and ${\bf k}$. The result is 
\begin{eqnarray}
\Gamma_{ex}({\bf k})&=&\frac{\mu^2}{2\epsilon_0a^2\hbar^2c}\frac{E_{ex}^2({\bf k})}{\sqrt{E_{ex}^2({\bf k})-E_{0}^2(k)}}\left\{\sin^2\theta\left(1-\cos^2\phi\frac{E_{0}^2(k)}{E_{ex}^2({\bf k})}\right)\right. \nonumber \\
&+&\left.\cos^2\theta\frac{E_{0}^2(k)}{E_{ex}^2({\bf k})}-2\sin\theta\cos\theta\cos\phi\frac{E_{0}(k)}{E_{ex}^2({\bf k})}\sqrt{E_{ex}^2({\bf k})-E_{0}^2(k)}\right\},
\end{eqnarray}
where the exciton dispersion is defined in Eq.~(\ref{TwoDis}), and $E_{0}(k)=\hbar ck$. The most important result here is that for $k\geq k_c$ we have $\Gamma_{\bf k}=0$, where $\hbar ck_c=E_{ex}({\bf k}_c)$, as beyond the singularity point the damping rate becomes imaginary. The results should be compared with the radiative damping rate of a single excited atom $\Gamma_A$.

\begin{figure}
\begin{center}
\leavevmode
\includegraphics[width=65mm]{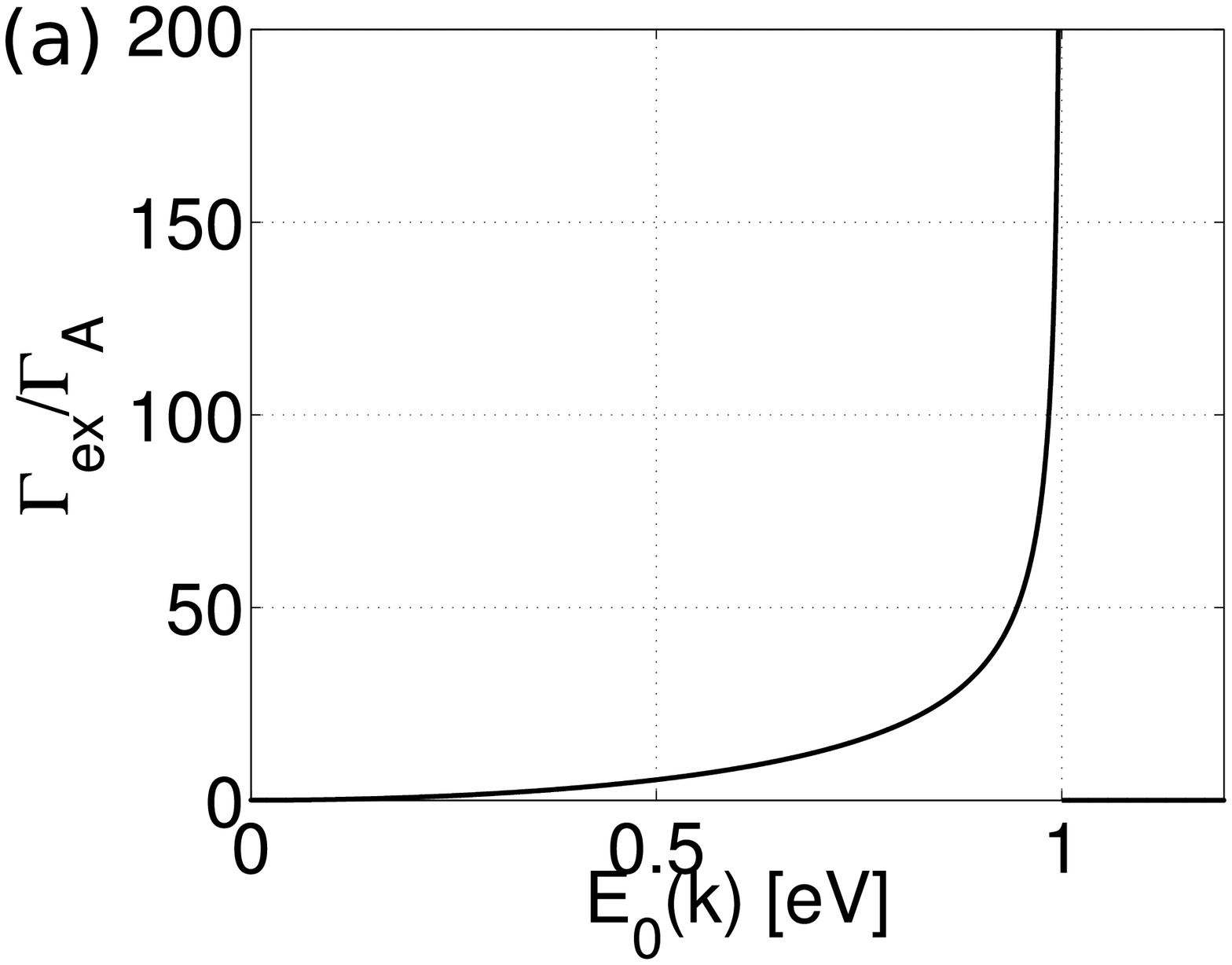}\ \ \ \includegraphics[width=65mm]{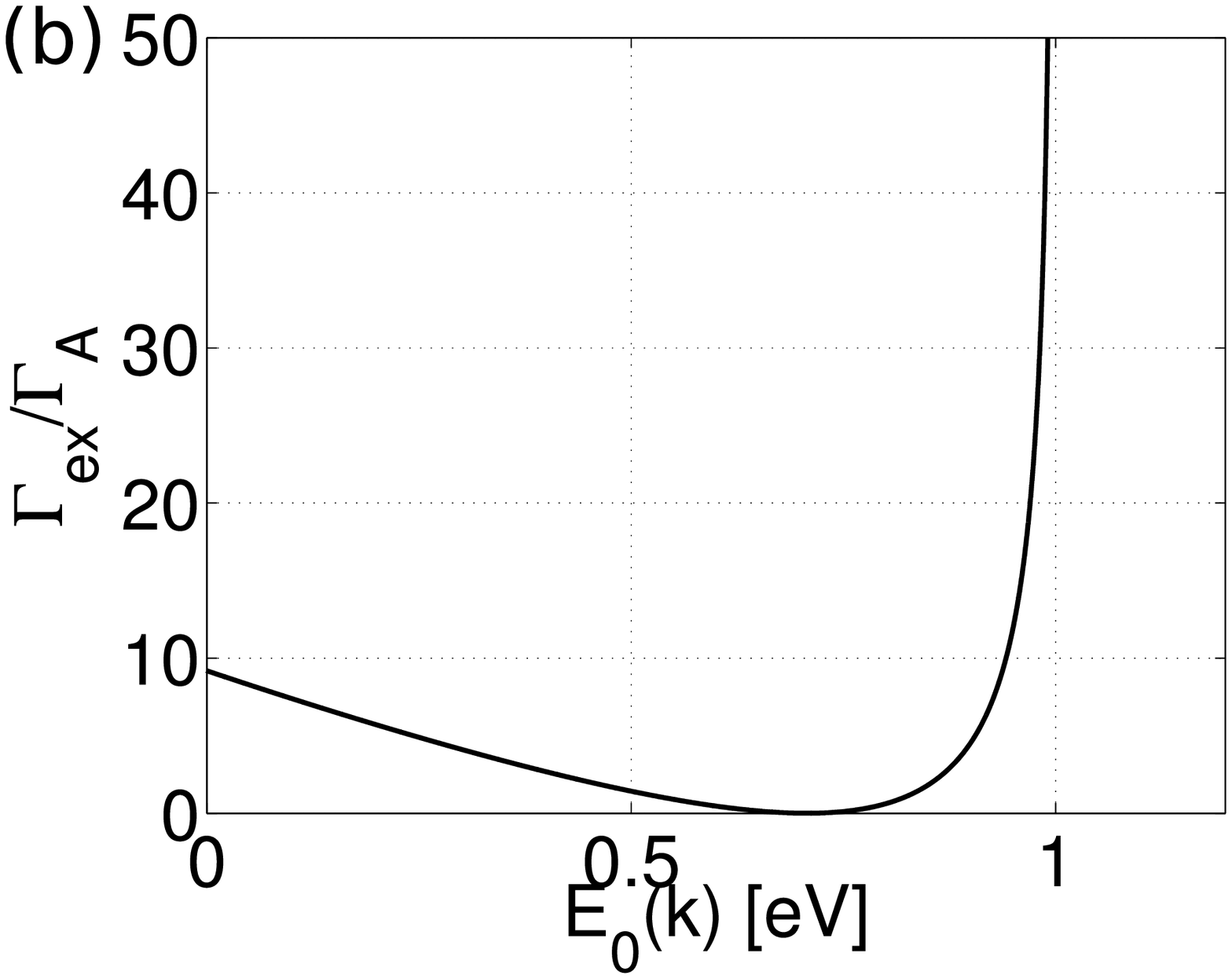}
\caption{The scaled damping rate $\Gamma_{ex}/\Gamma_A$ vs. $E_{0}(k)=\hbar ck$, for (a) $\theta=0$, which is $\phi$ independent, and for (b) $\theta=\pi/4$ with $\phi=0$. Here $\hbar c\pi/a\sim 6.2\ eV$. Figure reprinted with permission from~\citet{Zoubi2011a} Copyright 2011 by American Physical Society.}
\label{11}
\end{center}
\end{figure}

We present numerical results using the typical previous numbers for optical lattices, except for simplicity we neglect the ${\bf k}$ dependence of the exciton energy, negligible compared to the $E_{0}(k)$ one. Fig. \ref{11}(a) shows $\Gamma_{ex}/\Gamma_A$ vs. $E_{0}(k)$ for $\theta=0$, where the transition dipole has only component normal to the lattice plane. Here $\Gamma_{ex}({\bf k})$ is $\phi$ independent. The damping rate starts from zero, where the long wave length excitons are metastable, and increases with $k$ to become superradiant with damping rate larger than $\Gamma_A$. Close to $\hbar ck_c=1\ eV$ the damping rate diverges, and beyond that value it jumps back to zero, all excitons between $k_c$ and the Brillouin zone boundary at $\pi/a$ being metastable. Fig. \ref{11}(b) shows $\Gamma_{ex}/\Gamma_A$ vs. $E_{0}(k)$ for $\theta=\pi/4$, where the transition dipole has equal in-plane and normal components. We take the in-plane wave vector ${\bf k}$ to be parallel to the in-plane dipole, that is $\phi=0$. Very interesting behavior appears here for the damping rate. For small in-plane wave numbers the excitons are superradiant with damping rate larger than $\Gamma_A$, and decreases in increasing $k$ till it becomes zero at $E_0(k)=E_{ex}({\bf k})/\sqrt{2}$ where excitons around this point are metastable. For larger $k$ the damping rate start to increase and diverges at $E_0(k_c)$, and then jump back to zero beyond this point.

The dark and metastable states have an important potential use for the quantum information processing and the memory devices \citep{Bouwmeester2000} but their excitation is an experimental challenge. The dark state can be addressed by applying external fields or by adding a structural asymmetry. The metastable ones can be coupled to confined photons in using different optical resonators as shown in Subsection 4.5. The bright states are coupled directly to photons through the dipole coupling, and due to their strong dipole moment they can easily be tuned to achieve the strong coupling regime, making the atomic chain an optimal candidate as an active material for cavity QED.

\subsection{Excitons in Optical Lattices with Two-Atoms per site}

\begin{figure}
\begin{center}
\leavevmode
\includegraphics[width=85mm]{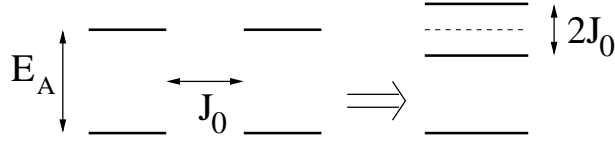}
\caption{(a) The two-level on-site two atoms with a transfer parameter $J_0$, and a transition energy $E_A$. (b) The two diagonal states with the splitting energy of $2J_0$.}
\label{12}
\end{center}
\end{figure}

Up to this point we have treated only the case of an optical lattice in the Mott insulator phase with one atom per site, but the case of two atoms per site is of large interest and can be easily realized in the present optical lattice experiments \citep{Spielman2007}. Furthermore, this case shows novel phenomena beyond that of one atom per site, e.g. the appearance of dark and localized excitons \citep{Zoubi2008b}. As before we consider two-level atoms, in which the ground and excited optical lattice potentials have minima at the same positions, and the atoms are located in the lowest Bloch band. Moreover, we neglect overlaps between electronic wave functions for on-site and off-site atoms. Namely, we neglect the on-site formation of molecules from the two atoms, and we consider only transfer of electronic excitations. For planar optical lattices, the electronic excitation Hamiltonian is as in Eq.~(\ref{Planar}), but with an additional index for the two on-site atoms 
\begin{equation} \label{TwoAtoms}
H_{ex}=\sum_{{\bf n},\alpha}E_A\ B_{\bf n}^{\alpha\dagger}B_{\bf n}^{\alpha}+\sum_{{\bf nm},\alpha\beta}J_{\bf nm}^{\alpha\beta}\ B_{\bf n}^{\alpha\dagger}B_{\bf m}^{\beta},
\end{equation}
where $\alpha$ and $\beta$ run over the two on-site atoms, which are denoted by $(1)$ and $(2)$. For simplicity, we assume energy transfer only among on-site atoms with average coupling parameter $J_{\bf nn}^{12}=J_{\bf nn}^{21}=J_0$, and among nearest-site atoms with coupling parameter $J_{\langle{\bf nm}\rangle}^{11}=J_{\langle{\bf nm}\rangle}^{22}=J_{\langle{\bf nm}\rangle}^{12}=J_{\langle{\bf nm}\rangle}^{21}=J$, and we consider a lattice of square symmetry. The above Hamiltonian can be diagonalized in two steps, first the diagonalization related to the two on-site atoms, and second related to the site positions. The first transformation is taken in introducing the on-site symmetric and antisymmetric states
\begin{equation}
B_{\bf n}^{s}=\frac{B_{\bf n}^{1}+B_{\bf n}^{2}}{\sqrt{2}},\ B_{\bf n}^{a}=\frac{B_{\bf n}^{1}-B_{\bf n}^{2}}{\sqrt{2}},
\end{equation}
to get
\begin{equation}
H_{ex}=\sum_{\bf n}\left\{E_{s}\ B_{\bf n}^{s\dagger}B_{\bf n}^{s}+E_{a}\ B_{\bf n}^{a\dagger}B_{\bf n}^{a}\right\}+\sum_{\langle{\bf nm}\rangle}2J\ B_{\bf n}^{s\dagger}B_{\bf m}^{s},
\end{equation}
with the energies $E_{s}=E_A+J_0$ and $E_{a}=E_A-J_0$. The on-site energy states are represented in Fig. \ref{12}. The second transformation is given by $B_{\bf n}^{\nu}=\frac{1}{\sqrt{N}}\sum_{\bf k}e^{i{\bf k}\cdot{\bf n}}B_{\bf k}^{\nu}$, which yields the diagonal Hamiltonian
\begin{equation}
H_{ex}=\sum_{\bf k}E_{s}({\bf k})\ B_{\bf k}^{s\dagger}B_{\bf k}^{s}+\sum_{\bf n}E_{a}\ B_{\bf n}^{a\dagger}B_{\bf n}^{a}.
\end{equation}
The on-site symmetric states are delocalized in the lattice and form excitons with energy dispersion $E_{s}({\bf k})=E_A+J_0+4J\left[\cos(k_xa)+\cos(k_ya)\right]$. The antisymmetric states are on-site localized and form no excitons in the lattice; they are dispersion-less with energy $E_{a}=E_A-J_0$. The on-site symmetric state is found to be superradiant with damping rate of two time a single excited atom, that is $\Gamma_s=2\Gamma_A$, and the antisymmetric state is dark with $\Gamma_a=0$. These results can be obtained from Eq.~(\ref{DampingFinite}) of Subsection~3.4 for a finite chain with $N=2$, as presented also in \citet{Ficek2002}.

Dark localized states can in principle also be excited through nonlinear optical processes or the decay of multiply excited atomic states \citep{Ostermann2012}. Moreover, the addition of asymmetry to the system through the application of external fields or the deformation of the lattices sites, e.g. in using different laser frequencies and intensities for different directions, can make the antisymmetric states to be slightly bright and to delocalize in the lattice. This asymmetry can be applied in a controllable way for the writing and reading of the antisymmetric states.

Another possibility is the use of an optical super-lattice with two on-site potential minima. A single atom is taken to be localized at each minimum, and the two on-site atoms are now separated by a given distance. This leads to the exciton Hamiltonian $H_{ex}=\sum_{{\bf k}\nu}E_{\nu}({\bf k})\ B_{\bf k}^{\nu\dagger}B_{\bf k}^{\nu}$, where $(\nu=a,s)$ stands for the two exciton branches, with the symmetric state dispersion $E_{s}({\bf k})=E_A+J_0+4J_s\left[\cos(k_xa)+\cos(k_ya)\right]$, where $J_s=J^{\prime}+J^{\prime\prime}$, here $J_{\langle{\bf nm}\rangle}^{11}=J_{\langle{\bf nm}\rangle}^{22}=J^{\prime}$ and $J_{\langle{\bf nm}\rangle}^{12}=J_{\langle{\bf nm}\rangle}^{21}=J^{\prime\prime}$. The antisymmetric dispersion $E_{a}({\bf k})=E_A-J_0+4J_a\left[\cos(k_xa)+\cos(k_ya)\right]$, where $J_a=J^{\prime}-J^{\prime\prime}$. The antisymmetric state can transfer among nearest neighbor sites with the parameter $J_a$, and to be delocalized and to form a second exciton branch, with a narrow energy band, and is slightly bright with a finite damping rate.

The formation of an on-site molecule out of the two on-site atoms significantly changes the optical properties, as the diatomic molecules can be oriented on the lattice due to the laser polarization or the external trapping fields. This makes the system anisotropic similar to uni-axial crystals \citep{Zoubi2009b}. As presented with more details in Subsection~4.3, the  localization of the anisotropic lattice between planar cavity mirrors results in cavity photon polarization mixing, which can be observed through linear spectra.

\subsection{Coherent Transfer of Excitons among Optical Lattices}

In this Subsection we discuss the exciton coherent transfer between two different optical lattices \citep{Zoubi2010c,Zoubi2011b,Zoubi2012b}. We start with two atoms that are separated by a vector distance ${\bf R}$ and have transition dipoles $\mbox{\boldmath$\mu$}_{\alpha}$ and $\mbox{\boldmath$\mu$}_{\beta}$. The transition dipoles are equal for identical atoms, but they can have in general different orientations in space. The dipole-dipole interaction is given by
\begin{equation}
J_{\alpha\beta}({\bf R})=\frac{1}{4\pi\epsilon_0}\left\{\frac{(\mbox{\boldmath$\mu$}_{\alpha}\cdot\mbox{\boldmath$\mu$}_{\beta})}{|{\bf R}|^3}-\frac{3(\mbox{\boldmath$\mu$}_{\alpha}\cdot{\bf R})(\mbox{\boldmath$\mu$}_{\beta}\cdot{\bf R})}{|{\bf R}|^5}\right\},
\end{equation}
which behaves as inverse cube of the inter-atomic distance. This electrostatic interaction holds for inter-atomic distances smaller than the atomic transition wave length $\lambda_A$, where $E_A=hc/\lambda_A$. This interaction can be used also for two finite atomic chains, with lengths and inter-chain distances smaller than $\lambda_A$, where the transition dipoles are taken to be of the collective electronic excitations. The energy transfer can be blocked among chains of different lengths as the collective state energies can be off resonance \citep{Zoubi2012b}.

\begin{figure}
\begin{center}
\leavevmode
\includegraphics[width=85mm]{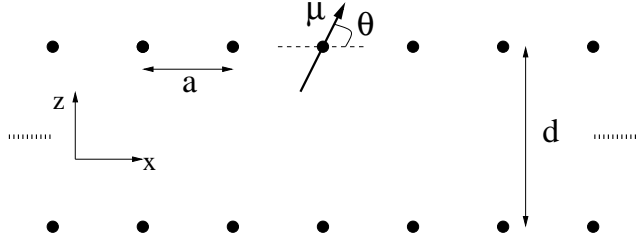}
\caption{Two parallel one-dimensional optical lattices which are separated by a distance $d$, with one atom per site and lattice constant $a$. The transition dipole makes an angle $\theta$ with the lattice direction.}
\label{14}
\end{center}
\end{figure}

Because things can be completely different for extended systems, first we present the coherent exciton transfer among two long parallel one dimensional optical lattices \citep{Zoubi2010c}, as schematized in Fig. \ref{14}. The two lattices are separated by a distance $d$, with one atom per site and lattice constant $a$.  We consider an exciton with a given wave number $k$ that propagates in one optical lattice with a fixed polarization $\theta$, and we derive the coupling parameter for the exciton transfer into the other lattice. We calculate this parameter in considering the long range dipole-dipole interaction among each pair of atoms in applying the Ewald's summation method of \citet{Born1954}. In the limit of small wave numbers, that is $ka\ll1$, and for large inter-lattices distance, that is $kd\gg1$ so that  $d\gg a$, we obtain
\begin{equation}\label{TransPara}
J(k,\theta)=\frac{\sqrt{2\pi}}{ad^2}\frac{(kd)^{3/2}}{4\pi\epsilon_0}\ e^{-kd}\mu^2\left\{2\cos^2\theta-1\right\}.
\end{equation}
The formation of excitons affect strongly the transfer parameter, as in place of the inverse cube dependence on the inter-atomic distance for the transfer parameter we get exponentially decay parameter behavior.

\begin{figure}
\begin{center}
\leavevmode
\includegraphics[width=85mm]{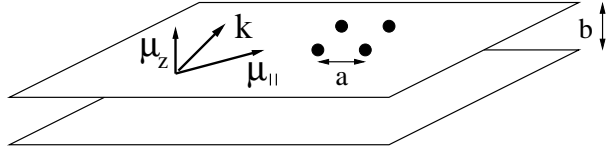}
\caption{Two parallel planar optical lattices that are separated by a distance $b$, with one atom per site and lattice constant $a$. The transition dipole includes in-plane and normal components.}
\label{15}
\end{center}
\end{figure}

Now we consider the case of two planar parallel optical lattices \citep{Zoubi2011b}, which are separated by a distance $b$, as schematized in Fig. \ref{15}. For an exciton with in-plane wave vector ${\bf k}$ that propagates in one lattice, we calculate the transfer parameter into the other lattice in applying once again the Ewald's summation method. The coupling parameter is
\begin{equation} \label{Interlattices}
J^{\prime}({\bf k})=\frac{1}{4\pi\epsilon_0}\frac{2\pi}{a^3}\left\{\left(\mbox{\boldmath$\mu$}_{\|}\cdot\hat{\bf k}\right)^2-\mu_z^2\right\}ka\ e^{-kb},
\end{equation}
where we defined $\mbox{\boldmath$\mu$}=(\mbox{\boldmath$\mu$}_{\|},\mu_z)$ with the in-plane dipole $\mbox{\boldmath$\mu$}_{\|}=(\mu_x,\mu_y)$, and the unit vector $\hat{\bf k}={\bf k}/k$. This result shows a strong anisotropic effect.

The most important feature here is the exponential decay behavior as a function of the inter-lattices distance of the exciton coupling among the two optical lattices. Even though the resonance dipole-dipole interaction has inverse cubic dependence of the distance between two localized dipoles, the appearance of excitons induce exponentially decrease dependence for the exciton collective dipole among the two lattices. For several parallel lattices the result allows the use of the assumption of energy transfer among only nearest neighbor planes.

Such coherent coupling of the exciton transfer among the two lattices can be responsible for the formation of hybrid excitons in the whole systems, in which an exciton can coherently oscillate between the two lattices. The excitation Hamiltonian for two parallel planar optical lattices is exactly as in Eq.~(\ref{TwoAtoms}), but here $\alpha$ and $\beta$ stand for the two lattices. The same diagonalization steps lead to the exciton Hamiltonian $H_{ex}=\sum_{{\bf k}\nu}E_{\nu}({\bf k})\ B_{\bf k}^{\nu\dagger}B_{\bf k}^{\nu}$, with two hybrid exciton branches. The symmetric excitons have the dispersion $E_{s}({\bf k})=E_A+J({\bf k})+J^{\prime}({\bf k})$, and the antisymmetric dispersion is $E_{a}({\bf k})=E_A+J({\bf k})-J^{\prime}({\bf k})$. Here $J({\bf k})$ is the intra-lattice coupling, and $J^{\prime}({\bf k})$ is for the inter-lattices, which is given in Eq.~(\ref{Interlattices}). Note that the two hybrid excitons have an energy split of $2J^{\prime}({\bf k})$. Also here the antisymmetric modes are dark with zero damping rate, and the superradiant ones are bright with two times the single exciton damping rate.

\section{Cavity QED with Excitons: Polaritons}

In the recent years few experiments implemented a dilute boson gas of ultracold atoms in an optical cavity \citep{Brennecke2007,Colombe2007,Slama2007}, and they reached the strong coupling regime because the collective coupling enhancement is enhanced by the square root of the atom number. In the experiment of \citet{Ritter2009} the optical lattice was formed by off resonance cavity photons, with the resulting force effects investigated in \citet{Maschler2005,Mekhov2007,Larson2008}. Optical lattices within an optical cavity operating in the resonant regime are not yet implemented. \citet{Zoubi2007,Zoubi2008b,Zoubi2009c} have introduced optical lattices as a new type of active materials for cavity QED. Beside the importance of such a system for fundamental physics and its wide range of applications, the appearance of superradiant collective states in optical lattices implies the confinement of the electromagnetic field within a cavity. Moreover, as large one and two dimensional lattices give rise for metastable states, which decouple to the free space photons, they appeal to confined photons for efficient coupling.

\subsection{Quantum Phases for an Optical Lattice within a Cavity}

\begin{figure}
\begin{center}
\leavevmode
\includegraphics[width=50mm]{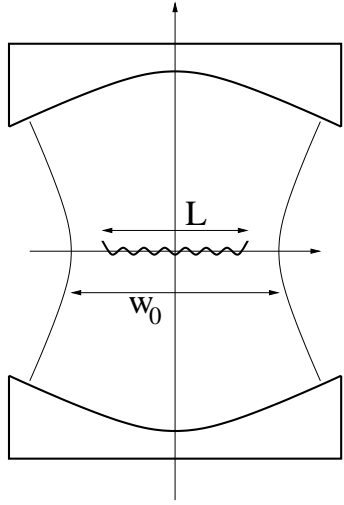}
\caption{An optical lattice localized along the waist of the Gaussian mode of a cavity with spherical mirrors. The lattice length $L$ is smaller than the mode waist $W_0$.}
\label{16}
\end{center}
\end{figure}

We begin by answering the crucial question concerning the influence of the cavity photons on the previously calculated phase diagrams \citep{Zoubi2009d}. Namely, how the different Mott insulator phases are affected due to the coupling to cavity photons. The optical lattice is taken to be localized between cavity mirrors, as in Fig.~\ref{16}. Two electromagnetic fields are required, the classical laser field which is off-resonance and used to form the optical lattice, and the quantized cavity field. A single cavity mode is assumed to be close to resonance with a single atomic transition. The previous two-component Bose-Hubbard model is generalize to include the coupling to cavity photons. This coupling changes the behavior of the  bosons through the emission and absorption of cavity photons, and that is expected to modify the phase diagram.

The two-component Bose-Hubbard model within a cavity in the resonant regime is represented by the Hamiltonian
\begin{eqnarray}
H&=&-J_g\sum_{\langle i,j\rangle}b_j^{\dagger}b_i-J_e\sum_{\langle i,j\rangle}c_j^{\dagger}c_i-\mu_g\sum_i n_i^g-\mu_e\sum_in_i^e \nonumber \\
&+&\frac{U_g}{2}\sum_in_i^g\left(n_i^g-1\right)+\frac{U_e}{2}\sum_in_i^e\left(n_i^e-1\right)+U_{eg}\sum_in_i^gn_i^e \nonumber \\
&+&\mu_c\ n^c+\sum_i\left(f\ c_i^{\dagger}b_i\ a+f^{\ast}\ a^{\dagger}\ b_i^{\dagger}c_i\right),
\end{eqnarray}
where here $\mu_g=\bar{\mu}_g-\varepsilon^g$ and $\mu_e=\bar{\mu}_e-\varepsilon^e$, with $\bar{\mu}_g$ and $\bar{\mu}_e$ are the ground and excited state atoms chemical potentials, respectively. $\varepsilon^g_i$ and $\varepsilon^e_i$ are taken to be site independent, hence we dropped their index $i$. We use also $\mu_c=\bar{\mu}_c-\varepsilon_c$, where $\varepsilon_c$ is the cavity mode energy and $\bar{\mu}_c$ is the photon chemical potential. Furthermore, the photon-excitation coupling $f$ is assumed to be site independent, which is the case for a homogeneous optical lattice parallel to the cavity mirrors. We also defined the mean atom and photon number operators, by $n_i^g=b_i^{\dagger}b_i$, $n_i^e=c_i^{\dagger}c_i$, and $n^c=a^{\dagger}a$. Here $a^{\dagger}$ and $a$ are the creation and annihilation operators of a cavity mode, and close to resonance one has $\varepsilon_c\sim\varepsilon_e-\varepsilon_g$.

The modified phase diagram is calculated in applying the consistent mean-field theory suggested in \citet{Sheshadri1993}, as the conventional Bogoliubov approximation does not predict the correct quantum phase transition \citep{vanOosten2001}. In this mean-field theory the atoms are subject to the mean-field of the neighboring sites and of the other kind of bosons. Hence the Hamiltonian can be separated into on-site terms and the rest, which can be written as a sum of a free part and an interacting part, and where we drop the site index in the following. Then the interaction part can be treated in perturbation theory. In this method atom-atom interactions are treated exactly while the hopping and the coupling to photons are treated approximately. More complex quantum phases are discussed in \citet{Bhaseen2009} beyond the mean-field theory.

We present an example of phase diagram by plotting the scaled chemical potentials as a function of the atom-atom interactions. The system in total has a large number of open parameters, so that we cannot represent the whole phase space. For our arguments here we choose the parameters in such a way to allow the existence the Mott insulator phase and to emphasize the modifications brought about by the coupling to the cavity photons. We concentrate on the case with two atoms per site, that is in the case $n_e=1,\ n_g=1$. We assume one photon in average per site, that is $\bar{n}_c=1$, where $\bar{n}_c=n_c/N$, although this assumption is not so crucial. For $\bar{n}_c=1$ we have a large number of photons in the cavity, so that photon number fluctuations are suppressed.

As we concentrate in the following on the influence of the coupling to the cavity photons, we assume $J_g=J_e=J$, and all the parameters will be scaled by $zJ$. Furthermore, we assume zero ground state energy, $\varepsilon_g=0$, and the cavity mode  in resonance with the electronic excitation, that is $\varepsilon_e=\varepsilon_c$. Moreover, by assuming a resonant pump-cavity mode  the chemical potential for cavity photons becomes zero, that is $\bar{\mu}_c=0$. In the following phase diagram plots we use $\varepsilon_c=100$ and for the excitation-photon scaled coupling we used $F=25$, where $F=|f|^2\bar{n}_c(z^2J_gJ_e)^{-1}$.

\begin{figure}
\begin{center}
\leavevmode
\includegraphics[width=65mm]{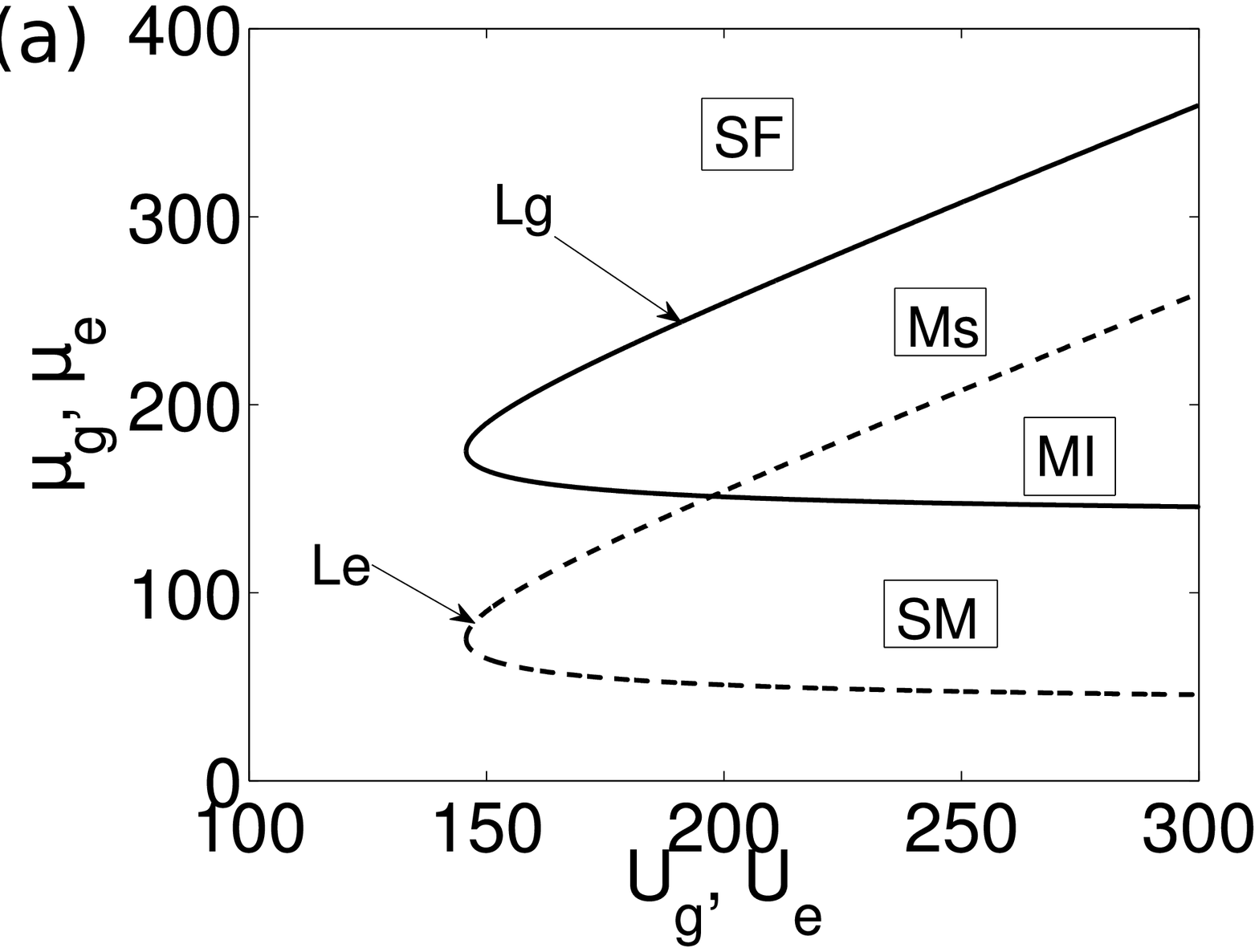}\ \ \ \includegraphics[width=65mm]{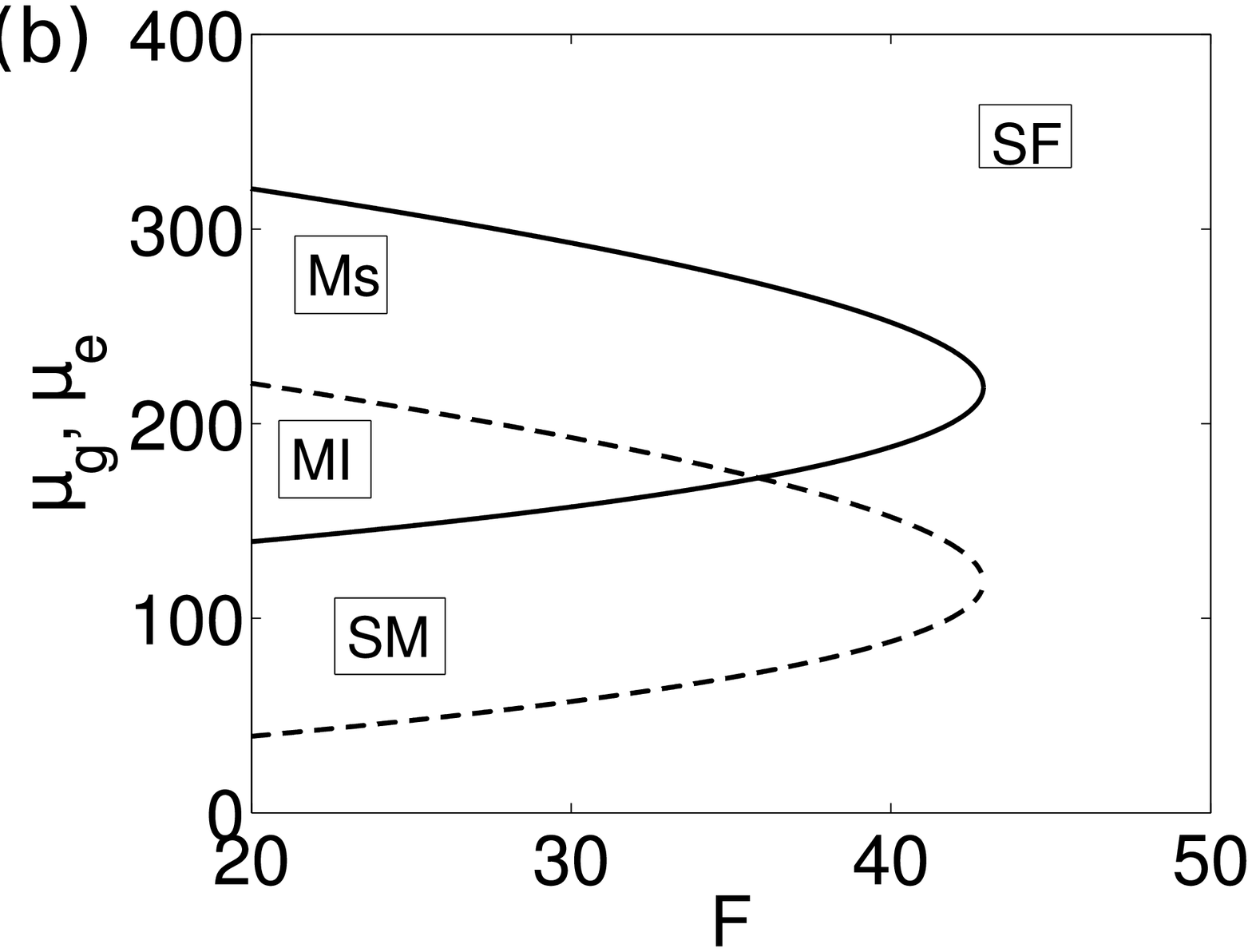}
\caption{(a) Scaled phase diagram: $\mu_g$ vs. $U_g$, and $\mu_e$ vs. $U_e$, for $n_g=n_e=1$. We have $\varepsilon_c=100$, $F=25$, $U_{eg}=15$, and $\bar{n}_c=1$. The dashed line is for the excited state atoms, and the full line is for the ground state atoms. (b) Scaled phase diagram: $(\mu_g,\mu_e)$ vs. $F$, for  $U_g=U_e=250$. Figure reprinted with permission from~\citet{Zoubi2009d} Copyright 2009 by American Physical Society.}
\label{17}
\end{center}
\end{figure}

Fig.\ref{17}(a) reports the phase diagram for the planes $\mu_g$ vs. $U_g$, and $\mu_e$ vs. $U_e$. We used $U_{eg}=15$ for the scaled atom-atom coupling. Similar to the case without a cavity, the plot shows the superfluid and the Mott insulator phases for one ground or excited state atom per site, that is $n_g=n_e=1$, with the Le and Lg transition lines. Beside the SF and MI Mott regions, we have as before the SM and MS regions. The results of Fig.~\ref{17}(a) should be compared  to those of Fig.~\ref{3} for the phase diagram of a two-component Bose-Hubbard model without coupling to cavity photons. We deduce that the effect of the coupling to cavity photons is to shift the Mott insulator phase to larger atom-atom interactions, by one order of magnitude in the present case. As we used $\bar{\mu}_c=0$, the ground state transition line is shifted relative to the excited state one.

The coupling to cavity photons tend to delocalize the excited atoms through the process in which an excited atom jumps to the ground state by emitting a cavity photon, in turn absorbed by another atom in the lattice located farther apart. In this situation the atom-atom interaction competes against the atom hopping and the coupling to cavity photons, and to recover the Mott insulator phase within a cavity one need to apply a deeper optical lattice.

Fig. \ref{17}(b) reports the scaled chemical potentials $\mu_g$ and $\mu_e$ as a function of $F$, for $U_{eg}=15$, and $U_g=U_e=250$. It is clear that the Mott insulator phase appears only for a limited range of coupling parameters, and for a limited range of mean cavity photon number per site, which can range from few cavity photons up to about one photon per lattice site. Beyond some critical values of $F$ the atom-atom interaction is not large enough to form the Mott insulator phase and  only the superfluid phase persists. More Mott insulator phases with other number of atoms per site are presented in \citet{Zoubi2009d}.

In all the above discussion we assumed long life times for both the excited atoms and the cavity photons. This assumption can be a good approximation, e.g., for Rydberg atoms in a high-Q cavity. The overlaps between the Mott insulator phases of different number of atoms per site make it possible to choose the quantum phase such that following the decay of an excited atom the systems may remain in one of the Mott insulator phases. The lifetime of the cavity photons has less influence on the Mott insulator phase, as the leak of a photon in the Mott insulator keeps the system inside the same phase zone.

The calculated phase diagrams provide us with a new tool for controlling the system quantum phases and their transitions. In place of controlling the external laser fields, which form the optical lattice, it is possible to control the cavity photons. For example we can chose the different quantum phases by fixing the average number of photons within the cavity, which can be easily achieved via controlling the external pump.

\subsection{Cavity Polaritons of Planar Two-Dimensional Optical Lattices}

\begin{figure}
\begin{center}
\leavevmode
\includegraphics[width=60mm]{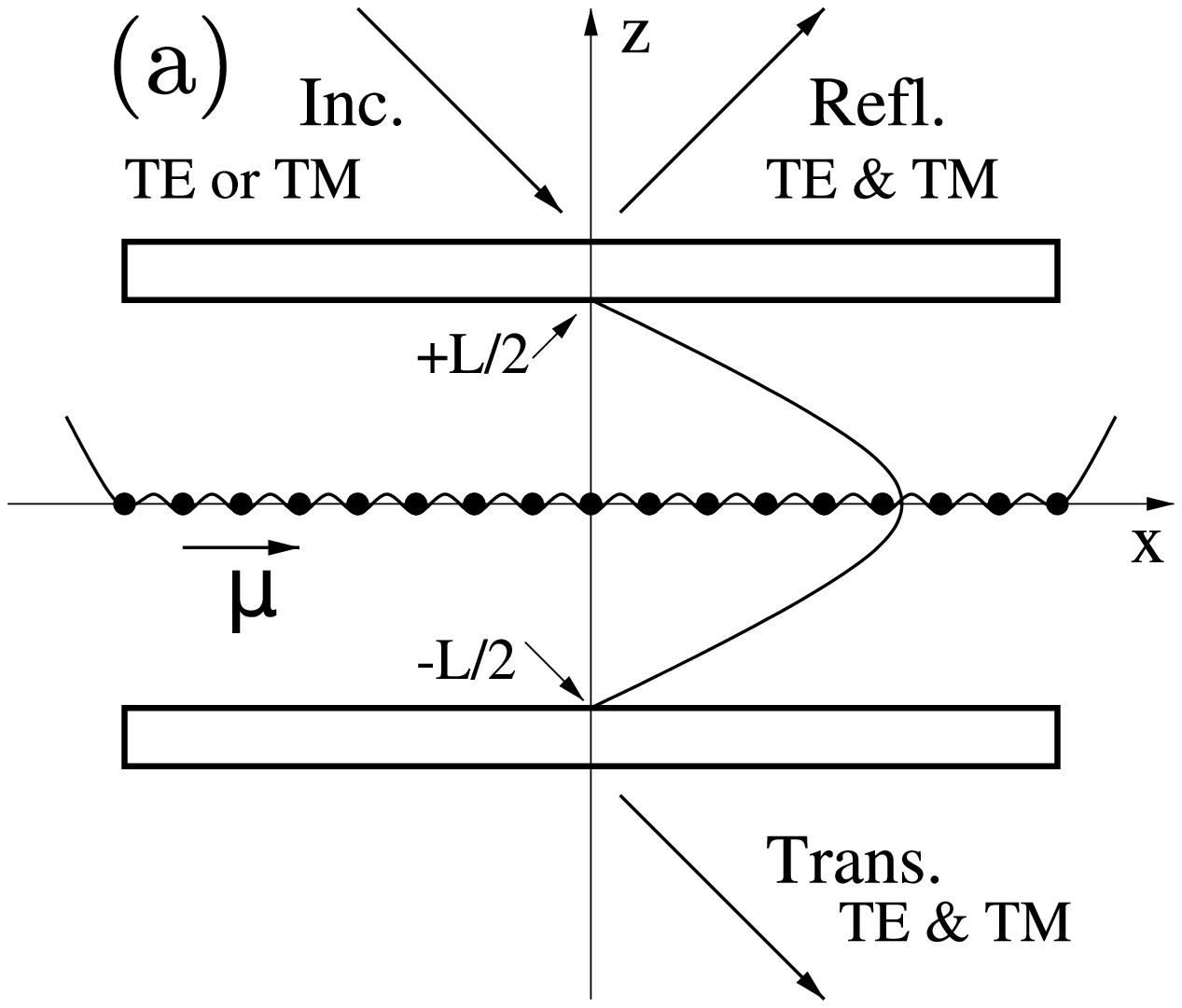}\ \ \ \ \includegraphics[width=60mm]{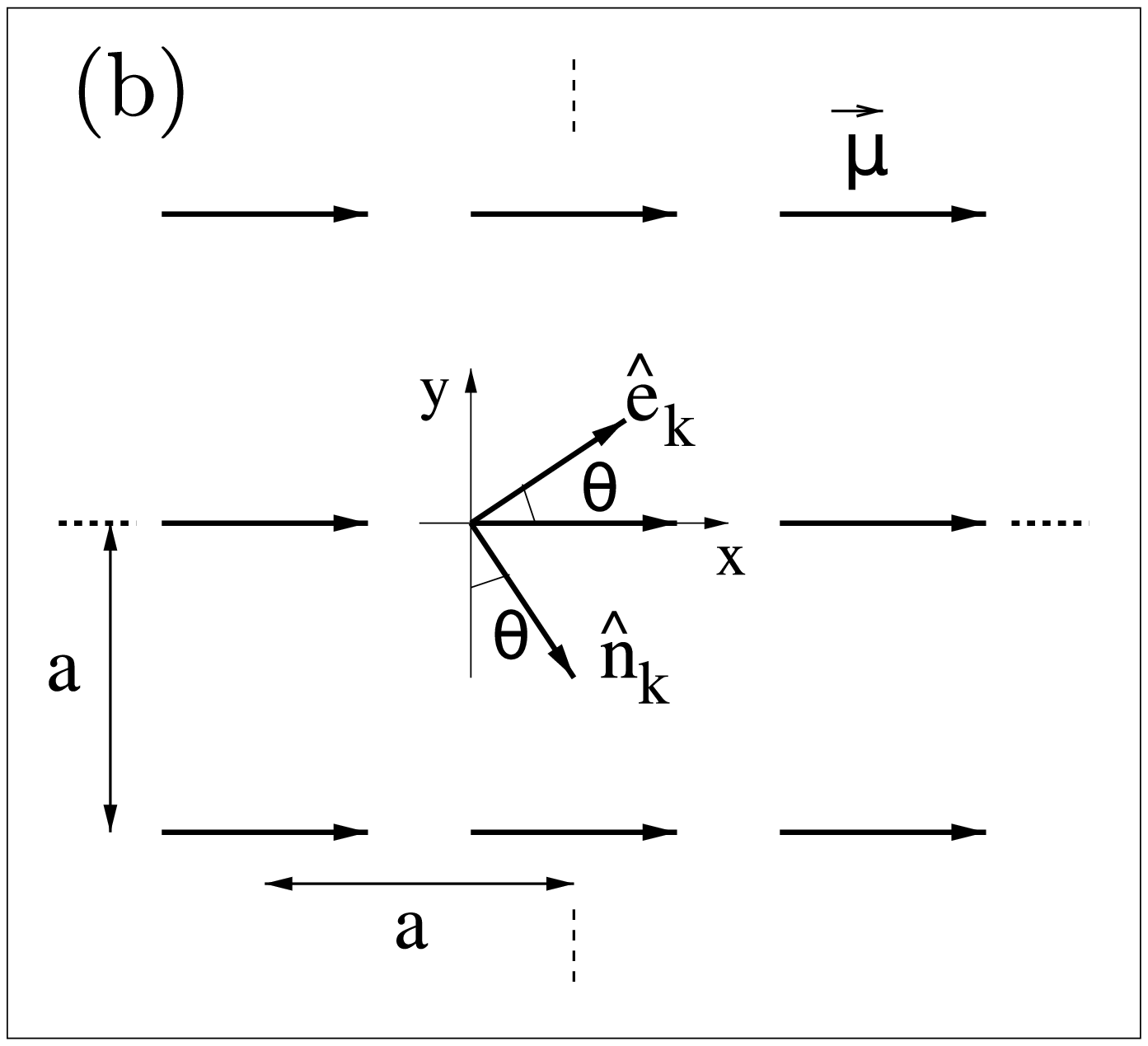}
\caption{ (a) Cold atom optical lattice located within a cavity of two parallel planar mirrors. The two-dimensional optical lattice created by external lasers is located in the middle between, and within a plane parallel to, the two cavity mirrors. The incident, transmitted, and reflected fields, with different polarisations are indicated. (b) Schematic plot of an optical lattice plane with oriented induced transition dipoles. The unitary vectors defining the photon polarization, $\hat{e}_{\bf k}$, and its orthogonal direction, $\hat{n}_{\bf k}$, are reported.}
\label{21}
\end{center}
\end{figure}

Here we introduce optical lattice ultracold atoms into an optical resonator as an active material in a cavity QED \citep{Zoubi2007}. To represent the main results of this set-up we consider a planar optical lattice that enclosed in the middle between planar cavity mirrors, as represented in Fig.~\ref{21}(a). We start by assuming an ideal cavity and later we include the damping of cavity photons into the outside radiation field. The cavity photons are free to propagate in the cavity plane with a given in-plane wave vector ${\bf k}$, and they have discrete modes in the perpendicular direction with wave numbers $k_z=m\pi/L$, where $L$ is the distance between the mirrors and $m$ is an integer $(m=1,2,\cdots)$. In the perpendicular direction we consider only a single mode, the one that close to resonance with the atomic transition, e.g. the one with $m=1$ with a maximum field at the lattice plane, the other modes having higher energies. In this part we consider isotropic optical lattice, hence cavity photons with a fixed polarization, i.e. TE and TM modes or any coherence superposition of them. In Subsection 4.3 we treat an anisotropic optical lattice of uni-axial symmetry that give rise to photon polarization mixing.

For the coupled exciton-photon system, in the dipole interaction and for the rotating wave approximation, the Hamiltonian is written as
\begin{equation}
H=\sum_{\bf k}\hbar\left\{\omega_{ex}(k)\ B_{\bf k}^{\dagger}B_{\bf k}+\omega_{ph}(k)\ a^{\dagger}_{\bf k}a_{\bf k}+f_k\ B^{\dagger}_{\bf k}a_{\bf k}+f^{\ast}_k\ a^{\dagger}_{\bf k}B_{\bf k}\right\}.
\end{equation}
Here $a^{\dagger}_{\bf k},\ a_{\bf k}$ are the creation and annihilation operators of a cavity photon with in-plane wave vector ${\bf k}$. The cavity photon dispersion reads
\begin{equation}
\omega_{ph}(k)=\frac{c}{\sqrt{\epsilon}}\sqrt{k^2+\left(\frac{\pi}{L}\right)^2},\end{equation}
where $L$ is the distance between the cavity mirrors, and $\epsilon \approx 1$ is the cavity medium dielectric constant. The coupling parameter, e.g. for a lattice of square symmetry, is given by
\begin{equation}
\hbar f_k=-i\sqrt{\frac{\hbar\omega_{ph}(k)\mu^2}{2\epsilon_0La^2}}.
\label{dispersion}
\end{equation}
In the case of excitons and photons close to resonance having small wave numbers, that is $\omega_{ex}(k)\approx\omega_{ph}(k)$ for $k\sim 0$, the excitons can taken to be dispersion-less with $\omega_{ex}(k)\sim \omega_A+4J$. Translational symmetry leads to coupling between excitons and photons with the same in-plane wave vectors, namely to a momentum conservation.

In the strong coupling regime where the exciton and photon damping rates are smaller than the coupling parameter, the Hamiltonian can be diagonalize by introducing polaritons, through the transformation
\begin{equation}
A_{{\bf k}\pm}=X_{k}^{\pm}\ B_{\bf k}+Y_{k}^{\pm}\ a_{{\bf k}},
\end{equation}
where the exciton and photon amplitudes are
\begin{equation}
X_{k}^{\pm}=\pm\sqrt{\frac{\Delta_k\mp\delta_k}{2\Delta_k}}\ ,\ Y_{k}^{\pm}=\frac{f_{k}}{\sqrt{2\Delta_k(\Delta_k\mp\delta_k)}},
\end{equation}
with $\Delta_k=\sqrt{\delta_k^2+|f_k|^2}$, and we defined the exciton-photon detuning by $\delta_k=(\omega_{ph}(k)-\omega_{ex})/2$. In this regime the real system excitations are polaritons, quasi-particles formed by a coherent superposition of excitons and photons, with the system coherently oscillating between those two. The polariton Hamiltonian is given by
\begin{equation}
H=\sum_{{\bf k}r}\hbar\Omega_r(k)\ A^{\dagger}_{{\bf k}r}A_{{\bf k}r},
\end{equation}
with the polariton dispersions
\begin{equation}
\Omega_{\pm}(k)=\frac{\omega_{ph}(k)+\omega_{ex}}{2}\pm\Delta_k.
\end{equation}
We get two polariton branches with a splitting of $2\Delta_k$, the intersection point appearing at $\delta_k=0$ and the vacuum Rabi splitting equal to $2f_k$. The polaritons are denoted as upper and lower branches.

\begin{figure}
\begin{center}
\leavevmode
\includegraphics[width=65mm]{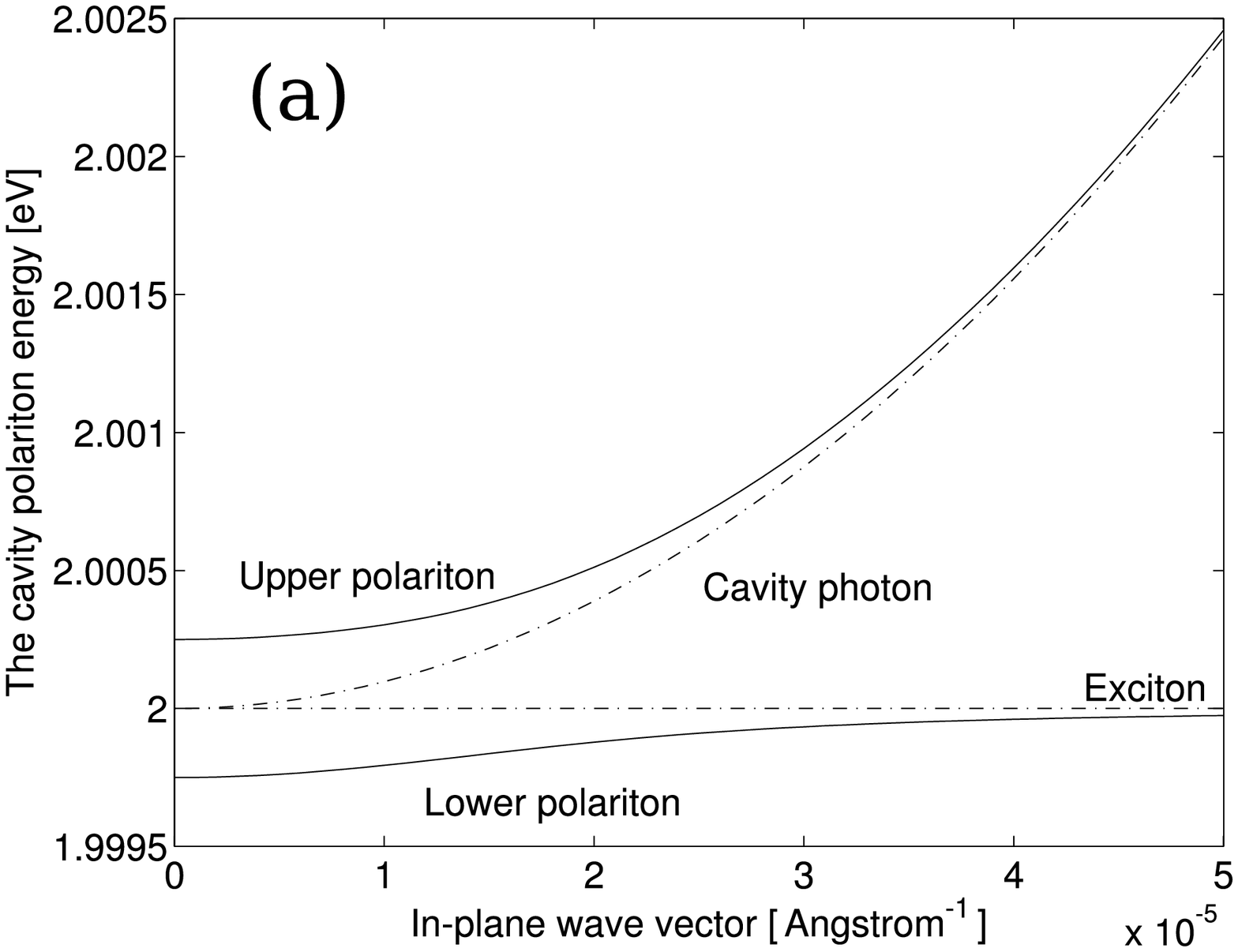}\ \ \ \includegraphics[width=65mm]{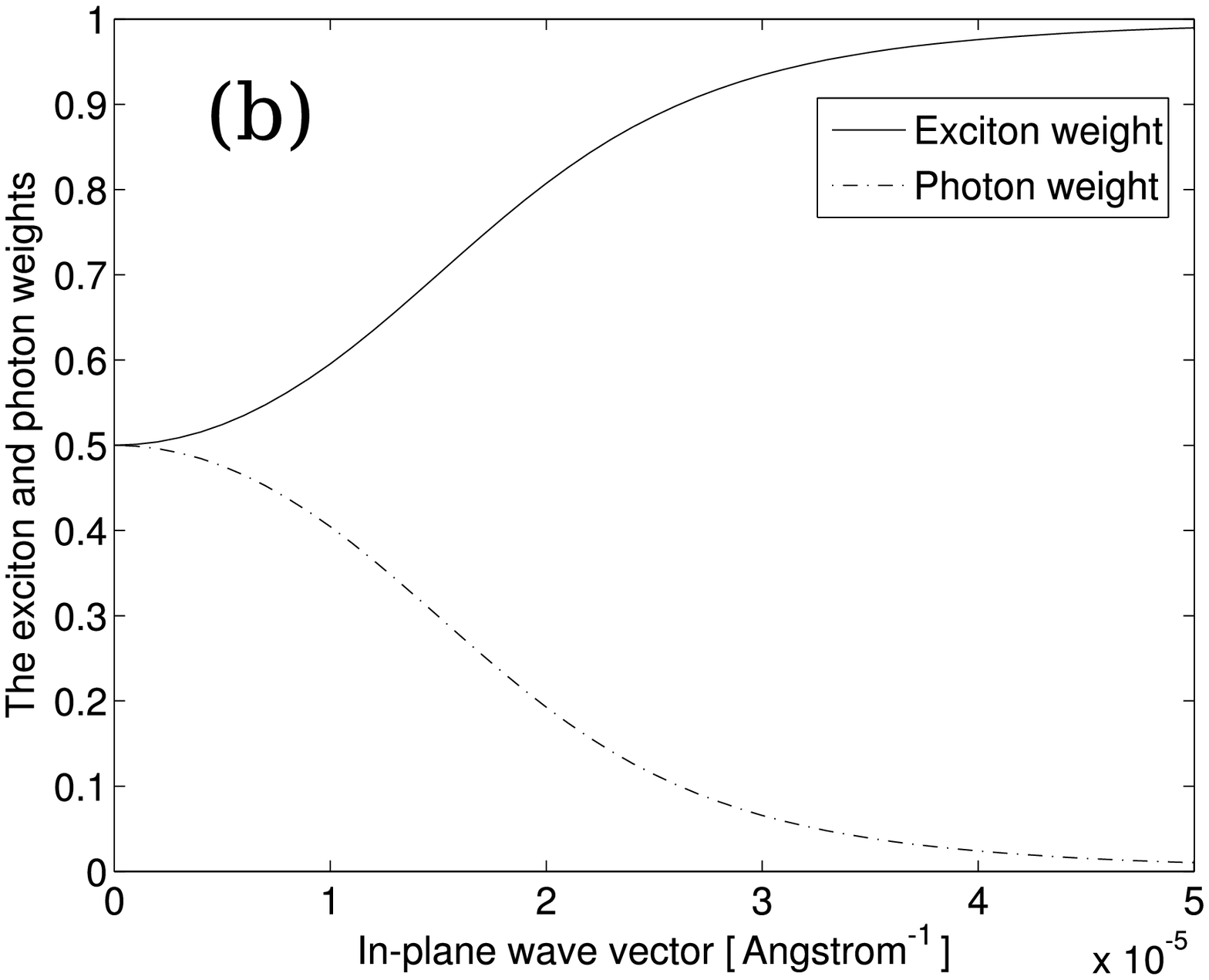}
\caption{(a) The upper and the lower polariton branches vs. the in-plane wave vector $k$. The dashed-dotted line denotes for the exciton dispersion, and the dashed-dotted parabola denotes for the cavity photon dispersion. (b) The exciton and photon weights vs. the in-plane wave vector $k$ for the lower polariton branch. Figure reprinted with permission from~\citet{Zoubi2007} Copyright 2007 by American Physical Society.}
\label{19}
\end{center}
\end{figure}

The plots of Fig.~\ref{19}(a) report  the results for the upper and lower polariton branches in addition to the cavity photon and exciton dispersions, using the typical parameters listed in the following. The  exciton energy around small in-plane wave vectors is taken to be $\hbar\omega_{ex}(0)=2\ eV$, and a zero detuning between the exciton and the cavity photon dispersions at zero in-plane wave vector. The transition dipole is $\mu=2\ e\AA$, and the optical lattice constant is $a=2000\ \AA$. The exciton-photon coupling parameter is $|f|\approx 4\times 10^{11}\ Hz$, by neglecting the $k$ dependence for small in-plane wave vectors. At zero in-plane wave vector the splitting  is clearly given by the Rabi frequency. Fig.~\ref{19}(b) reports the $|X^-_k|^2$ and $|Y^-_k|^2$ exciton and photon weights for the lower polariton branch. At the exciton-photon intersection point the polaritons are half exciton and half photon, that is $|X_{k}^{\pm}|^2=|Y_{k}^{\pm}|^2=1/2$. At large wave vectors the lower branch becomes excitonic, that is $|X_{k}^{-}|^2\approx1,\ |Y_{k}^{-}|^2\approx0$, and the upper branch becomes photonic, that is $|X_{k}^{+}|^2\approx0,\ |Y_{k}^{+}|^2\approx1$.

The above discussion considered an ideal cavity with perfect mirrors. But in order to observe the system eigenmodes, we need to couple the internal cavity modes to the external world. This is done using the standard input-output formalism, which is well proved for high-Q cavities \citep{Gardiner2000}. In this approach the internal cavity field modes and the external free radiation field are quantized separately and weakly coupled via the cavity mirrors. This leads to the cavity field damping but also allows to calculate consistently the relation between the cavity input and output fields. To include the atomic spontaneous emission as an energy loss term, we add a phenomenological effective exciton damping. This approach finally allows to calculate the transmission $T$, reflection $R$, and absorption $A$ spectra for a given incident field in the cavity configuration of Fig.~\ref{21}(a).

\begin{figure}
\begin{center}
\leavevmode
\includegraphics[width=70mm]{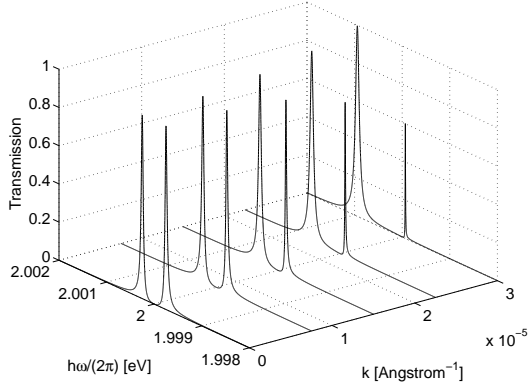}
\caption{The transmission spectrum vs. the angular frequency $\omega$ for different in-plane wave vectors. The two peaks correspond to the upper and lower polariton branches at each in-plane wave vector. Figure reprinted with permission from~\citet{Zoubi2007} Copyright 2007 by American Physical Society.}
\label{20}
\end{center}
\end{figure}

For the previously investigated system we plot in Fig.~\ref{20} the transmission spectra for different wave vectors. Here we choose the following parameters for the damping rates: exciton damping rate $\Gamma_{ex}=1.5\times 10^{9}\ Hz$, upper and lower mirror damping rates $\gamma_U=\gamma_L=7.5\times10^{10}\ Hz$. The peaks of the transmission correspond to the two polariton branches, which are the real eigenmodes of the system. The plot indicates that high transmission peaks are obtained for polaritons which are much more photonic than excitonic. At zero wave vector, as the polariton is half exciton and half photon, the upper and the lower spectra are identical. As the wave vector increase the upper branch becomes more photonic than excitonic, and the opposite for the lower one. Hence, the line-width of the upper branch becomes wider, which equals the cavity line-width, and the lower one becomes narrower, which equals the atom line-width. The reflection and absorption spectra can be easily derived in the same manner. The present results for the Mott insulator with one atom per site are expected to be significantly different from the other cases with other number of atoms per site and also from the superfluid phase.

The widely used time-of-flight method for observing the different quantum phases in optical lattices with their quantum phase transitions is a destructive observation tool \citep{Greiner2002,Bloch2008}. As the laser  creating the lattice is switched off and after waiting some time a picture is taken, the system is destroyed and  in each observation one need to prepare a new system. We suggest to use the optical linear spectra as a non-destructive observation tool for the quantum phases.  The appearance of exciton and polariton provides a direct observation tool that can be repeated without modifying the quantum phase. Later we present other optical spectra, and the comparison between them allows us to extract information concerning the quantum phases.

\subsubsection{Two-Atoms per site Optical Lattice in a Planar Cavity}

The case of a two  dimensional optical lattice in the Mott insulator phase with two atoms per site was investigated in Subsection 3.5. Here this system is introduced into a cavity where the optical lattice is taken to be localized in the middle and parallel to two planar cavity mirrors \citep{Zoubi2008b}, as in Fig.~\ref{21}(a). The main objective is to examine how the previously derived dark and bright excitons behave within a cavity in the strong coupling regime. We start by considering the case of asymmetric optical lattice in which the two on-site atoms are separated by an average distance. The electronic transitions are coupled to the cavity photons with the interaction Hamiltonian $H_I=\sum_{{\bf k},i,\alpha}\left(f_{{\bf k},i}^{\alpha}\ B_i^{\alpha\dagger}a_{\bf k}+f_{{\bf k},i}^{\alpha\ast}\ a^{\dagger}_{\bf k}B_i^{\alpha}\right)$, with $(\alpha=1,2)$ for the two atoms. The coupling parameter $f_{{\bf k},i}^{\alpha}$ between atom $\alpha$ at site $i$ and a photon with wave vector ${\bf k}$ is given by $f _{{\bf k},i}^{\alpha}=-i\sqrt{\hbar\omega_{ph}(k)\mu^2(2LNa^2\epsilon_0)^{-1}}e^{i{\bf k}\cdot{\bf r}_i^{\alpha}}$, where ${\bf r}_i^{\alpha}$ is the atom position. In terms of on-site symmetric and antisymmetric operators we get $H_I=\sum_{{\bf k},i,\nu}\left(f_{{\bf k},i}^{\nu}\ B_i^{\nu\dagger}a_{\bf k}+f_{{\bf k},i}^{\nu\ast}\ a^{\dagger}_{\bf k}B_i^{\nu}\right)$, with $(\nu=s,a)$ for the two modes, where we defined the symmetric and antisymmetric coupling parameters $f_{{\bf k},i}^{s}=(f_{{\bf k},i}^{1}+f_{{\bf k},i}^{2})\sqrt{2}^{-1}$, and $f_{{\bf k},i}^{a}=(f_{{\bf k},i}^{1}-f_{{\bf k},i}^{2})\sqrt{2}^{-1}$. As the on-site interatomic distance is much smaller than the atomic transition wavelength, the symmetric state coupling is much larger than the antisymmetric one. The coupling can be controlled through the external laser fields creating the optical lattice in using different lasers at different directions. It is possible to switch the antisymmetric coupling on and off by converting the lattice site from asymmetric into symmetric one.

For symmetric optical lattices we have approximately $f_{{\bf k},i}^{1}=f_{{\bf k},i}^{2}=f_{{\bf k},i}$, where $f_{{\bf k},i}^{s}=\sqrt{2}f_{{\bf k},i}$ and $f_{{\bf k},i}^{a}=0$. Then the antisymmetric excitation states are almost decoupled from the cavity-photons, and they can be considered as dark states. Only the symmetric excitation states are strongly coupled to the cavity-photons. In this case by using the symmetric and antisymmetric excitonic picture in the momentum space, we get the total Hamiltonian
\begin{equation}
H=\sum_{{\bf k}}\hbar\left\{\omega^{s}_{ex}(k)\ B_{\bf k}^{s\dagger}B_{\bf k}^{s}+\omega_{ph}(k)\ a^{\dagger}_{\bf k}a_{\bf k}f_{{k}}\ B_{\bf k}^{s\dagger}a_{\bf k}+f_{{k}}^{\ast}\ a^{\dagger}_{\bf k}B_{\bf k}^{s}\right\}+\sum_{i}\hbar\omega^{a}_{ex}\ B_i^{a\dagger}B_i^{a},
\end{equation}
where $f_{{\bf k}}^{s}\equiv f_{k}=-i\sqrt{\frac{\hbar\omega_{ph}(k)\mu^2}{La^2\epsilon_0}}$, and $f_{{\bf k}}^{a}=0$. In the strong coupling regime the symmetric excitons and the cavity photons are coherently mixed to form the new system excitations which are the polaritons. The above Hamiltonian first part can be easily diagonalize to give the following polariton Hamiltonian:
\begin{equation}
H=\sum_{{\bf k}r}\hbar\Omega_r(k)\ A^{\dagger}_{{\bf k}r}A_{{\bf k}r}+\sum_i\hbar\omega^a_{ex}\ B_i^{a\dagger}B_i^{a}.
\end{equation}
Here we can use exactly the definition of polaritons by replacing the exciton energy with the symmetric one and the coupling with the symmetric one. The antisymmetric states are on-site localized and dark, but they  can be written and read in a controllable way, this fact making them useful as  memory devices.

\subsection{Optical Lattices with Oriented Dipoles}

In the present Subsection we assume the transition dipole to have a fixed direction \citep{Zoubi2009b}. This case can occur for optically pumped atoms in a fixed Zeeman sub-level of a given angular momentum. Also molecules can have oriented transition dipole, e.g. diatomic molecules, where the electronically excited states have transition dipoles oriented along the molecule axis. In combination with external traps and fixed optical lattice laser polarizations the atomic and molecular transition dipole are aligned along a specific direction.This orientation that makes the lattice to be anisotropic with uniaxial symmetry, as represented in Fig \ref{21}(b) for a planar optical lattice. The anisotropic properties can be implemented as a signature for the formation of on-site molecules in the optical lattice. For example the formation of homonuclear diatomic molecules out of the two on-site atoms in the Mott insulator phase can be manifested through the anisotropic optical spectra. We will show that the electronic excitations and the cavity photons of both TE and TM polarizations are coherently mixed to form polariton branches, which can be directly observed in the linear optical spectra. For example, for an incident field of TE polarization we obtain both TE and TM transmitted and reflected fields.

The planar optical lattice is taken to be localized in the middle and parallel to cavity mirrors, as presented in Fig.~\ref{21}(a). Here we generalize the previous presentation of the cavity photons in order to include the polarizations. The electromagnetic field is free in the cavity plane with in-plane wave-vector ${\bf k}$, and is quantized in the perpendicular, $z$, direction with wave numbers $k_z=\frac{m\pi}{L}$, where $L$ is the distance between the cavity mirrors, and $m$ takes integer numbers, ($m=0,1,2,3,\cdots$). Thus for each cavity mode $({\bf k}, m)$ we have two possible polarizations TE and TM, which will be denoted by the indexes $(s)$ and $(p)$, respectively. The two cavity photon polarizations, $(\nu=s,p)$, are assumed to be degenerate. Because  we will consider only cavity modes with $m=1$ to be close to resonance to the atomic transition,  we have  for the two polarizations $\omega_{{\bf k},s}=\omega_{{\bf k},p}=\omega_{ph}(k)$, with the cavity photon dispersion $\omega_{ph}(k)$ given by Eq.~\eqref{dispersion}. The cavity photon Hamiltonian is $H_{ph}=\sum_{k,\nu}\hbar\omega_{ph}(k)\ a_{{\bf k},\nu}^{\dagger}a_{{\bf k},\nu}$, where $a_{{\bf k},\nu}^{\dagger}$ and $a_{{\bf k},\nu}$ are the creation and annihilation operators of a cavity photon in the mode $({{\bf k}\nu})$, respectively. The electric field operators and  the field spatial functions are reported by \citep{Haroche1992}. We introduce the following unit vectors $\hat{e}_{z}$ along the $z$ axis, $\hat{e}_{\bf k}$ along the ${\bf k}$ in-plane wave vector, that is $\hat{e}_{\bf k}={\bf k}/k$, and $\hat{n}_{\bf k}=\hat{e}_{\bf k}\times\hat{e}_{z}$, as illustrated in Fig.~\ref{21}(b).

The optical lattice is located at $z=0$, the cavity middle, and parallel to the cavity mirrors. The coupling between the atomic transition and the cavity modes is given by the Hamiltonian $H_{i}=\hbar \sum_{{\bf k},\nu}\left\{f_{\bf k}^{\nu}\ B_{\bf k}^{\dagger}a_{{\bf k}\nu}+f_{\bf k}^{\nu\ast}\ a_{{\bf k}\nu}^{\dagger}B_{\bf k}\right\}$, where the coupling parameter is $\hbar f_{\bf k}^{\nu}=i\sqrt{\frac{\hbar\omega_{\bf k}}{La^2\epsilon_0}}\ \left(\vec{\mu}\cdot{\bf u}_{\nu}({\bf k})\right)$. For example, we take $\vec{\mu}$ to be real, e.g. for $\vec{\mu}=\mu\hat{x}$,  see Fig. 21(a), and also we have $\hat{e}_{\bf k}=\cos\theta\ \hat{x}+\sin\theta\ \hat{y}$ and $\hat{n}_{\bf k}=\sin\theta\ \hat{x}-cos\theta\ \hat{y}$, to get $\hbar f_{\bf k}^{s}=iC_{\bf k}\ \sin\theta$ and $\hbar f_{\bf k}^{p}=C_{\bf k}\left(\frac{\omega_0}{\omega_{\bf k}}\right)\ \cos\theta$, where $C_{\bf k}=\sqrt{\frac{\hbar\omega_{\bf k}\mu^2}{La^2\epsilon_0}}$, and $\omega_0=c\pi/L$.

\begin{figure}
\begin{center}
\leavevmode
\includegraphics[width=85mm]{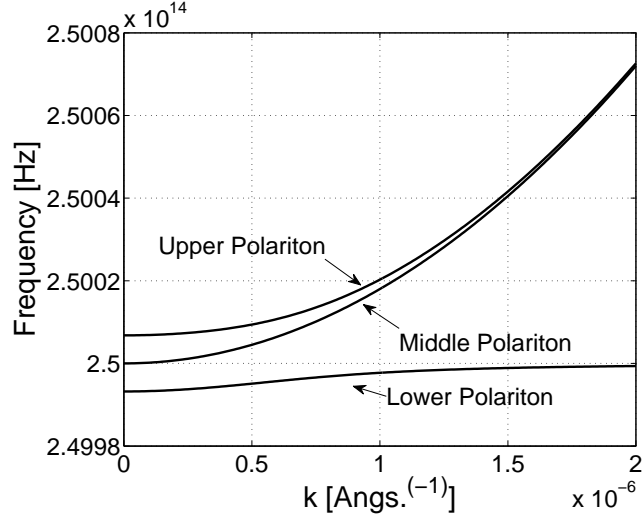}
\caption{The three polariton frequency dispersions $\Omega_{r}/2\pi$ vs. $k$, at $\theta=\pi/4$ and other parameters reported in the text. Figure reprinted with permission from~\citet{Zoubi2009b} Copyright 2009 by Institute of Physics.}
\label{22}
\end{center}
\end{figure}

The total Hamiltonian of the coupled excitations and photons is given by
\begin{equation}
H=\hbar\sum_{\bf k}\left\{\omega_{ex}\ B_{\bf k}^{\dagger}B_{\bf k}+\sum_{\nu}\omega_{\bf k}\ a_{{\bf k}\nu}^{\dagger}a_{{\bf k}\nu}+\sum_{\nu}\left(f_{\bf k}^{\nu}\ B_{\bf k}^{\dagger}a_{{\bf k}\nu}+f_{\bf k}^{\nu\ast}\ a_{{\bf k}\nu}^{\dagger}B_{\bf k}\right)\right\}.
\end{equation}
In the strong coupling regime the real system eigenmodes are cavity polaritons, which are obtained by diagonalizing the above Hamiltonian. We obtain the polariton Hamiltonian $H = \sum_{{\bf k},r} \hbar\Omega_{{\bf k}r}\ A_{{\bf k}r}^{\dagger}A_{{\bf k}r}$, characterized by three polariton branches, with the eigenfrequencies $\Omega_{{\bf k}\pm}=\frac{\omega_{\bf k}+\omega_{ex}}{2}\pm\Delta_{\bf k}$ and $\Omega_{{\bf k}0}=\omega_{\bf k}$, where $\Delta_{\bf k}=\sqrt{\delta_{\bf k}^2+|f_{\bf k}|^2}$ and $|f_{\bf k}|^2=\sum_{\nu}|f_{\bf k}^{\nu}|^2$, with the excitation-photon detuning $\delta_{\bf k}=(\omega_{\bf k}-\omega_{ex})/2$. For the case of exciton frequency of $\omega_{ex}/2\pi=2.5\times10^{14}\ Hz$, with a distance between the cavity mirrors of $L=c\pi/\omega_0\approx 3.77\ \mu m$,  zero detuning between the atomic transition and the cavity photon at $k=0$, at the angle $\theta=\pi/4$, transition dipole of $\mu=2\ e\AA$, and lattice constant $a=2000\ \AA=0.2\ \mu m$,  Fig.~\ref{22}  reports the three polariton frequency dispersions $\Omega_{{\bf k}r}/2\pi$ as a function of $k$.

In the limit of large detuning, where $\delta_{\bf k}\gg|f_{\bf k}|$, we get $\Omega_{{\bf k}+}\approx\omega_{\bf k}+\frac{|f_{\bf k}|^2}{2\delta_{\bf k}}$, $\Omega_{{\bf k}-}\approx\omega_{ex}-\frac{|f_{\bf k}|^2}{2\delta_{\bf k}}$, and $\Omega_{{\bf k}0}=\omega_{\bf k}$. The upper $(0)$ branch is a photon with a shift due to the coupling to the excitation, and the lower $(+)$ one is an excitation with a light shift due to the coupling to the cavity photon. The middle polariton branch is pure photonic. In that limit we get birefringence \citep{Born1997}, determined by two refracted cavity fields, the ordinary field of the $(0)$ branch, and the extraordinary field of the $(+)$ branch. The polariton operators are in general a coherent superposition of excitations and photons of both polarizations, namely we have $A_{{\bf k}r}=X_{{\bf k}r}\ B_{\bf k}+\sum_{\nu}Y_{{\bf k}r}^{\nu}\ a_{{\bf k}\nu}$, with the relation $|X_{{\bf k}r}|^2+\sum_{\nu}|Y_{{\bf k}r}^{\nu}|^2=1$, where the excitation and photon amplitudes of the upper $(+)$ and the lower $(-)$ polariton branches are $X_{{\bf k}\pm}=\pm\sqrt{\frac{\Delta_{\bf k}\mp\delta_{\bf k}}{2\Delta_{\bf k}}}$ and $Y_{{\bf k}\pm}^{\nu}=\frac{f_{\bf k}^{\nu}}{\sqrt{2\Delta_{\bf k}(\Delta_{\bf k}\mp\delta_{\bf k})}}$, and the excitation and photon amplitudes of the middle $(0)$ polariton branch are $X_{{\bf k}0}=0$ and  $Y_{{\bf k}0}^{\nu}=\frac{f_{\bf k}^{\nu}}{|f_{\bf k}|}$. For each photon polarization,  in the strong coupling regime the longitudinal component along the transition dipole coupled to the excitation contributes to  the upper and lower polariton branches. Instead the transverse component orthogonal to the dipole and decoupled from the excitation, creates the photonic middle branch.  For each ${\bf k}$, a direction exists where the polarization direction of a photon is orthogonal to the transition dipole, and hence no interaction between this photon and the material is obtained.

\begin{figure}
\begin{center}
\leavevmode
\includegraphics[width=65mm]{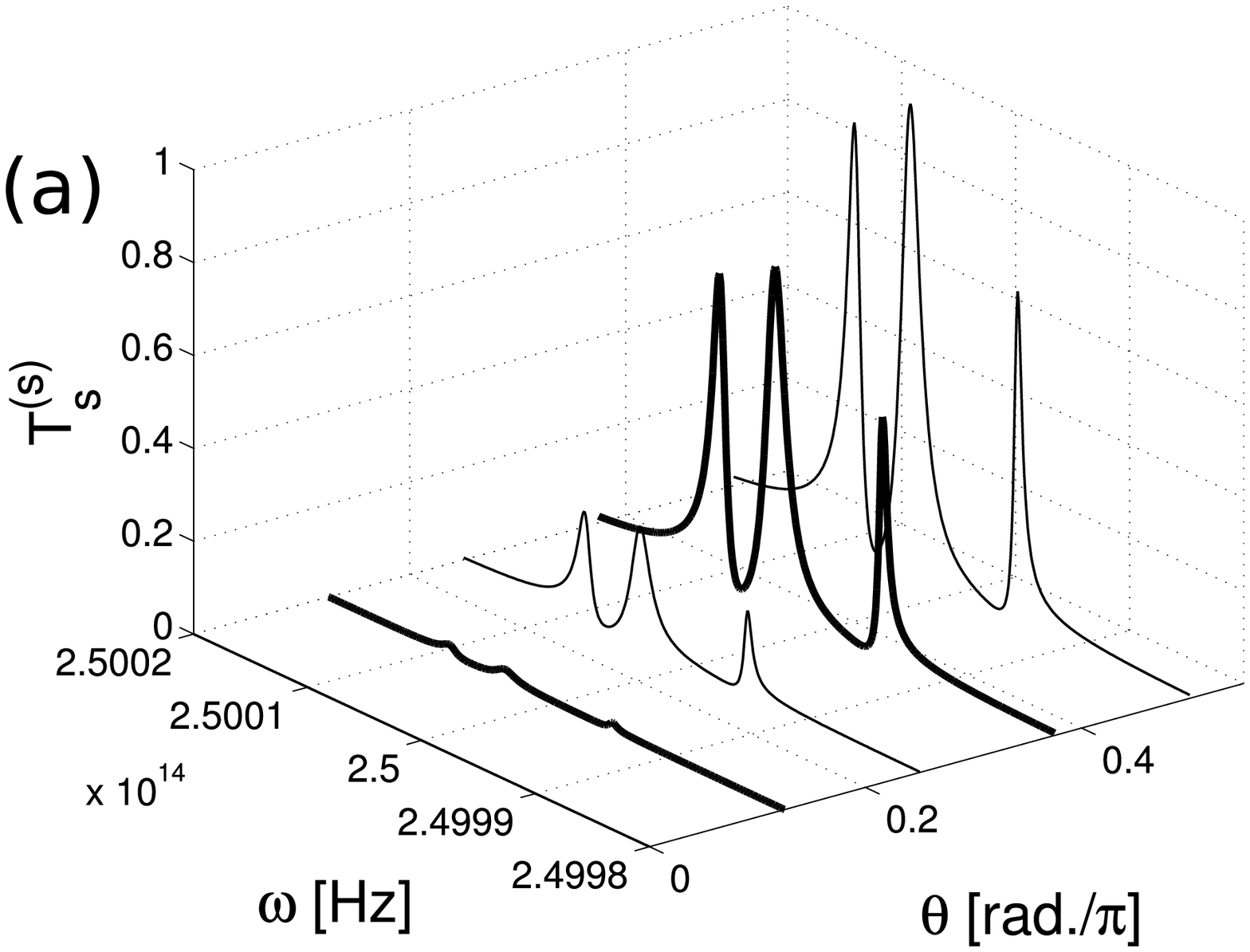}\ \ \ \includegraphics[width=65mm]{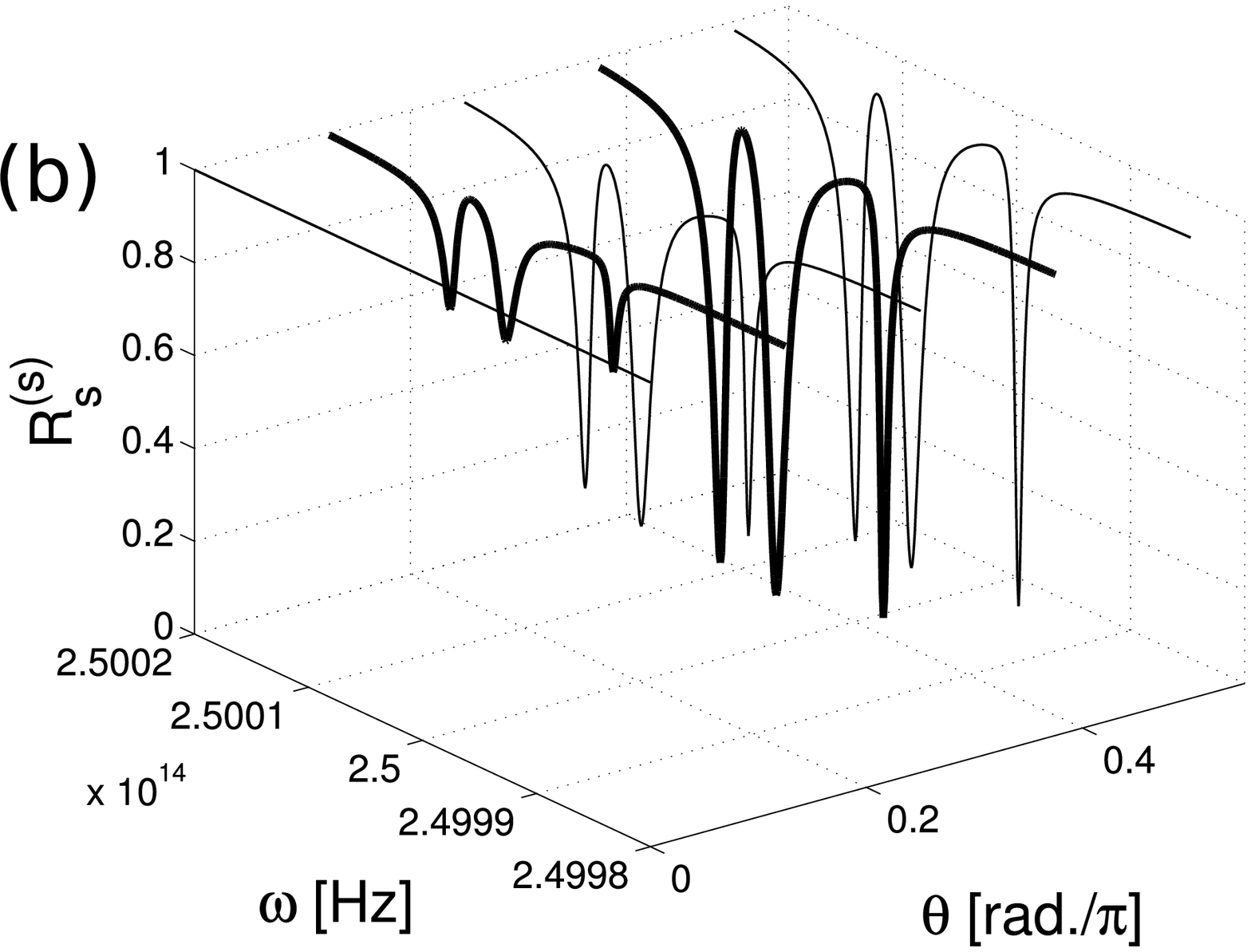}
\includegraphics[width=65mm]{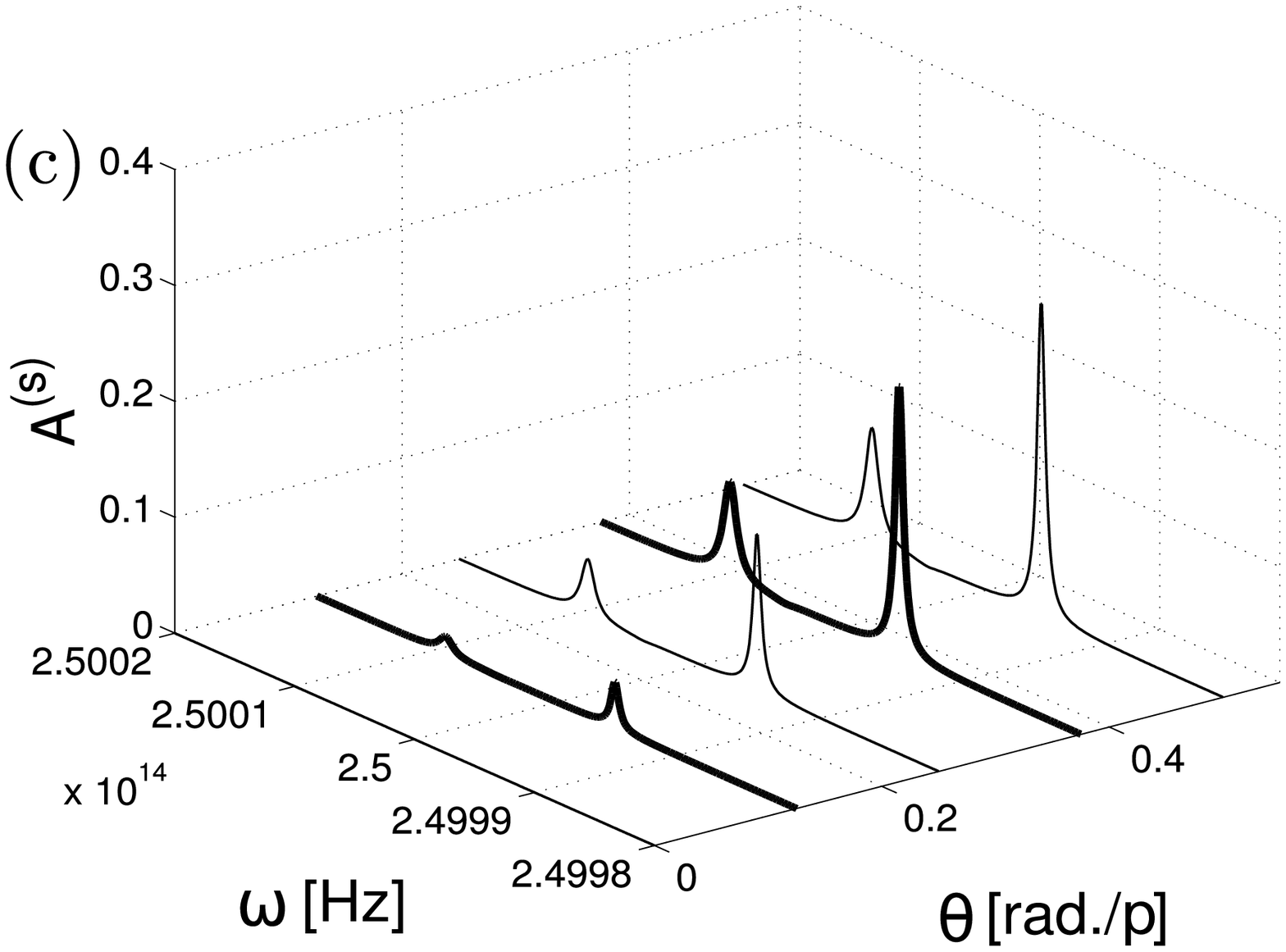}\ \ \ \includegraphics[width=65mm]{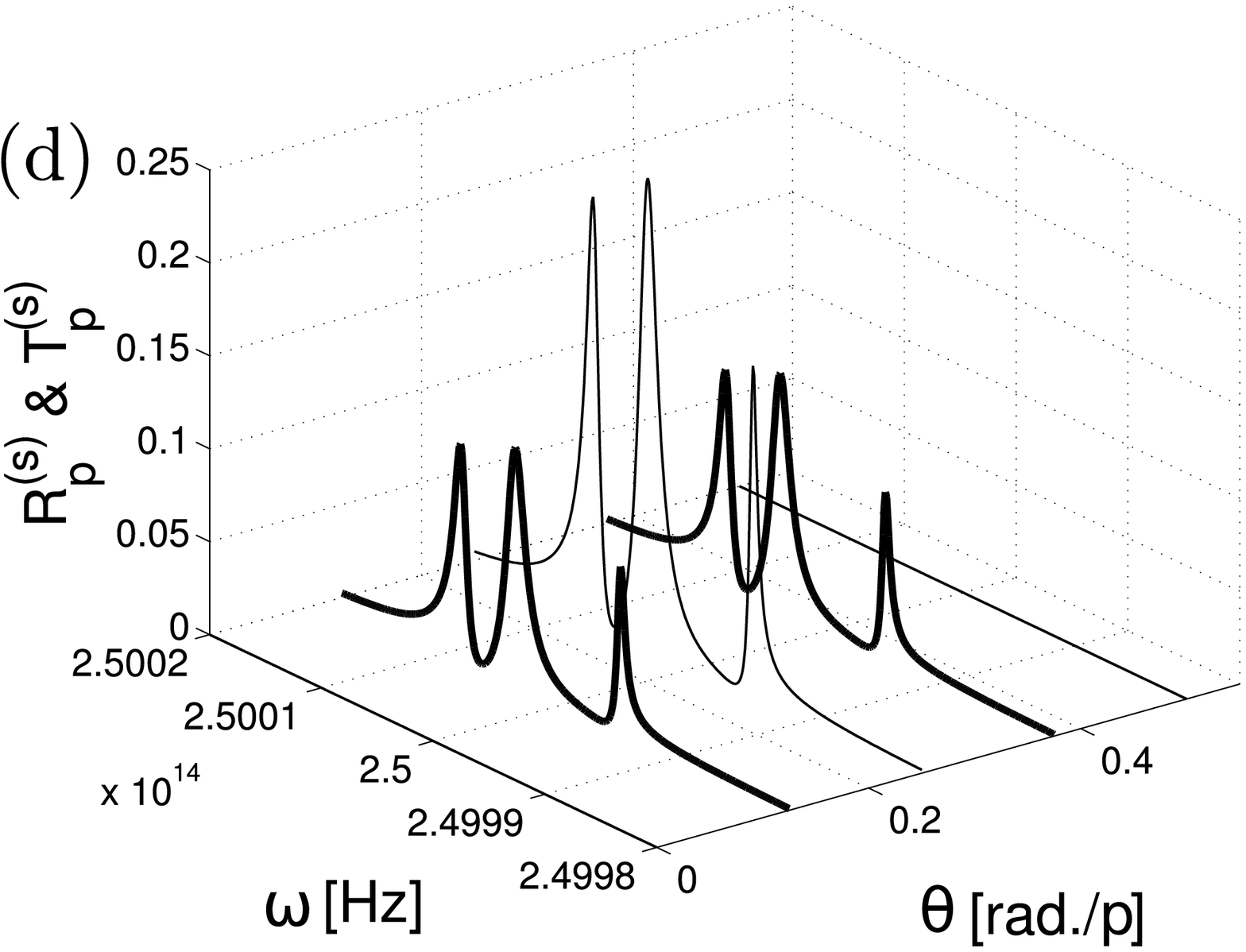}
\caption{Optical spectra vs the frequency in Hz for different angles, $\theta$, at $k=5\times10^{-7}\ \AA^{-1}$. In  (a) the $(s)$ polarized field transmission spectra, $T_s^{(s)}$ vs, of $(s)$ polarized incident field; in (b) the $(s)$ polarized field reflection spectra, $R_s^{(s)}$; in  (c) the absorption spectra, $A^{(s)}$;  in (d) the $(p)$ polarized field transmission and reflection spectra, $T_p^{(s)}$ and $R_p^{(s)}$. Figure adapted with permission from~\citet{Zoubi2009b} Copyright 2009 by Institute of Physics.}
\label{23}
\end{center}
\end{figure}

To observe the system eigenmodes we need to couple the internal system to the external world. Our observation tool is linear optical spectra. In assuming non-perfect mirrors, the internal system get coupled to their environment. For an incident external field with a fixed in-plane wave vector and polarization we calculate the transmission, reflection, and absorption spectra, as shown in Fig.~\ref{21}(a). The lifetime of the excited state is included phenomenologically. For a TE or $(s)$ polarized incident field, Fig.~\ref{23} report the transmission, reflection and absorption spectra of the $(s)$ polarized fields, $T_s^{(s)}$, $R_s^{(s)}$, and $A^{(s)}$, respectively,  as a function of frequency, $\omega\rightarrow\omega/2\pi$, at different angles $\theta$ and  $k=5\times10^{-7}\ \AA^{-1}$. For the excitation and photon damping rates we used $\gamma/2\pi= 10^9$ Hz and $\Gamma_{ex}/2\pi= 10^8$ Hz. For  $T_s^{(s)}$ and $R_s^{(s)}$ the three peaks and three dips correspond to the three polariton branches. At $\theta=0$ we get zero transmission and complete reflection. The largest transmission peaks and the deepest reflection dips are obtained at $\theta=\pi/2$. For the absorption spectra $A^{(s)}$ only two peaks are obtained correspond to the upper and lower branches, as the middle branch is pure photonic and no absorption take place, where the absorption is only for the polariton excitation part. Also here zero absorption at $\theta=0$, and maximum absorption at $\theta=\pi/2$, are obtained.  Fig.~\ref{23}(d) reports the transmission and reflection spectra of the $(p)$ polarized fields, $T_p^{(s)}$ and $R_p^{(s)}$, which are equal for identical cavity mirrors. Even though the incident field is $(s)$ polarized we get transmission and reflection of $(p)$ polarized fields, with maximum at $\theta=\pi/4$, and zeros at $\theta=0$ and $\theta=\pi/2$.

\subsection{Finite Atomic Chain in an Optical Cavity}

In Subsection 3.2 we studied collective electronic excitations in a finite linear atomic chain, and in Subsection 3.4 we derived their damping rate into free space radiation. Now we localize the finite chain in the middle between two identical spherical cavity mirrors such that the chain axis is perpendicular to the cavity axis \citep{Zoubi2009c}, as shown in Fig.~\ref{16}. Only the lowest Gaussian mode in the cavity is assumed to be close to resonance to the atomic transition. The electric field operator across the cavity waist, chosen along the optical lattice axis at $z=0$ and as a function of the distance from the cavity axis $r$, is given by \citep{Haroche2006}
\begin{equation}
\hat{\bf E}(r)=i\sqrt{\frac{E_c}{2\epsilon_0V}}\ e^{-r^2/w_0^2}\ \left\{{\bf e}\ a-{\bf e}^{\ast}\ a^{\dagger}\right\},
\end{equation}
where ${\bf e}$ is the photon polarization unit vector, $w_0$ the beam waist, and $V$ the mode volume given by $V=\pi w_0^2L_0/4$. Here $E_c$ is the cavity photon energy, and $L_0$ is the distance between the cavity mirrors.

The collective states are divided into two groups of symmetric and antisymmetric modes. The antisymmetric modes are found to be dark and decoupled from the cavity photons, while the symmetric ones are bright. For odd $k$-s, in the dipole interaction approximation, the coupling parameter is given by
\begin{equation}
f_k=-i\sqrt{\frac{E_c\mu^2}{\epsilon_0V(N+1)}}\ \cot\left(\frac{\pi k}{2(N+1)}\right).
\end{equation}
The lowest mode, $k=1$, the one without nodes, has the highest coupling parameter, that is nine times larger than the next bright mode of $k=3$. This mode dominates the optical properties of the system and can easily achieve the strong coupling regime within a cavity. It includes $81\%$ of the oscillator strength, and is termed superradiant mode. Hence, in the following we consider only the coupling of such superradiant mode to the cavity photons. The Hamiltoian of the coupled superradiant mode and the cavity mode is given by the Jaynes-Cumming model $H=E_{ex}\ B^{\dagger}B+E_c\ a^{\dagger}a+f\ B^{\dagger}a+f^{\ast}\ a^{\dagger}B$, where now the creation and annihilation operators $B^{\dagger}$ and $B$ stand for the superradiant mode respectively, and $a^{\dagger}$ and $a$ for the cavity photon. The coupling parameter is $f=-i\sqrt{\frac{E_c\mu^2}{\epsilon_0V(N+1)}}\ \cot\left(\frac{\pi}{2(N+1)}\right)$, and the superradiant mode energy is $E_{ex}=E_A+2J\ \cos\left(\frac{\pi}{N+1}\right)$. The Hamiltonian can be easily diagonalized using the transformation $A_{\pm}=X^{\pm}\ B+Y^{\pm}\ a$, to get $H=\sum_r E_p^r\ A_r^{\dagger}A_r$, with the two dressed eigenenergies $E_p^{\pm}=(E_c+E_{ex})/2\pm\Delta$, where $\Delta=\sqrt{\delta^2+|f|^2}$ and the detuning is $\delta=(E_c-E_{ex})/2$. The eigenstates are coherent superpositions of the superradiant mode and the cavity photon. The superradiant mode amplitudes are $X^{\pm}=\pm\sqrt{(\Delta\mp\delta)/2\Delta}$, and the cavity photon amplitudes are $Y^{\pm}=f/\sqrt{2\Delta(\Delta\mp\delta)}$. At the intersection point, with $\delta=0$, the dressed state is half a superradiant state and half s photon. But for large positive detuning the $(-)$ branch tends to the superradiant state and the $(+)$ branch becomes photon, and vice versa for the negative detuning.

\begin{figure}
\begin{center}
\leavevmode
\includegraphics[width=85mm]{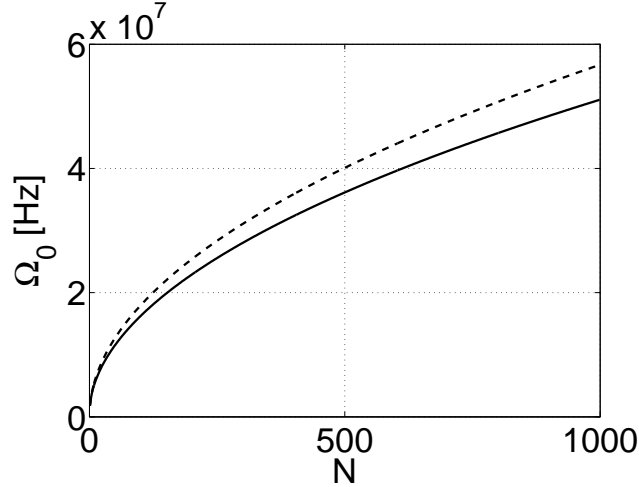}
\caption{The vacuum Rabi splitting frequency $\Omega_0$ vs. the atom number $N$, for interacting (full line), and non-interacting (dashed line), optical lattice ultracold atoms. The difference increases with the atom number. Figure reprinted with permission from~\citet{Zoubi2009c} Copyright 2009 by European Physical Society.}
\label{25}
\end{center}
\end{figure}

To emphasize the energy transfer influence within a cavity, we compare the results to those without dipole-dipole interactions. The electronic excitation is described now by the Hamiltonian $H_{ex} = \sum_nE_A\ B_n^{\dagger}B_n$. The excitation-photon coupling Hamiltonian reads $H_I=\sum_{n}\left(f\ B_n^{\dagger}a+f^{\ast}\ a^{\dagger}B_n\right)$, where the coupling parameter is $f=-i\sqrt{\frac{h\nu_c\mu^2}{2\epsilon_0V}}$. We define the collective excitation operator $B_n=\frac{1}{\sqrt{N}}\ B$ with $B=\frac{1}{\sqrt{N}}\sum_nB_n$, to get the Hamiltonian $H = E_A\ B^{\dagger}B+E_c\ a^{\dagger}a+\bar{f}\ B^{\dagger}a+\bar{f}^{\ast}\ a^{\dagger}B$, where $\bar{f}=-i\sqrt{\frac{h\nu_c\mu^2N}{2\epsilon_0V}}$. In Fig.~\ref{25} we compare the vacuum Rabi splittings, $\Omega_0=2|f|/h$, as a function of the atom number for the two cases, with and without the resonance dipole-dipole interaction. It appears that for a large atom number the vacuum Rabi splitting for independent atoms is larger than that of interacting atoms. We conclude that the vacuum Rabi splitting is reduced by the dipole-dipole interactions and deviates from the square-root behavior. This result can be used as a signature for the formation of collective electronic excitations in atomic chains.

\subsection{One-Dimensional Optical Lattice coupled to a Tapered Nanofiber}

\begin{figure}
\begin{center}
\leavevmode
\includegraphics[width=75mm]{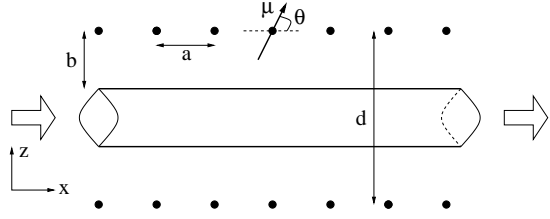}
\caption{Two one-dimensional optical lattices with one atom per site, lattice constant $a$ and distance $d$ between them, are  located parallel to a fiber, at the  two opposite sides. An incident field is injected from the left side of the fiber and the transmitted field is observed on the right side. The transition dipole $\mu$ makes an angle $\theta$ with the lattice direction.}
\label{26}
\end{center}
\end{figure}

Recently new directions have been opened for the light-matter coupling in using optical fibers \citep{Nayak2007}. In tapered optical fibers the atoms are trapped outside the fiber and couple to the evanescent field surrounding the fiber. The guided modes of ultrathin optical fibers exhibit pronounced evanescent field that give rise to an array of optical microtraps. The system was realized recently for cesium atoms localized in one dimensional optical lattice parallel to the nanofiber \citep{Vetsch2010,Goban2012}. We study here a simplified model related to this setup, based on a tapered optical nanofiber and two optical lattices \citep{Zoubi2010c}. We consider two identical parallel one-dimensional optical lattices with one atom per site, separated by a distance $d$ and  located at the two opposite sides of the fiber, at a distance $b$ from the fiber surface,  as schematized in Fig.~\ref{26}.

The electronic excitation Hamiltonian in the two optical lattices is given by the one-dimensional version of the planar Hamiltonian of Eq.~(\ref{TwoAtoms})
\begin{equation}
H_{ex}=\sum_{n,\alpha}E_A\ B_{n\alpha}^{\dagger}B_{n\alpha}+\sum_{nm,\alpha\beta}J_{nm}^{\alpha\beta}\ B_{n\alpha}^{\dagger}B_{m\beta}.
\end{equation}
Here $(n,m)$ run over all the lattice sites, and $(\alpha,\beta)$ stand for the two lattices $(1,2)$, with $B_{n\alpha}^{\dagger},\ B_{n\alpha}$ are the creation and annihilation operators of an electronic excitation at site $n$ in lattice $\alpha$, respectively. The coupling parameter $J_{nm}^{\alpha\beta}$ is for resonant dipole-dipole interactions, and gives rise to the excitation transfer among two lattice sites. The transfer term includes transfers among atoms in the same lattice, $J_{nm}^{11}$ and $J_{nm}^{22}$, and transfers among atoms at different lattices, $J_{nm}^{12}$ and $J_{nm}^{21}$. As before, the Hamiltonian can be diagonalized in two steps. First, in the lattice sites in using the transformation $B_{n\alpha}=\frac{1}{\sqrt{N}}\sum_{k}e^{ikx_n^{\alpha}}B_{k\alpha}$. Here $x_n^{\alpha}$ is the position of site $n$ in lattice $\alpha$. Second, we diagonalize the Hamiltonian relative to the two lattice indexes, $(\alpha,\beta)$, by applying the transformation $B_{k\nu}=\frac{B_{k1}\pm B_{k2}}{\sqrt{2}}$. The new states represent entangled states between excitons from the two lattices, which are symmetric and antisymmetric states. The symmetric state is denoted by $(\nu=s)$ and takes the plus sign $(+)$, the antisymmetric state is denoted by $(\nu=a)$ and takes the minus sign $(-)$. The Hamiltonian casts into the diagonal form $H_{ex}=\sum_{k,\nu}E_{ex}^{\nu}(k)\ B_{k\nu}^{\dagger}B_{k\nu}$, where the eigenenergies are $E_{ex}^s(k)=E_A+J(k)+J'(k)$ and $E_{ex}^a(k)=E_A+J(k)-J'(k)$. We define the exciton dynamical matrix by $J^{\alpha\beta}(k)=\sum_{L}e^{ikL}J^{\alpha\beta}(L)$, using $J_{nm}^{\alpha\beta}=J^{\alpha\beta}(L)$, with $L=x_m^{\beta}-x_n^{\alpha}$, where the dipole-dipole interaction is a function of the distance between the two atoms. For identical lattices, inside the same lattice we have $J(k)=J^{11}(k)=J^{22}(k)$, and among different lattices we have $J'(k)=J^{12}(k)=J^{21}(k)$. Excitons in one dimensional optical lattices are presented in Subsection 3.2, and the coherent transfer parameter of excitons among two parallel lattices is discussed in Subsection 3.6.

For the nanofiber photons we use a simplified picture. The one-dimensional optical fiber modes are given by the Hamiltonian $H_{ph}=\sum_k\hbar\omega_{ph}(k)\ a_k^{\dagger}a_k$. The photon dispersion is taken to be of the form $\omega_{ph}(k)=\frac{c}{\sqrt{\epsilon}}\sqrt{k_0^2+k^2}$, where the wave number $k_0$ results of the fiber transverse confinement, and $\epsilon$ is the fiber average dielectric constant. The electric field operator outside the fiber is
\begin{equation}
\hat{\bf E}({\bf r})=i\sum_k\sqrt{\frac{\hbar\omega_{ph}(k)}{2\epsilon_0 V}}\ {\bf e}\ u(r)\left\{a_k\ e^{ikz}-a_k^{\dagger}\ e^{-ikz}\right\},
\end{equation}
where ${\bf e}$ is the photon linear polarization unit vector, $V$ the normalization volume, and $u(r)$ the mode function which includes the fiber field complexity. The exciton-photon coupling Hamiltonian is given by $H_{in}=\sum_{k}\{\hbar f_k\ a_kB_{ks}^{\dagger}+\hbar f_k^{\ast}\ a_k^{\dagger}B_{ks}\}$, with the coupling parameter $\hbar f_k=-iu(b)\sqrt{\frac{\hbar\omega_{ph}(k)\mu^2}{\epsilon_0 Sa}}$. Only the symmetric excitons are coupled to the fiber photons, where the antisymmetric ones are dark and decoupled from the photons. The total Hamiltonian reads
\begin{eqnarray}
H&=&\sum_{k}\hbar\left\{\omega_{ex}^a(k)\ B_{ka}^{\dagger}B_{ka}+\omega_{ex}^s(k)\ B_{ks}^{\dagger}B_{ks}+\omega_{ph}(k)\ a_k^{\dagger}a_k\right. \nonumber \\
&+&\left.f_k\ a_kB_{ks}^{\dagger}+f_k^{\ast}\ a_k^{\dagger}B_{ks}\right\},
\end{eqnarray}
where due to translational symmetry along the lattice and fiber axis, the Hamiltonian is separated for each $k$. In the strong coupling regime we define the fiber polaritons by diagonalizing the Hamiltonian to get
\begin{equation}
H=\sum_{kr}\hbar\omega^r_{pol}(k)\ A_k^{r\dagger}A^r_k+\sum_{k}\hbar\omega_{ex}^a(k)\ B_{ka}^{\dagger}B_{ka}.
\end{equation}
We can use the polaritons definition of Subsection 4.2 in replacing the exciton and photon dispersions with the present one dimensional ones.

\begin{figure}
\begin{center}
\leavevmode
\includegraphics[width=65mm]{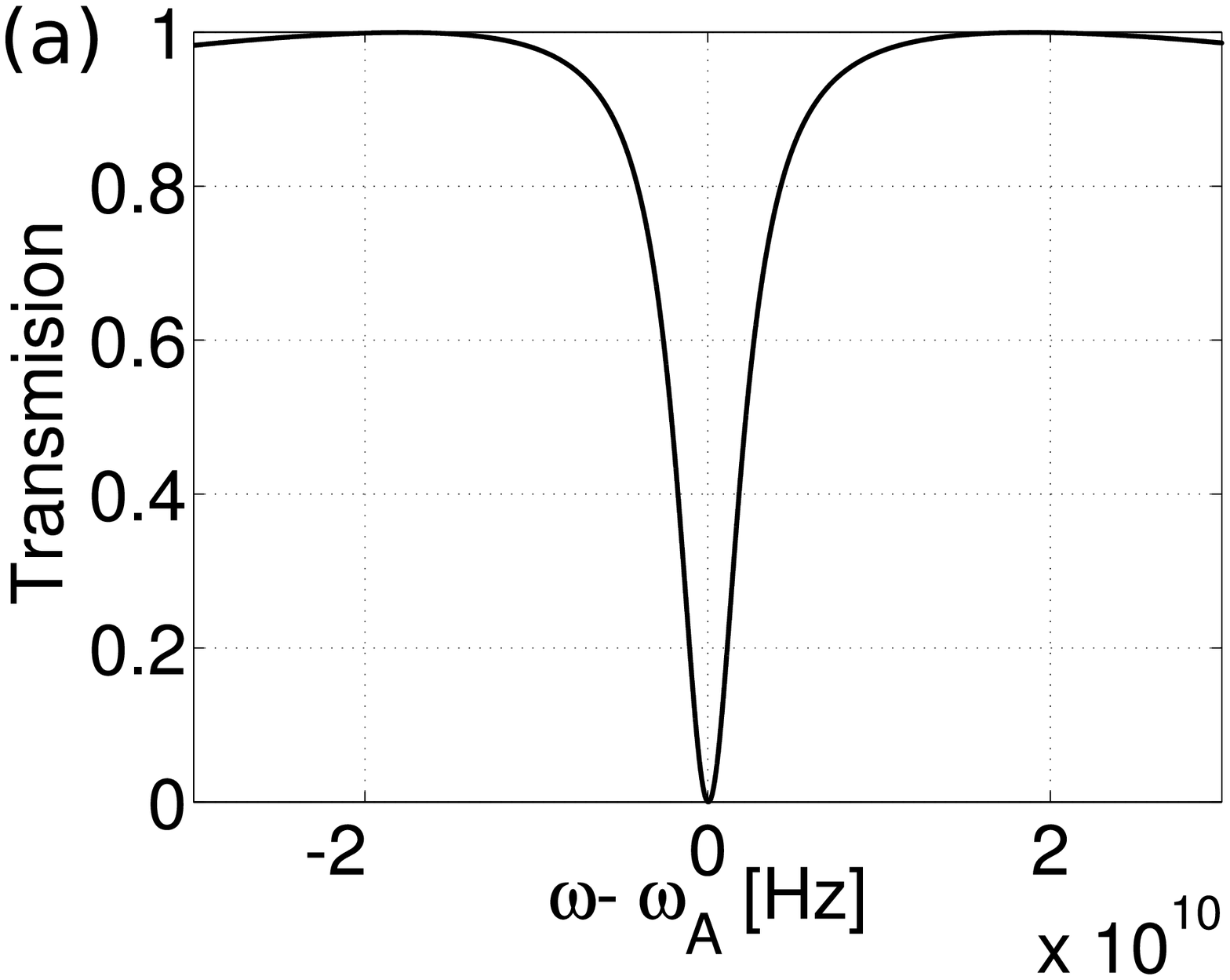}\ \ \ \includegraphics[width=65mm]{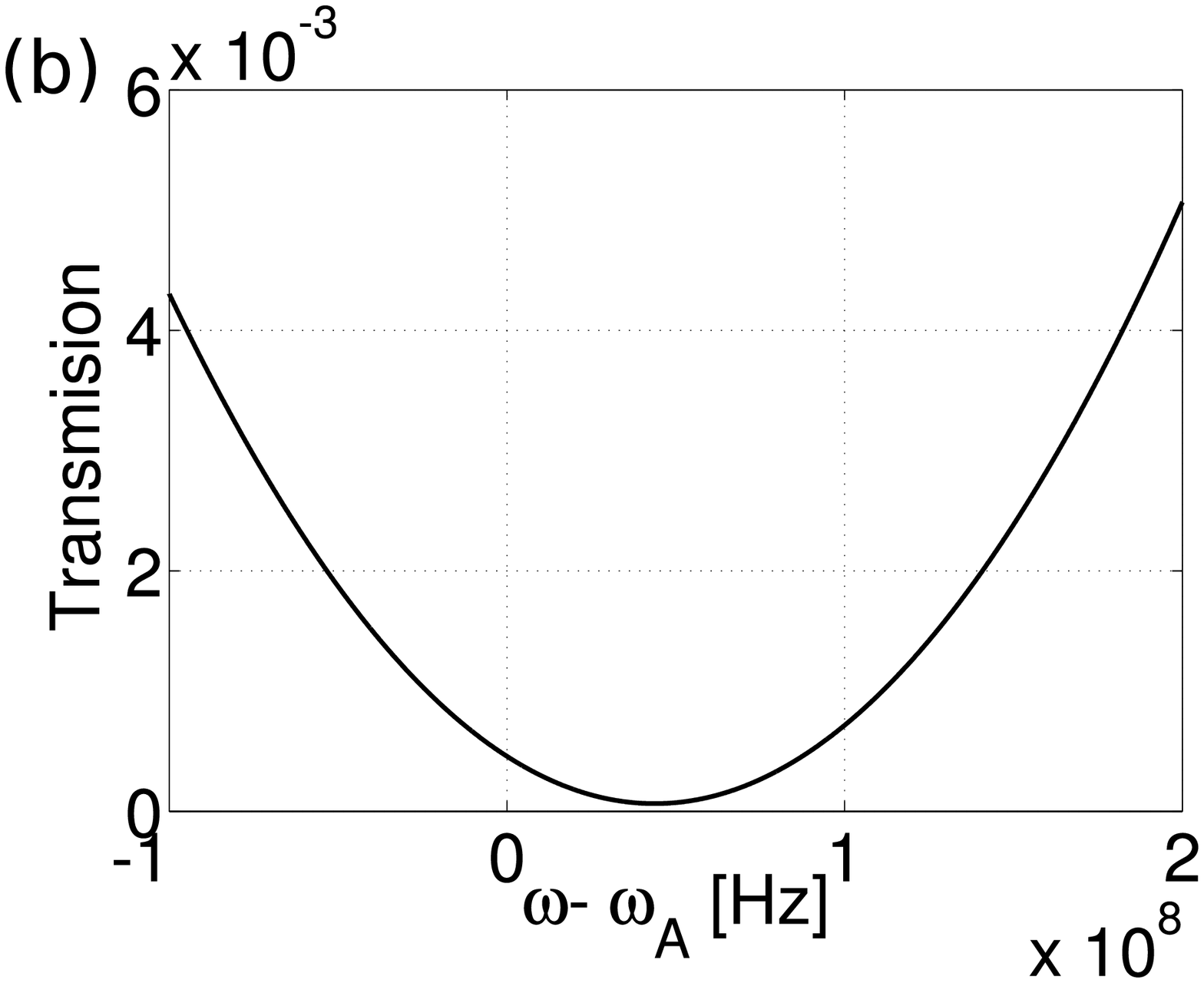}
\caption{(a) The transmission spectrum, for $k=10^{-6}\ \AA^{-1}$, and $\hbar\gamma=10^{-4}\ eV$. (b) A close look around the minimum. Figure reprinted with permission from~\citet{Zoubi2010c} Copyright 2010 by Institute of Physics.}
\label{27}
\end{center}
\end{figure}

Here we present the results for a system with the following parameters: lattice constant $a=1000\ \AA$; transition energy $E_A=1\ eV$; transition dipole $\mu=1\ e\AA$ at an angle $\theta=90^o$ with the lattice axis; distance between the two optical lattices $d=10a$; fiber dielectric constant is $\epsilon=3$; mode cross area $S=4\pi a^2$, and mode function at the lattice position $u(b)=0.1$. The cavity mode energy at $k=0$ is taken to be in resonance with the free atom transition energy, that is $E_{ph}(k=0)=E_A$, where $k_0=E_A\sqrt{\epsilon}/\hbar c$. The polariton dispersions are similar to those plotted in Figs.~\ref{19}. The exciton-photon intersection point is here to at $k=0$, where the polaritons are split by the Rabi splitting. To get the linear optical spectra we consider an incident field from the far left side of the fiber, and we calculate the transmission spectra. The coupling of the fiber field to the external field at the two far edges of the fiber is included in the parameter $\gamma$, and the fiber photon damping rate is taken to be $\Gamma_{ph}$ which is assumed to be $k$ independent. The symmetric exciton damping rate is included phenomenologically, and is taken to be $\Gamma_{ex}^s(k)$ as discussed in Subsection 3.4. The polariton damping rate is assumed to be $\Gamma_{pol}^{\pm}(k,\theta)=\Gamma_{ex}^s(k)\ |X^{\pm}(k)|^2+\Gamma_{ph}\ |Y^{\pm}(k)|^2$;  a small damping rate of $\hbar\Gamma_{ph}=10^{-10}\ eV$ is introduced for the photon, and the fiber edge coupling parameter is assumed to produce a bandwidth  $\hbar\gamma=10^{-4}\ eV$. Fig.~\ref{27}(a) reports the transmission spectra for a wavenumber $k=10^{-6}\ \AA^{-1}$, where at resonance we get  a dip in the transmission spectrum. Fig.~\ref{27}(b) representing a close look at the minimum shows a small shift relative to the free atom transition, produced by  the dipole-dipole interaction. For this case the symmetric exciton line width is $\hbar\Gamma_{ex}^s=2.32\times 10^{-8}\ eV$, larger than the free atom line width of $\hbar\Gamma_A=2.5\times 10^{-9}\ eV$.

\section{Optical Lattices with Defects: beyond the Mott Insulator State}

Up to this point we treated ideal optical lattices in the Mott insulator phase with a fixed number of atoms per site and in which all atoms are localized in the lowest Bloch band. In this section we partly relax this idealization and include different kinds of disorders into the system. In the following we introduce mainly two kinds of disorders. First, we exploit the excitation of on-site atoms into higher Bloch bands, where we concentrate in the lowest bands \citep{Zoubi2009a,Zoubi2010a}. Second, in the Mott insulator phase we consider localized defects that appear at several separated sites in the lattice \citep{Zoubi2008a}. We treat defects of vacancies distributed randomly in the lattice in the Mott insulator phase with one atom per site.

\subsection{Collective States of Atoms excited into higher Bloch Bands}

For deep optical lattices in the Mott insulator the lowest Bloch bands are well separated and can be represented by vibrational modes of a quantum harmonic oscillator. The transition of atoms between different vibrational states can be induced, e.g., by inelastic Raman scattering of light, or by on-site atom-atom interactions and atom hopping among nearest neighbor sites. We present an alternative mechanism for the excitation and de-excitation of atoms between different Bloch bands through the coupling of the internal electronic excitation to the external atomic motion via electrostatic interactions \citep{Zoubi2009a}. We consider here the Mott insulator with one atom per site where the atoms are treated as two-level systems. In general ground and excited atoms in optical lattices experience different optical lattice potentials, and can have minima at different positions, the fact that induces the coupling of electronic excitations to the vibrational modes. We show that the on-site electronic excitation is dressed by a cloud of virtual vibrational modes that renormalize the excitation energy. In the appropriate regime, the excitation transfers among nearest neighbor sites due to resonance dipole-dipole interactions and can be accompanied by the emission or absorption of vibrational modes.

We start from the electronic excitation Hamiltonian $H=\sum_i\hbar(\omega^e_i-\omega^g_i)\ B_i^{\dagger}B_i+\sum_{i,j}\hbar J_{ij}\ B_i^{\dagger}B_j$. The internal atomic transition frequency at site $i$ of position ${\bf n}_i$ is $\omega^e_i-\omega^g_i$, where we use the approximation $\omega^{\lambda}_i=\omega^{\lambda}+D^{\lambda}_{m}({\bf u}^{\lambda m}_i)$ for the ground and excited state energies $(\lambda=e,g)$. Here $D^{\lambda}_{m}({\bf u}^{\lambda m}_i)$ is the change in the local transition energy due to an on-site atomic vibration excitation with $(m>0)$ vibrational quanta, and $\omega^{\lambda}$ is vibration-free transition frequency with zero quanta $(m=0)$. We take ${\bf u}^{\lambda m}_i$ to be the average spatial growth of the atomic size at site $i$ in the $\lambda$ internal state due to the excitation of the atom to a higher vibrational state with $m$ quanta. $D^{\lambda}_{m}({\bf u}^{\lambda m}_i)$ is calculated relative to the local lattice ground state $(m=0)$, where $D^{\lambda}_{0}({\bf u}^{\lambda 0}_i)$ is included in $\omega^{\lambda}$, and ${\bf u}^{\lambda m}_i$ is measured relative to ${\bf u}^{\lambda 0}_i$ of the vibration-less state. The excitation can exchange among atoms at different sites $i$ and $j$ due to dipole-dipole interactions, and it parametrized by the coupling parameter $J_{ij}$. This coupling depends on the average atomic vibration shift ${\bf u}^{\lambda m}_i$ and thus on the local vibrational excitation. In a deep lattice, ${\bf u}^{\lambda m}_i$ is a small deviation relative to the lattice constant $a$ for the lowest vibrational modes. To lowest orders in this small perturbation we then can split the Hamiltonian into the form $H=H_{ex}+H_{vib}+H_{ex-vib}$. Here $H_{ex}$ is the internal excitation Hamiltonian, which is obtained for atoms in the ground vibrational states, $H_{vib}$ is the atom vibration Hamiltonian, for excited and ground state atoms, and the coupling Hamiltonian $H_{ex-vib}$ is derived perturbatively for atoms excited to higher vibrational states. To the zero order, for atoms in the ground vibrational states, we thus get $H_{ex}=\sum_i\hbar\omega_a\ B_i^{\dagger}B_i+\sum_{i,j}\hbar J^0_{ij}\ B_i^{\dagger}B_j$, where the transition frequency $\omega_a=\omega^e_i-\omega^g_i$ is the same at each site, and $J^0_{ij}$ is the transfer parameter among atoms in the lowest vibrational states. This is exactly the Hamiltonian we treat before in Subsection 3.1, and that results in excitons for optical lattices.

Vibrational excitations of atoms in the ground and excited state potentials are approximately described by the harmonic Hamiltonian
\begin{equation}
H_{vib}=\sum_i\hbar\omega_v^g\ b_i^{\dagger}b_i+\sum_i\hbar\omega_v^e\ c_i^{\dagger}c_i,
\end{equation}
where $\omega_v^g$ and $\omega_v^e$ are the vibration frequency for ground and excited state atoms, respectively. $b_i^{\dagger},\ b_i$ and $c_i^{\dagger},\ c_i$ are the creation and annihilation operators of a vibrational mode at site $i$ for ground and excited state atoms, respectively. The corresponding atomic displacement operators are
\begin{equation}
\hat{x}_i^g=\sqrt{\frac{\hbar}{2\bar{m}\omega_v^g}}\left(b_i+b_i^{\dagger}\right),\ \hat{x}_i^e=\sqrt{\frac{\hbar}{2\bar{m}\omega_v^e}}\left(c_i+c_i^{\dagger}\right),
\end{equation}
where $\bar{m}$ is the atomic mass. Here, ground and excited state atoms are considered as two different kinds of bosons, where each has its own optical lattice potential, as in Subsection 2.2. In the transition between the ground and the excited state one kind of bosons is destroyed and another created.

\begin{figure}
\begin{center}
\leavevmode
\includegraphics[width=80mm]{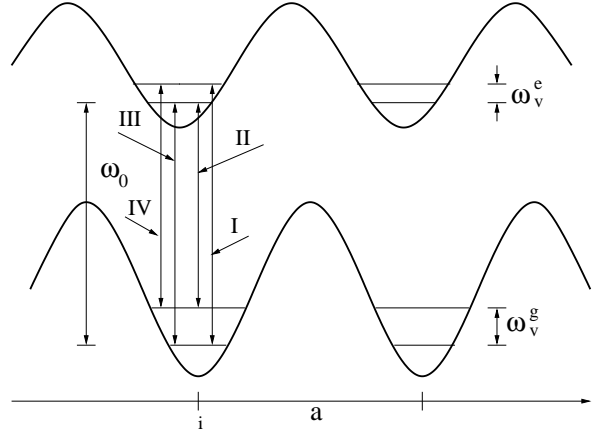}
\caption{The four possible transitions between the ground and excited electronic states including only single ground and excited vibrational states. Transition I includes a single excited vibrational quantum, and transition II includes a single ground vibrational quantum. Transition III is without any vibrational quanta, and transition IV includes ground and excited vibrational quanta.}
\label{28}
\end{center}
\end{figure}

To first order in the perturbation series with respect to ${\bf u}^{\lambda}_i$, we get excitation-vibration coupling by the Hamiltonian $H_{ex-vib}=H_{ex-vib}^I+H_{ex-vib}^{II}$, where $H_{ex-vib}^I$ is for the on-site part, and $H_{ex-vib}^{II}$ for the transfer part. Processes in the first order terms include only a single vibrational quanta. To treat processes of more than a single vibrational quanta one needs to consider higher order terms of the perturbation series. As we treat here only the first order terms we drop the index $(m)$ off $D^{\lambda}_{m}({\bf u}^{\lambda m}_i)$. The on-site part is given by
\begin{equation}
H_{ex-vib}^I=\sum_i\hbar\left\{M^e_i\ \left[B_i^{\dagger}\left(B_ic_i\right) +\left(B_i^{\dagger}c_i^{\dagger}\right)B_i\right]-M^g_i\ \left[\left(B_i^{\dagger}b_i\right)B_i+B_i^{\dagger}\left(B_ib_i^{\dagger}\right)\right]\right\},
\end{equation}
where the coupling parameter is
\begin{equation}
M^{\lambda}_i=\sqrt{\frac{\hbar}{2\bar{m}\omega_v^{\lambda}}}\left\{\frac{\partial D^{\lambda}({\bf u}^{\lambda}_i)}{\partial {\bf u}^{\lambda}_i}\right\}_{{\bf u}^{\lambda}_i=0},
\end{equation}
related to the slope of $D^{\lambda}({\bf u}^{\lambda}_i)$. As $D^{\lambda}({\bf u}^{\lambda}_i)$ is a function of the atomic wave function at site $i$ and in a fixed Bloch band, the parameter $M^{\lambda}_i$ is related to the correlations between atomic wave functions at the initial and the final Bloch bands. The four possible transitions are presented in Fig.~\ref{28}. The transfer part is given by
\begin{equation}
H_{ex-vib}^{II}=\sum_{i,j}\hbar\left[F^{ei}_{ij}\ c_i^{\dagger}+F^{gi}_{ij}\ b_i+F^{ej}_{ij}\ c_j+F^{gj}_{ij}\ b_j^{\dagger}\right]\ B_i^{\dagger}B_j,
\end{equation}
where the coupling parameter is
\begin{equation} \label{TranCoup}
F^{\lambda i}_{ij}=\sqrt{\frac{\hbar}{2\bar{m}\omega_v^{\lambda}}}\left\{\frac{\partial J_{ij}}{\partial {\bf u}^{\lambda}_i}\right\}_{{\bf u}^{\lambda}_i=0},
\end{equation}
which is related to the derivative of $J_{ij}$.

\begin{figure}
\begin{center}
\leavevmode
\includegraphics[width=65mm]{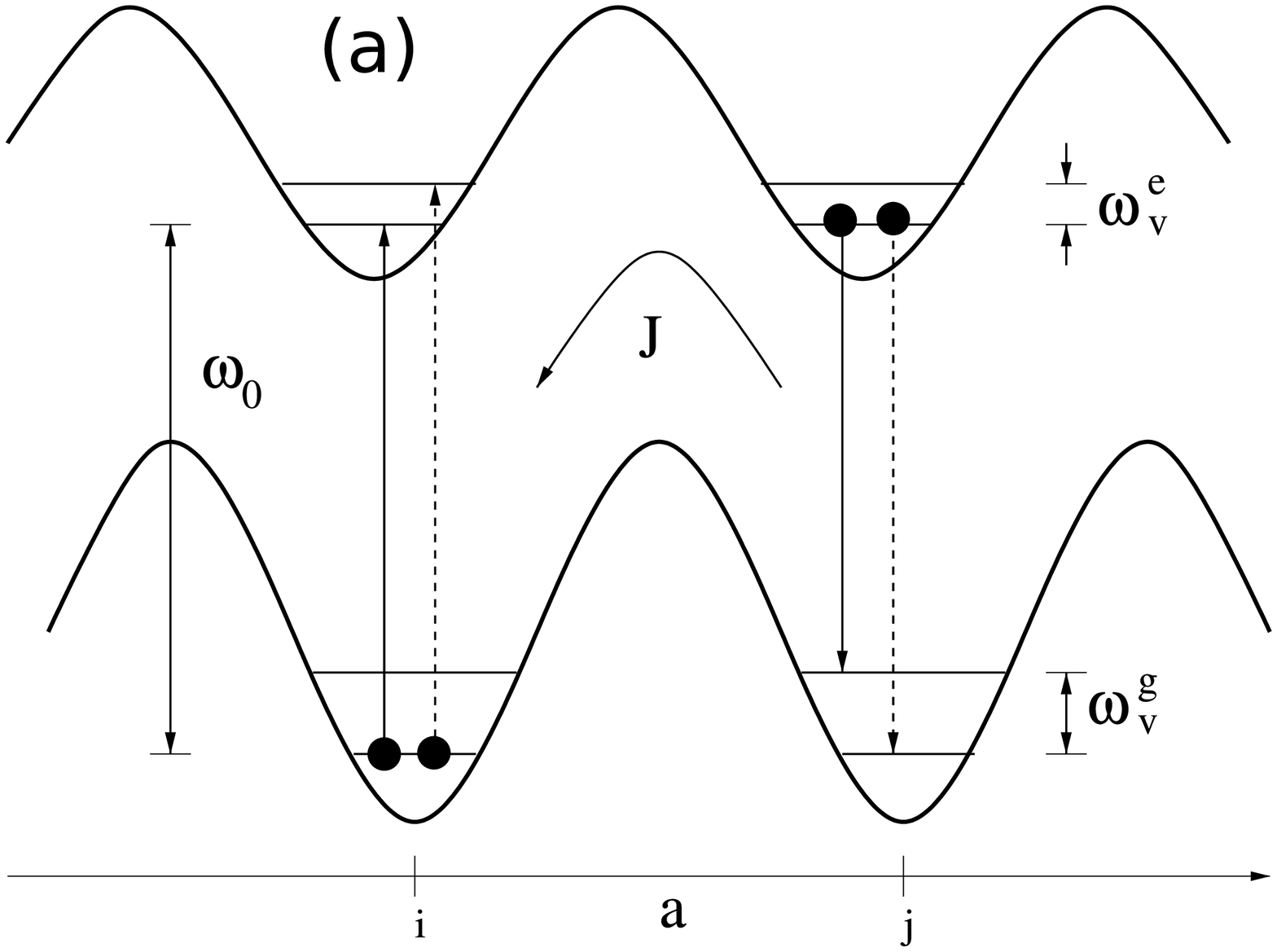}\ \ \ \includegraphics[width=65mm]{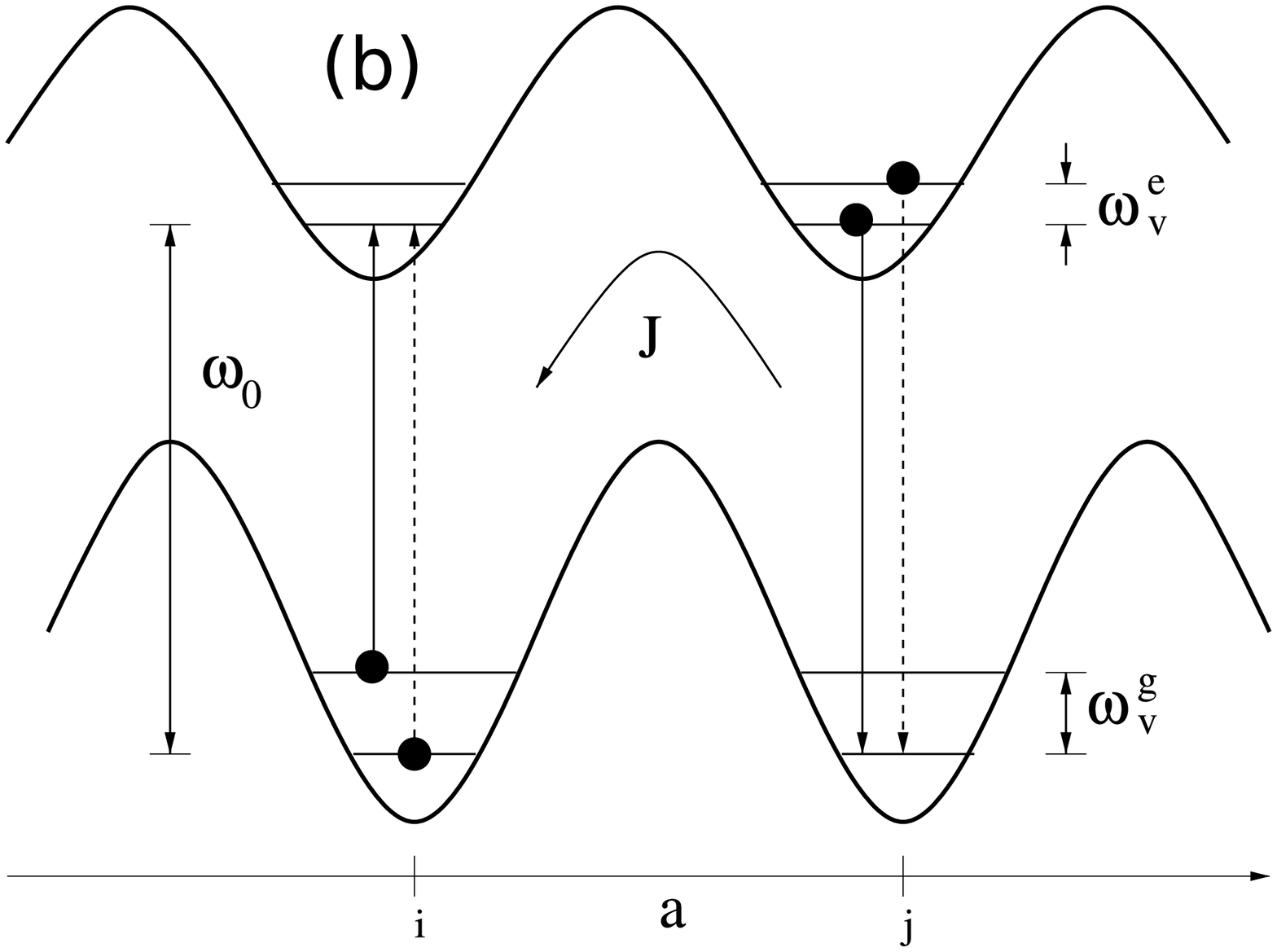}
\caption{(a) Process I (full-line) for $F^{g}_{ij}\ \tilde{B}_i^{\dagger}\tilde{B}_j\ \tilde{b}_j^{\dagger}$ with an energy transfer accompanied by the emission of ground state vibration at site $j$. Process II (dashed-line) for $F^{e}_{ij}\ \tilde{B}_i^{\dagger}\tilde{B}_j\ \tilde{c}_i^{\dagger}$ with an energy transfer accompanied by the emission of excited state vibration at site $i$. (b) Process III (full-line) for $F^{g}_{ij}\ \tilde{B}_i^{\dagger}\tilde{B}_j\ \tilde{b}_i$ with an energy transfer accompanied by the absorption of ground state vibration at site $i$. Process IV (dashed-line) for $F^{e}_{ij}\ \tilde{B}_i^{\dagger}\tilde{B}_j\ \tilde{c}_j$ with an energy transfer accompanied by the absorption of excited state vibration at site $j$.}
\label{29}
\end{center}
\end{figure}

In the limit of on-site excitation-vibration coupling which is stronger than the transfer coupling, namely in the limit of $M_i^{\lambda}\gg F^{\lambda i}_{ij}$, we can apply the following canonical transformation \citep{Mahan1990} of $\tilde{O}=e^{\hat{\sigma}}{O}e^{-\hat{\sigma}}$ and $\hat{\sigma}=\hat{s}\ B_i^{\dagger}B_i$, with $\hat{s}=\left[\frac{M_i^g}{\omega_v^g}\left(b_i^{\dagger}-b_i\right)-\frac{M_i^e}{\omega_v^e}\left(c_i^{\dagger}-c_i\right)\right]$, to get $\tilde{b}_i=b_i-\frac{M_i^g}{\omega_v^g}\ B_i^{\dagger}B_i$ and $\tilde{c}_i=c_i+\frac{M_i^e}{\omega_v^e}\ B_i^{\dagger}B_i$, and with $\tilde{B}_i=B_i\hat{X}\ ,\ \hat{X}=e^{-\hat{s}}$. The new effective local excitation represents an electronic excitation dressed by a cloud of on-site virtual vibrations and can be considered as an excitation-polaron. Namely, any electronic transition is accompanied by excitation and annihilation of ground and excited state vibrational modes. Here we assume the excitation-vibration coupling parameters to be site independent, by defining $M_{\lambda}\equiv M_i^{\lambda}$ and $F^{\lambda}_{ij}\equiv F^{\lambda i}_{ij}$. In terms of the new operators, the excitation Hamiltonian reads $H_{ex}=\sum_i\hbar\omega_0\ \tilde{B}_i^{\dagger}\tilde{B}_i+\sum_{i,j}\hbar J^0_{ij}\ \tilde{B}_i^{\dagger}\tilde{B}_j$, where $\omega_0=\omega_a-\Delta$, with $\Delta=\frac{M^{g\ 2}}{\omega_v^g}+\frac{M^{e\ 2}}{\omega_v^e}$. In this limit the effect of the on-site excitation-vibration coupling is a simple renormalization  of the electronic excitation energy by a shift of $\Delta$, where $\omega_a\gg \Delta$. This shift might be quite small but it needs to be considered for precision measurements as in a lattice clock transition energy \citep{Takamoto2005}. The vibration Hamiltonian is now given by $H_{vib}=\sum_i\hbar\omega_v^g\ \tilde{b}_i^{\dagger}\tilde{b}_i+\sum_i\hbar\omega_v^e\ \tilde{c}_i^{\dagger}\tilde{c}_i$. The excitation-vibration coupling Hamiltonian of the transfer part is written as $H_{ex-vib}=\sum_{i,j}\hbar\left[F^{e}_{ij}\left(\tilde{c}_i^{\dagger}+\tilde{c}_j\right)+F^{g}_{ij}\left(\tilde{b}_i+\tilde{b}_j^{\dagger}\right)\right]\ \tilde{B}_i^{\dagger}\tilde{B}_j$. This term describes vibration changes induced by excitation transfer among different sites. Four possible processes are plotted in Fig.~\ref{29}. They show the emission and absorption of a vibration due to the excitation transfer among nearest neighbor sites.

In terms of transformed operators the interaction Hamiltonian  can then be transformed into a momentum space representation in terms of exciton operators, with the exciton Hamiltonian $H_{ex}=\sum_{\bf k}\hbar\omega({\bf k})\ \tilde{B}_{\bf k}^{\dagger}\tilde{B}_{\bf k}$, for two dimensional optical lattices. The vibration operators, even though representing on-site localized vibrations, can be rewritten equally in the momentum space by applying $\tilde{b}_i=\frac{1}{\sqrt{N}}\sum_{\bf q}e^{i{\bf q}\cdot{\bf n}_i}\tilde{b}_{\bf q}$ and $\tilde{c}_i=\frac{1}{\sqrt{N}}\sum_{\bf q}e^{i{\bf q}\cdot{\bf n}_i}\tilde{c}_{\bf q}$, to get $H_{vib}=\sum_{\bf q}\hbar\omega_v^g\ \tilde{b}_{\bf q}^{\dagger}\tilde{b}_{\bf q}+\sum_{\bf q}\hbar\omega_v^e\ \tilde{c}_{\bf q}^{\dagger}\tilde{c}_{\bf q}$, which have flat dispersions, namely $q$-independent. The exciton-vibration coupling now reads
\begin{eqnarray}
H_{ex-vib}&=&\sum_{\bf k,q}\hbar\left\{F^{e}({\bf k+q})\ \tilde{c}_{\bf q}+F^{g}({\bf k})\ \tilde{b}_{\bf q}+F^{e}({\bf k})\ \tilde{c}_{\bf -q}^{\dagger}\right. \nonumber \\
&+&\left.F^{g}({\bf k+q})\ \tilde{b}_{\bf -q}^{\dagger}\right\}\ \tilde{B}_{\bf k+q}^{\dagger}\tilde{B}_{\bf k},
\end{eqnarray}
where $F^{\lambda}({\bf k})=\frac{1}{\sqrt{N}}\sum_{\bf L}F^{\lambda}({\bf L})e^{i{\bf k}\cdot{\bf L}}$. The Hamiltonian $H_{ex-vib}$ describes scattering of excitons between different wave vectors involving the emission and absorption of a ground or excited state vibrational quantum. Such a process is possible in the limit of $J\gg \omega_v^{\lambda}$, where the exciton band width is larger than the vibration energy. The scattering conserves energy and momentum, and as the vibration dispersion is a flat one, the vibration can absorb any amount of momentum from the exciton. The exciton transition rate within a spontaneous emission is $w^{\lambda}_{\bf k}\propto|\hbar F^{\lambda}({\bf k})|^2$, which results of the Fermi golden rule. The exciton-vibration interaction will serve as an important source for relaxation of excitons toward lower energies of the exciton band and therefore a thermal equilibrium in optical lattices.

The exciton-vibration coupling yields polariton-vibration coupling \citep{Zoubi2010a}. In using the inverse previous transformation $\tilde{B}_{\bf k}=\sum_rX_{k}^{r\ast}\ A_{{\bf k}r}$, we get the polariton-vibration interaction by
\begin{eqnarray}
H_{pol-vib}&=&\sum_{\bf k,q}\sum_{r,s}\hbar\left\{F^{e}({\bf k+q})\ c_{\bf q}+F^{g}({\bf k})\ b_{\bf q}+F^{e}({\bf k})\ c_{\bf -q}^{\dagger}\right. \nonumber \\
&+&\left.F^{g}({\bf k+q})\ b_{\bf -q}^{\dagger}\right\}\left(X_{k}^{r\ast}X_{(k+q)}^{s}\right)\ A_{({\bf k+q})s}^{\dagger}A_{{\bf k}r},
\end{eqnarray}
where the exciton amplitudes indicate that the interaction is due to the polariton excitonic parts. This interaction describes scattering of polaritons between states with different wavevectors by the emission or absorption of vibrational quanta. As the vibrations are dispersionless, one need to take care only of the energy conservation. A scattering takes place between different momentum states in the lower and the upper polariton branches. But if the vibration energy equals the splitting energy between the two branches, the polaritons can jump between the two branches by the emission or absorption of vibrations. The polariton-vibration interaction can represent a significant source for polaritons relaxation toward the minimum energy at $k=0$. Using the Fermi golden rule, the damping rate for the spontaneous emission of vibrations at zero temperature $T=0$ off polaritons at the lower branch is given by $w_{\bf k,q}^{--}\propto\left(\left|F^{e}({\bf k})\right|^2+\left|F^{g}({\bf k-q})\right|^2\right)\ \left|X_{k}^{-}\right|^2\left|X_{(k-q)}^{-}\right|^2$. The appearance of the exciton amplitudes indicate that the process is much more efficient in regions where the polaritons are more excitonic than photonic.

\subsection{Excitons and Polaritons scattering by Defects in the Mott Insulator}

The appearance of some defects is unavoidable due to imperfections in the dynamical formation of the Mott insulator phase, in particular there can be missing or extra atoms at some sites. We investigate here the effect of such defects on the dynamical properties of excitons and cavity polaritons \citep{Zoubi2008a}. We concentrate on the case of a low defect number, where the exciton and cavity polariton picture still holds and these quasi-particles are only scattered by single defects. Namely, the distance between each two defects is large enough for the formation of coherent excitons and cavity polaritons, which propagate as free quasi-particles between each two scattering processes. In principle the defects can move by hopping among the lattice sites, but the corresponding time scale is so long that they can be considered frozen. Here we concentrate on one type of defect which is induced by a missing atom, that in the case of one atom per site corresponds to the presence of vacant sites. In calculating the exciton and cavity polariton elastic scattering amplitude of such defects, we show that vacancies behave like hard sphere scatterers. We suggest cavity polaritons as a tool to detect such defects and show how the scattering can be controlled by changing the exciton-photon detuning.

\begin{figure}
\begin{center}
\leavevmode
\includegraphics[width=80mm]{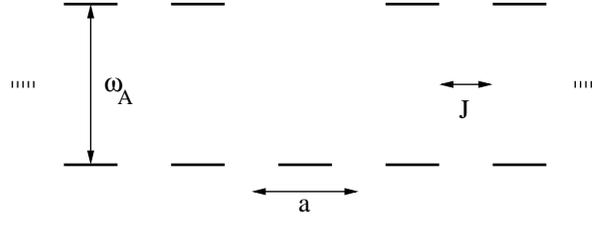}
\caption{Schematization of a vacancy in one dimensional optical lattice.}
\label{30}
\end{center}
\end{figure}

\begin{figure}
\begin{center}
\leavevmode
\includegraphics[width=65mm]{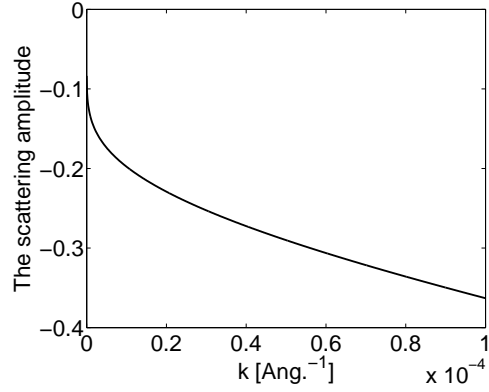}
\caption{The scattering amplitude vs. wave vector $k$, for zero detuning and a lattice constant $a=2000\ \AA$. Figure reprinted with permission from~\citet{Zoubi2008a} Copyright 2008 by Institute of Physics.}
\label{31}
\end{center}
\end{figure}

We treat a two dimensional optical lattice filled with one atom per site in the Mott-insulator phase. A single atom missing at the origin (${\bf r}_i={\bf 0}$) then creates an impurity in the artificial lattice of ultracold atoms, which for sufficient lattice depth stays localized and will not hop among the lattice sites as shown in Fig.~\ref{30}. We study the scattering of excitons off such a missing atom. The addition of  the impurity to the ideal atom crystal needs only a small change to the system Hamiltonian $H=H_0+V$ presented in Subsection 3.3, whose components now read $H_0=\sum_{\bf k}\hbar\omega_{ex}(k)\ B_{\bf k}^{\dagger}B_{\bf k}$ and $V=-\hbar\omega_A\ B_0^{\dagger}B_0$, where $H_0$ represents free excitons, and $\omega_{ex}(k)$ is the exciton dispersion with in-plane wave vector ${\bf k}$. $V$ is the impurity Hamiltonian at the origin. To get the above form, we add and subtract the impurity Hamiltonian $V$ to the whole Hamiltonian $H$, that is $H=H-V+V$, and then we define the ideal case Hamiltonian by $H_0=H-V$. For an exciton the impurity thus appears as a potential well located at the origin, with depth $\hbar\omega_A$ and radius $a$. We will not try to find the self-consistent eigenstates of $H$ and consider only the scattering problem. Because we can neglect trapping of an exciton in the impurity potential as a vacant site cannot absorb the trapping energy, we consider only the scattering process of an exciton off the impurity and  calculate its scattering amplitude. The initial exciton initially with wavevector ${\bf k}$ very far from the impurity, is scattered elastically into the  wavevector ${\bf k'}$ also very far, with $|{\bf k'}|=|{\bf k}|$. As the impurity is very deep with depth $E_A=\hbar\omega_A$,  in order to calculate the scattering amplitude we need to proceed beyond the Born approximation even though the perturbation is confined to a radius $a$. We apply the Schwinger-Lippmann equation \citep{Newton1982} in the case of small wave vector scattering excitons, that is $ka\ll 1$. For this case we can use the parabolic approximation for the dispersion of the excitons, $\omega_{ex}(k)=\omega_{ex}(0)+\frac{\hbar k^2}{2m_{ex}}$, with $m_{ex}$ the exciton effective mass, and for isotropic atoms in a square lattice of cubic symmetry as in Subsection 3.3  $m_{ex}=-\hbar/(2Ja^2)$. We obtain the scattering amplitude
\begin{equation}
f(k)=\frac{\frac{\pi E_A}{2\Delta_{ex}}}{1+\frac{\pi E_A}{2\Delta_{ex}}\left[\ln\left(\frac{ka}{\pi}\right)-i\frac{\pi}{2}\right]},
\end{equation}
where the effective exciton band width is defined by $\Delta_{ex}=\frac{\hbar^2\pi^2}{2m_{ex}a^2}$. The dipole-dipole interaction energy between different sites in the optical lattice is small, and hence the excitons have a small band width. In the present case we have the limit of $ E_A\gg\Delta_{ex}$ so that the scattering amplitude is $f(k)\approx\frac{1}{\ln\left(\frac{ka}{\pi}\right)}$, which exactly reproduces the result for the scattering off a hard disk of radius $a$ with the scattering cross section defined by $\sigma(k)=2\pi|f(k)|^2$. We thus conclude that an impurity generated by a missing atom in an optical lattice acts effectively like a hard disk of radius $a$.  Hence if a large number of incident excitons with identical wave vector are scattered elastically off the impurity, we get a ring of radius $k$ of scattered excitons. The scattering amplitude is plotted in Fig.~\ref{31} as a function of wave vectors, $k$, for lattice constant $a=2000\ \AA$. The singularity at $k=0$ is the two dimensional signature.

We now add cavity mirrors to our lattice as described in Subsection 4.2, where in the strong coupling regime, the system eigenstates are the cavity polaritons. As a polariton is a coherent superposition of an exciton and a photon it can be scattered off an impurity due its excitonic part. The corresponding Hamiltonian then reads $H_0=\sum_{{\bf k}r}\hbar\Omega_r(k)\ A_{{\bf k}r}^{\dagger}A_{{\bf k}r}$ and $V=-E_A\ B_0^{\dagger}B_0$, where the upper and lower polariton branches are presented explicitly in Subsection 4.2. We calculate the polariton scattering off such impurity. Again the scattering is elastic so that an incident polariton in branch $r$ with wave vector ${\bf k}$ will scatter into a polariton with wave vector ${\bf k'}$ in the same branch. We consider the scattering of small wave vector polaritons, that is in the limit $ka\ll1$ in the lower branch. In this limit the polariton dispersion can considered approximately parabolic with a polariton effective mass of $m_p$. This is of the order of the cavity photon effective mass $m_p\approx (\hbar\pi)/(cL)$. Hence, the lower polariton branch dispersion is taken to be $E_-(k)=E_-(0)+\frac{\hbar^2k^2}{2m_p}$, and we have in general $m_p\ll m_{ex}$. The scattered states will also have small wavevectors due to the energy conservation. As the upper branch has higher energies its contributions are negligibly small. In addition the upper branch for larger wavevectors is mainly photonic with small excitonic part,  and thus it contributes only weakly to the impurity scattering. We consider the case around zero detuning between the excitons and photons. Therefore the excitonic weight $|X_{k}^-|^2$ will change from half around zero wave vector up to one for large wave vectors, where the lower branch became excitonic. Because we neglect the contribution of the upper polariton branch, we drop the branch index, and all the parameters will be only for the lower branch. For simplicity, we use a model for the lower branch dispersion in place of the real one. The lower branch is divided into two parts: the first part between $0\leq k\leq k_0$ is taken to be of a parabolic dispersion with a polariton effective mass, where $E(k)=E(0)+(\hbar^2k^2)/(2m_p)$; and the second part between $k_0\leq k\leq \pi/a$, where $\pi/a$ is the Brillouin boundary, is taken to be  dispersionless with energy equal to the exciton energy at zero wave vector, with $k_0\ll\pi/a$. The intersection point between the two parts is fixed by $E_0=\hbar^2k_0^2/(2m_p)+E(0)$. The calculation yields the scattering amplitude
\begin{equation}
f(k)=\frac{X_k^2\left(\frac{\pi E_A}{2\Delta_p}\right)}{1-\left(\frac{\pi E_A}{4\Lambda_k}\right)},
\end{equation}
where we defined the polariton effective band width $\Delta_p=\frac{\hbar^2\pi^2}{2m_{p}a^2}$, and $\Lambda_k=E_0-E(k)$. As $E_A>\Lambda_k$, we have $(\pi E_A)/(4\Lambda_k)>1$, and the scattering amplitude is negative, that is $f(k)<0$. Thus the above impurity effective potential is repulsive for the polaritons. In the case of zero detuning, that is $\delta_0=0$, for small wavevectors $|X_k|^2= 1/2$, $\Lambda_k$ of the order of the exciton-photon coupling, $|g_k|$, where $E_A\gg\Lambda_k$, we get $f(k)\approx-\Lambda_k/\Delta_p$. This scattering amplitude equals that of the scattering from an effective potential of a square barrier potential of height $\Lambda_k$ and width $a$. For the case of negative detuning, in the limit of $E_A\gg\delta_0$, where $\Lambda_k$ is of the order of the exciton-photon detuning, the scattering amplitude is $f(k)\approx-(2X_k^2\Lambda_k)/\Delta_p$. This scattering amplitude equals that of the scattering from an effective potential of a square barrier potential of height $2X_k^2\Lambda_k$ and width $a$. Hence we find that the scattering amplitude can be controlled by changing the exciton-photon detuning.

\begin{figure}
\begin{center}
\leavevmode
\includegraphics[width=65mm]{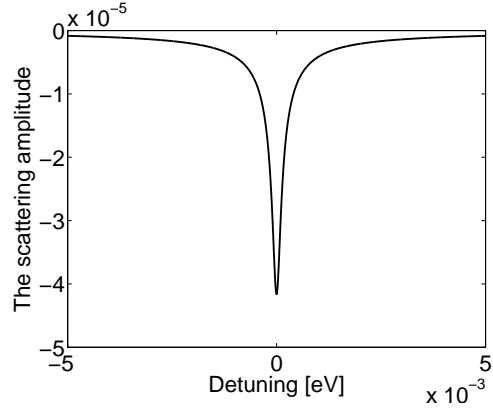}
\caption{The scattering amplitude vs. exciton-photon detuning, for $k\approx 0$ polaritons. Figure reprinted with permission from~\citet{Zoubi2008a} Copyright 2008 by Institute of Physics.}
\label{32}
\end{center}
\end{figure}

Using the following parameters: atomic energy $E_A=2\ eV$, exciton-photon coupling $\hbar|g|=0.0001\ eV$, lattice constant $a=2000\ \AA$, approximately normal incident waves with $k=10^{-6}\ \AA$, the scattering amplitude is calculated as a function of the exciton-photon detuning, as reported in Fig.~\ref{32}. It is clear that maximum scattering is obtained for zero detuning. As the detuning increases, either positive or negative, the $k\approx 0$ polaritons become more photonic and they are  much less scattered off the impurity. As the atomic energy $E_A$ is much larger than the exciton-photon coupling energy $\hbar g_k$, the scattering amplitude shows only a negligibly small asymmetry for negative and positive detunings, but we expect a clear resonance of the scattering amplitude around a resonant value $\delta_0$. Considering a stream of incident polaritons with the same wave vector ${\bf k}=k\hat{\bf k}$ we then should see a ring of radius $k$ of scattered polaritons. The effect of defects  on the dynamics of excitons in optical lattices can be generalized to the case of the Mott insulator phase with several atoms per site that contain defects of extra or less atoms at specific sites, as we presented for the case of two atoms per site with a defect of a singly occupied site. 

\section{Conclusions}

Excitons and polaritons in an optical lattice filled with ultracold atoms in the Mott insulator phase represent an idealized form of an optically excitable solid state crystal. Modeling by a two component Bose-Hubbard model allows us to derive the quantum phase diagram identifying the regions of Mott insulator phases. Excitons occur as natural elementary excitations in such a system including energy transfer induced by resonant dipole-dipole interaction. We extract the condition for the appearance of excitons in such artificial crystals by comparing between the transfer parameter and the excited atom decay. Excitons in the Mott insulator phase are collective delocalized electronic excitations in the lattice with their momentum distributed among many lattice sites. Propagating excitons are associated to large systems, while systems with boundaries require standing wave excitons. In general their damping rates are large and deviate from single atom decay, where excitons can be metastable or superradiant depending on their wave vectors and polarizations. Dark or metastable excitons open the door for storage and manipulation of photonic qubits with a wide range of applications for quantum information processing. Coherent transfer of excitons among different optical lattices differs from hopping of electronic excitations among two separated atoms. Here the transfer amplitudes decay exponentially with the inter-lattices distance.

Lattice excitons can be used as active medium for a cavity QED in the resonant regime. Cavity photons affect the boundaries between quantum phases and in these systems the Mott insulator phases appear only for deeper optical lattices. This gives new tools to observe and control the quantum phase transitions via the average number of cavity photons. In the strong coupling regime the excitons and photons are coherently coupled to form new quasi-particles termed polaritons. Optical linear spectra are presented as a nondestructive observation tool for the system different quantum phases. Systems of various dimensionality and geometries are investigated, ranging from a finite chain between spherical cavity mirrors up to a two-dimensional optical lattice between planar cavity mirrors. Moreover, the formation of molecules in optical lattices results in an anisotropic crystal that can be manifested through cavity a photon polarization mixing in the linear spectra. We presented also a set-up of two parallel one-dimensional optical lattices formed by an optical nanofiber and that provides a tool for a direct excitation of one-dimensional excitons through the coupling to fiber photons. Moreover the system gives rise to one-dimensional polaritons that dominate the optical properties in the strong coupling regime. We suggested a new mechanism for the excitation and de-excitation of atoms into higher Bloch bands through the energy transfer process, and we presented it as a significant source for the thermalization of excitons and polaritons. The elastic scattering of excitons and polaritons off vacancies are used for verifying the presence of various defects that can appear naturally in optical lattices.

The discussion in the present review though presented for optical lattice ultracold atoms can be adapted directly for various systems containing lattices of active materials. For example, one can imagine a lattice of semiconductor nano-structures, e.g., chains of quantum dots, arrays of color centers in solids, organic molecules ordered on a matrix, and Rydberg atoms in a lattice. The results can be extended into various directions, just to mention the implementation of excitons and polaritons for the physical realization of quantum information processing. The parameter controllability and the wide range of lifetimes of excitons make them a strong candidate as qubits. But efficient quantum processing implies entangled qubits, which are achievable via interaction among such qubits. Therefore our results should be extended to exploit exciton-exciton interactions through saturation and Coulomb forces, which also can find application for photonics using optical lattices. We expect this research to guide future experiments of electronically excited optical lattice ultracold atoms both in free apace and for cavity QED.

\section{Acknowledgments}
The work was supported by the Austrian Science Fund (FWF) through the Lise-Meitner Program (M977) and the Stand-Alone project (P21101), and by the DARPA QUASAR program.


\end{document}